\documentclass[prd,aps,a4paper,floatfix,amsmath,amssymb,twocolumn,nofootinbib]{revtex4-1}
\usepackage{amssymb,amsmath,amsthm,xcolor,graphicx}
\usepackage[margin=2cm]{geometry}
\usepackage[pdftex,breaklinks,colorlinks,
linkcolor=blue,
citecolor=teal,
anchorcolor=red,
urlcolor=cyan]{hyperref}
\usepackage{float}
\usepackage{orcidlink}  
\usepackage{multirow}
\usepackage{makecell}

\begin{document}


\title{Critical and extremal gravitational collapse of a
spherical charged scalar field in 4+1 spacetime dimensions}

\author{Satwik Mittal\orcidlink{0009-0001-2009-8805}} 
\email{S.Mittal@physik.lmu.de}
\affiliation{Universitäts-Sternwarte, Faculty of Physics, Ludwig-Maximilians-Universität München, Scheinerstr. 1, 81679
  Munich, Germany}
\author{Carsten Gundlach\orcidlink{0000-0001-9585-5375}}
\email{C.J.Gundlach@soton.ac.uk}
\author{Laetitia Martel\orcidlink{0009-0004-7949-1865}}
\email{L.Martel@soton.ac.uk}
\affiliation{Mathematical Sciences, University of Southampton,
  Southampton SO17 1BJ, United Kingdom} 
\date{15 September 2026}


\begin{abstract}
We numerically investigate the gravitational collapse of a charged
scalar field coupled to the Einstein and Maxwell equations in
spherical symmetry, in 4+1 spacetime dimensions. At the threshold of
collapse we find type-II critical phenomena very similar to what is
seen in 3+1 dimensions, including self-similarity of the critical
solution and power-law scaling of the black hole mass and charge. Well
inside the collapse region, we also identify initial data that
collapse to an extremal black hole. By varying all parameters of the
initial data by up to $\sim0.5\%$, we give evidence that there is a
small neighbourhood in the space of initial data where an exactly
extremal charged black hole is formed in the evolution of a
codimension-1 set of initial data. We believe this is the first such
evidence for scalar field collapse from regular initial data.
\end{abstract}


\maketitle

\tableofcontents


\section{Introduction}
\label{sec:introduction}


\subsection{Type-II critical collapse}


The work presented in~\cite{GundlachMartel26} established a new
formulation of the Einstein equations in single-null gauge for
investigating gravitational collapse, the eG formulation. The
Einstein-Maxwell-charged scalar field system in spherical symmetry in
3+1 dimensions was considered in that paper, but mainly as an example
for the formulation. The physics of charged collapse in 3+1 dimensions
will be examined in more detail elsewhere
\cite{MartelGundlachMittal26}. In the present paper, we use this
framework to extend the charged critical collapse programme to 4+1
dimensions, and to study extremal black hole formation from regular
initial data.

Type-II critical collapse in 4+1 dimensions has been studied in a
cohomogeneity-1 ansatz for gravitational waves in vacuum in
\cite{BizonChmajSchmidt05}, with an uncharged real scalar field in
spherical symmetry in \cite{BlandEtAl05}, and with both fields
competing in \cite{PortoGundlach2022}. These works confirmed that
there exists a 4+1 counterpart to the 3+1 Choptuik critical solution,
with different mass scaling exponent and echoing period. In 3+1
dimensions, the paper \cite{GundlachMartin96} showed from perturbation
theory that, when the system is enlarged to a charged scalar field,
the (real) Choptuik solution, up to a constant complex rotation,
remains the critical solution, and charge is a subdominant decaying
mode of the Choptuik solution, with Lyapunov exponent $\mu$, yielding
an independent power law for the black hole charge. The power-law
scaling of the charge was confirmed numerically
in~\cite{HodPiran97,Petryk05,GundlachMartel26}, and its fine-structure
was first computed in~\cite{GundlachMartel26}. Whether the same
qualitative picture holds in 4+1 is the first main question of this
paper.

For Type-II critical collapse, we extract the curvature scaling
exponent $\gamma$, black hole mass scaling exponent $\gamma_M$ and
echoing period $\Delta$ from the evolution of 1-parameter families of
initial data. To determine the charge scaling exponent
$\delta_Q$\footnote{Note that \cite{GundlachMartel26} uses the symbol
$\mu$ for the charge scaling exponent.} and the related Lyapunov
exponent $\mu$, we work in the small-imaginary-perturbation regime of
\cite{GundlachMartin96}, initialising the scalar field with a small
imaginary component so that the charge perturbation propagates
essentially linearly on the dominant Choptuik background. We then test
whether the scaling laws are universal by varying the coupling
constant $q$, the scalar field imaginary part $\chi$, and the family
of initial data, checking in each case whether the rescaled charge
fine structures collapse onto a common curve. We find that this
universality breaks down, for the black-hole charge only, in the
large-$q$ and large-$\chi$ regimes.


\subsection{Extremal (critical) collapse}


The formation of extremal black holes in gravitational collapse, and
in particular at the threshold of collapse, is an active area of
mathematical and numerical research in the usual 3+1 dimensions. Here
we explore this question numerically in 4+1 dimensions. In the
following literature review we use the notation of the present
paper. In particular, the parameter of any 1-parameter family of
initial data will be called $p$. Its value at the threshold of
collapse will be called $p_*$, but $p_1$ at an extremal threshold that
is not also the threshold of collapse.

Kehle and Unger~\cite{KehleUnger22} proved, for the spherical
Einstein-Maxwell-charged scalar field system in 3+1 dimensions, that the
surface gravity of a dynamically formed black hole can be driven
exactly to zero in finite advanced time, thereby violating the
classical third law of black hole mechanics.

In subsequent work~\cite{KehleUnger24}, they proved (in Thm.~1) the
existence of smooth 1-parameter families of solutions to the spherical
Einstein-Maxwell-charged {\em Vlasov} system, with asymptotically flat
initial data, which interpolate between dispersal and gravitational
collapse. As the critical parameter value is reached from the collapse
side, an extremal Reissner-Nordstr\"{o}m (from now, eRN) black hole
forms at late times, a phenomenon they term extremal critical
collapse.

The conjecture (in Conj.~3 and Rem.~1.12) that extremal critical
collapse is stable in the following sense: the moduli space of
solutions is (teleologically) foliated by $C^1$ hypersurfaces of
constant parameter $\sigma$, such that a black hole forms for
$|\sigma|\le 1$, with final charge-to-mass ratio $Q/{\cal M}=\sigma$,
while no black hole forms for $|\sigma|>1$.

Stability means that extremal critical collapse can be explored in
generic smooth 1-parameter families of initial data. Let the parameter
be $p$, with $Q/{\cal M}=1$ (say) achieved as $p\to p_*$ from above
(say). Then, as $\sigma$ and the family are both smooth, and assuming
the family crosses $\sigma=1$ transversally, we must have
$1-\sigma\sim p-p_*$ to leading order for $p\ge p_*$, so that
power-law behaviour of any quantity in $1-|\sigma|$ is equivalent to
power-law behaviour in $|p-p_*|$.

In Thm.~3, still of \cite{KehleUnger24}, Kehle and Unger prove the
existence of 1-parameter families with a jump between an extremal
black hole in the continuous limit $p\to p_1$ on one side, and a
subextremal black hole in the discontinuous limit from the other
side. A key qualitative feature is \textit{event horizon jumping at
  extremality}: as the tuning parameter $p$ which governs the initial
configuration crosses the critical value, the final mass ${\cal M}$,
the final charge $Q$, the area-radius of the event horizon $R$, and
its retarded time coordinate $u$ all jump. In particular, the final
charge-to-mass ratio jumps from $|Q|/{\cal M}=1$ to $|Q|/{\cal M}< 1$,
and $u_\text{EH}$ jumps to a larger value. The stability of this
behaviour, in the sense above, is conjectured (text after Thm.~3), and
it is called ``local critical behaviour''. A clearer term for Thm.~3
(plus the unnumbered conjecture) could be (codimension-1) extremal
collapse, in contrast to (stable) extremal critical collapse for
Thm.~1 (plus Conj.~3).

In contrast to these results for the Einstein-Maxwell-charged {\em
  Vlasov} system, this picture has not yet been proved, or
demonstrated numerically, for the spherical Einstein-Maxwell-charged
{\em scalar} field system {\em in genuine collapse}. The latter is
defined in Rem.~1.11 of \cite{KehleUnger22} as $\Sigma\in
\mathcal{J}^-(\mathcal{I}^+)$, that is, there is a Cauchy surface
$\Sigma$ that lies entirely outside the black hole. Here we extend
this useful terminology to our null cone setup to mean that initial
data for genuine collapse can be posed on an outgoing null cone that
emanates from a regular centre, extends to $\mathcal{I}^+$, and again
lies entirely outside the black hole.

Rather, work in the spherical Einstein-Maxwell-charged scalar field
system has so far focused on a setting where initial data are given on
the union of an ingoing and an outgoing null cone, with non-zero mass
and charge present already on the corner 2-sphere where the two
initial cones intersect. In this setting, the initial data consists of
the complex scalar field on the two null cones, and the values of the
local mass ${\cal M}$ and local charge $Q$ on the corner sphere. We
note already that the resulting spacetime cannot be extended to a
genuine collapse spacetime, but the formation of an event horizon in
this setting can serve as a local toy model for gravitational
collapse. We shall refer to it as the ``null rectangle'' setting.

The null rectangle setting was first explored numerically by Murata,
Reall, and Tanahashi~\cite{MurReaTan13} in the restriction to a {\em
  real} (uncharged) scalar field. The local charge function $Q=Q_0\ne
0$ is then constant everywhere. With the scalar field and $Q_0$ fixed,
they fine-tuned the corner mass ${\cal M}_0$ to a critical value,
where the spacetime settles locally to an eRN solution, and called
this a ``dynamical extremal black hole''. For ${\cal M}_0>{\cal M}_*$,
the event horizon is sub-extremal, while for {${\cal M}_0<{\cal
    M}_*$}, no horizon forms.  (As explained in \cite{MurReaTan13}
and reviewed below, this dichotomy follows from the fact that
the local charge function $Q(u,v)$ is constant for any uncharged
matter). On the subextremal side, the spacetime settles to a
non-extremal Reissner-Nordstr\"om (from now, RN) black hole over a
``decay (advanced) time'' that scales as $\Delta v \propto ({\cal M}_0
- {\cal M}_*)^{-1/2}$, and the surface gravity scales as $\kappa
\propto ({\cal M}_0 - {\cal M}_*)^{1/2}$. 

The same qualitative picture was subsequently observed in the
spherical Einstein-Maxwell-{\em charged} scalar field system by Gelles
and Pretorius~\cite{GellesPretorius26}. They set $Q_0 > M_0$ at the
corner, and fine-tune $Q_0$ toward a critical value $Q_*$ at which an
extremal horizon forms.  They track $v_\text{trap}$, defined as the
smallest advanced time at which a trapped surface is present. In fact,
the trapped region has the form of a cigar, bounded by two finite
values of $u$ and with $v_\text{trap}<v<\infty$. For $Q_0 < Q_*$ a
sub-extremal black hole forms and $v_\text{trap}$ is finite, while for
$Q_0 > Q_*$ no trapped region forms within the null rectangle
numerical domain. (The formation of a later trapped region is not
excluded a priori when $Q$ is not constant.) As $Q_0 \to Q_*$ from the
black-hole-forming side, the cigar-shaped trapped region recedes to
ever-later advanced times, with $v_\text{trap}\sim |Q_0 -
Q_*|^{-1/2}$.

The threshold phenomena in the null-rectangle setup were recently
proved for the spherical Einstein-Maxwell-{\em real uncharged} scalar
field system by Angelopoulos, Kehle, and
Unger~\cite{AngelopoulosKehleUnger26}, thus confirming the numerical
results of \cite{MurReaTan13}. The moduli space of solutions is
parameterised by ${\cal M}_0$, $Q_0$ and the real scalar field on the
two null cones, here taken to be in a weighted $C^2$ Banach
space. Again there is a teleologically defined function $\sigma$ on
the moduli space such that a black hole with $Q/{\cal M}=\sigma$ forms
for $|\sigma|\le 1$. As we will review in more detail below, on the
black-hole side $|\sigma|\to 1_-$ of the threshold, the surface
gravity $\kappa$, and the area and retarded time of the horizon minus
their threshold value, all scale as $(1-|\sigma|)^{1/2}$.

In this paper, we will show, we believe for the first time (in any
spacetime dimension), that extremal charged black holes can be created
in the {\em genuine} collapse of a spherical charged scalar field. We
will also give evidence that extremal black holes are codimension-1 in
a small neighbourhood of the solution space that we have explored,
that is, for what we have above provisionally named codimension-1
extremal collapse. In contrast to the mechanism for charged Vlasov set
out in \cite{KehleUnger24}, the local mechanism for reaching
extremality is the ``receding cigar'' of \cite{GellesPretorius26}
reviewed above.  On the other hand, in contrast to those papers, the
codimension-1 surface in the space of solutions where extremality
occurs is not the threshold of collapse but lies inside the collapse
region of the solution space. As the cigar disappears, the event
horizon jumps from it to a second trapped region that is always
present at larger $u$. In the process, $|Q|/{\cal M}$ of the black
hole jumps from extremal to sub-extremal. On the extremal side of the
jump, we find the scaling laws predicted in
\cite{AngelopoulosKehleUnger26}, as well as one observed in
\cite{GellesPretorius26}.

The remainder of this paper is organised as follows. In
Sec.~\ref{sec:fieldequations} we present the field equations in
covariant form and then in null coordinates in spherical
symmetry. Sec.~\ref{sec:algorithm} describes our numerical
scheme. Sec.~\ref{sec:typeII} presents our Type-II critical collapse
results, including the extraction of $\gamma$, $\Delta$, $\delta_Q$,
and $\mu$, and the universality tests. Sec.~\ref{sec:extremal} reports
the extremal black hole formation results, the evidence for
codimension-1 behaviour, our theoretical understanding of this, and
evidence for the near-extremal scaling laws. We conclude in
Sec.~\ref{sec:conclusions}. Appendixes give details of the code, the
4+1 RN solution, and an attempt at a derivation of the scaling of
$v_\text{trap}$.


\section{Field equations}
\label{sec:fieldequations}


\subsection{Field equations in covariant form}


The field equations in $4+1$ spacetime dimensions are very similar to
those in 3+1, given in \cite{GundlachMartel26}. We parameterize
possible conventions by the constants $c_J$, $c_M$, and $c_Q$, but we
always use gravitational units where $c=G_N=1$. In our numerical
examples, $c_J=c_M=c_Q=1$. In this convention, a RN black hole is
extremal if and only if $|Q|=\mathcal{M}$, see also
Appendix~\ref{RN}. The matter is a complex scalar $\phi=:\psi+i\chi$
field coupled to electromagnetism. The action is
\begin{equation}
S=\int \Biggl( {R\over {16\pi}} - {{1\over 2}}D_a\phi (D^a\phi)^* \\
 - {3\over {16\pi c_J}} F_{ab}F^{ab} \Biggr) \sqrt{-g}\,d^5x,
\end{equation}
where a star denotes the complex conjugate,
the field strength tensor $F$ in terms of the potential $A$ is
\begin{equation}
F_{ab}:=\nabla_aA_b-\nabla_bA_a,
\end{equation}
and we have introduced the charge-covariant derivative
\begin{equation}
D_a:=\nabla_a+iqA_a.
\end{equation}

$\phi$ obeys the wave equation
\begin{equation}
D^aD_a\phi=0.
\end{equation}
The Maxwell equations are
\begin{equation}
\label{divF}
{\nabla_bF^{ab}={4\pi \over 3}c_J j^a},
\end{equation}
where the charge current is
\begin{eqnarray}
j_a&=&-{iq\over 2}(D_a\phi)^*\phi+c.c. \\
&=&q(\chi\nabla_a\psi-\psi\nabla_a\chi)-q^2A_a(\psi^2+\chi^2).
\end{eqnarray}
It is conserved, $\nabla_a j^a=0$, because $F^{ab}$ is antisymmetric.
The Einstein equations are
\begin{equation}
R_{ab} - \frac{1}{2} R g_{ab}=8\pi T_{ab}, \\
\end{equation}
where the stress-energy tensor is 
\begin{equation}
\label{stressenergytensor}
\begin{split}
T_{ab}={} & D_{(a}\phi D_{b)}\phi^* -{{1\over
    2}}g_{ab}D^c\phi(D_c\phi)^* \\ & +{{3\over
    4\pi c_J}}\left[F_{ac}{F_b}^{c}-{1\over
    4}g_{ab}F_{cd}F^{cd}\right].
\end{split}
\end{equation}
An important distinction from the $3+1$ dimensional case is that the
electromagnetic part of the stress-energy tensor is not traceless in
$4+1$ dimensions. The equations admit the gauge freedom $\phi\to
e^{-iq\alpha}\phi$, $A_a\to A_a+\nabla_a \alpha$ for an arbitrary
scalar function $\alpha$, leaving $F_{ab}$, $j^a$ and $T_{ab}$
invariant.


\subsection{Field equations in spherical symmetry in null coordinates}


We now restrict to spherical symmetry and, following
  \cite{GundlachMartel26}, we introduce (single)
  null coordinates, where the line element takes the form 
\begin{equation}
ds^2=-2G\,du(dx+B\,du) +R^2\,d\Omega_3^2.
\end{equation}
Here $G$, $B$ and $R$ are functions of $(u,x)$, and $d\Omega_3^2$ is
the line element on the round unit 3-sphere. It can be written in
  coordinates, for example, as in \cite{BizonChmajSchmidt05}, namely
\begin{equation}
d\Omega_3^2:=\frac{1}{4}\Big( d\theta^2 +
d\varphi^2 - 2\sin\theta\, d\varphi\, d\vartheta + d\vartheta^2
\Big).
\end{equation}
Surfaces of constant $u$ are outgoing null
hypersurfaces. The tangent vector
to the affinely parameterized generators of these surfaces is given by
$U^a := -\nabla^a u=G^{-1}(\partial_x)^a$, corresponding to the
directional derivative
\begin{equation} \label{affine_parameter}
   U:=G^{-1}\partial_x=\frac{d}{d\lambda},
\end{equation}
where $\lambda$ is an affine parameter along these null hypersurfaces.
The radial coordinate $x$, which parametrizes the null cone
generators, will be fixed by a choice of the metric coefficient
$B$. We also define the ingoing null vector field (or derivative
operator)
\begin{equation}
\label{Xidef}
\Xi:=\partial_u-B\partial_x,
\end{equation}
which is normalised relatively to $U^a$ as
$U^a\Xi_a=-1$.

We use the same electromagnetic gauge as used in
\cite{GundlachMartel26}, that is
\begin{equation}
A_x(u,x)=0, \qquad A_u(u,0)=0.
\end{equation}
Note that $A_x=0$ can be written geometrically as $U^aA_a=0$. From now
on, we work exclusively in this gauge, and for conciseness we rename
$A_u$ to $A$.

For the scalar field, we also introduce the derivative operator
\begin{equation}
\hat\Xi\phi:=(\Xi+iqA)\phi,
\end{equation}
with real and imaginary parts
\begin{eqnarray}
\label{Xihatpsi}
\hat\Xi\psi&:=&{\rm Re}\,\hat\Xi\phi=\Xi\psi-qA\chi, \\
\label{Xihatchi}
\hat\Xi\chi&:=&{\rm Im}\,\hat\Xi\phi=\Xi\chi+qA\psi.
\end{eqnarray}

In terms of the local charge function
\begin{equation}
\label{Qdef}
Q:={R^3A_{,x}\over c_Q G},
\end{equation}
the Maxwell equations become
\begin{eqnarray}
\label{Qeqn}
Q_{,x}&=&{4 \pi c_J\over3c_Q}qR^3(\psi\chi_{,x}-\chi\psi_{,x}), \\
\label{XiQeqn}
\Xi Q&=&{4\pi c_J \over3c_Q}qR^3(\chi\hat\Xi\psi-\psi\hat\Xi\chi).
\end{eqnarray}
Hence we have a conserved charge current $\nabla_aj^a=0$ with
\begin{eqnarray}
\label{chargecurrent}
j^u&=&{3c_Q\over 4\pi c_J GR^3}Q_{,x}, \\
j^x&=&-{3c_Q \over 4\pi c_J GR^3}Q_{,u},
\end{eqnarray}
where an overall constant has been fixed from (\ref{divF}), and
we have used $\sqrt{-g}=\frac{1}{8}GR^3\cos\theta$.

We find $A$ by integrating (\ref{Qdef}), written as
\begin{equation}
\label{Aeqn}
A_{,x}={c_Q G Q\over R^3},
\end{equation}
starting from $A=0$ at $x=0$.

In $4+1$ dimensions, in our geometrized units ($c=G_N=1$), mass
carries dimensions of length squared. Consequently, in spherical
symmetry we define the Hawking mass in terms of the compactness
function, similar to \cite{PortoGundlach2022}, as
\begin{equation}
\label{Mdef}
M:={c_M CR^2 \over 2},
\end{equation}
where we define the Hawking compactness $C$ of a symmetry sphere
(equivalent to a point in the reduced spacetime) as
\begin{equation}
\label{Cdef}
C:=1-|\nabla R|^2.
\end{equation}
In our coordinates, $C$ is given by
\begin{equation}
\label{Cexpr}
C(u,x)=1+{2R_{,x}\Xi R\over G}.
\end{equation}

The derivatives of the Hawking (or Misner-Sharp) mass are
\begin{eqnarray}
\label{Mxexpr}
M_{,x}&=&-{8c_M \pi R^3 \Xi R\over 3G}
  (\psi_{,x}^2+\chi_{,x}^2) \nonumber \\ &&
  {}+{c_M c_Q^2 Q^2 R_{,x}\over c_J R^3}, \\
\label{XiMexpr}
\Xi M&=&-{8c_M \pi R^3 R_{,x}\over 3G}
  (\hat\Xi\psi^2+\hat\Xi\chi^2) \nonumber \\ &&
  {}+{c_M c_Q^2 Q^2\Xi R\over c_J R^3}.
\end{eqnarray}
Hence there is a conserved mass current analogous to
(\ref{chargecurrent}). $M$ obeys $M_{,x}\ge 0$ when the
  symmetry 3-spheres are neither trapped nor antitrapped, that is for
  $R_{,x}>0$ and $\Xi R<0$. Moreover, a surface is marginally trapped
  or marginally antitrapped if and only if $C=2M/c_M R^2=1$.

We also define the augmented mass
\begin{equation}
\label{calMdef}
{\cal M}:=M + \frac{c_Mc_Q^{2} Q^2}{2c_J R^{2}}.
\end{equation}
(Other authors call it the renormalised mass.) As $Q$ and $M$ are
scalars on the reduced spacetime, so is ${\cal M}$.  Its derivatives
are
\begin{eqnarray}
\label{calMxexpr}
{\cal M}_{,x}&=&-{8c_M \pi R^3 \Xi R\over 3G}
  (\psi_{,x}^2+\chi_{,x}^2)
  \nonumber \\ &&
  {}+{\tfrac{4}{3}\pi qc_Mc_Q}\,QR(\psi\chi_{,x}-\chi\psi_{,x}),
  \\
\Xi {\cal M}&=&-{8c_M\pi R^3 R_{,x}\over 3G}
  (\hat\Xi\psi^2+\hat\Xi\chi^2)
  \nonumber \\ &&
  {}+{\tfrac{4}{3}\pi qc_Mc_Q}\,QR(\chi\hat\Xi\psi-\psi\hat\Xi\chi).
\label{XicalMexpr}
\end{eqnarray}
Hence there is another conserved mass current analogous to
(\ref{chargecurrent}). 

As established in \cite{GundlachMartel26}, the electrovacuum regime is
characterized in our electromagnetic gauge $A_x=0$ by the vanishing of
the scalar field ($\phi=0$). Then both the charge $Q$ and the
augmented mass $\mathcal{M}$ are strictly constant, and the spacetime
is locally isomorphic to a member of the RN family (see
Appendix~\ref{RN}) with parameters $Q_0$ and $\mathcal{M}_0$. By
contrast, the Hawking mass $M$ is generically non-constant in vacuum
unless $Q=0$.

We note that $M$ can be computed algebraically from ${C}$ via
(\ref{Cexpr}), or equivalently by integrating (\ref{Mxexpr}) outward
from $M=0$ at $R=0$; we call this integrated version
$\tilde{M}$. Similarly, $\mathcal{M}$ can be obtained from $M$ via
(\ref{calMdef}) or by integrating (\ref{calMxexpr}) from $\mathcal{M}
= 0$ at $R=0$; we call that $\tilde{\mathcal{M}}$. In the continuum,
$M=\tilde{M}$ and $\mathcal{M}=\tilde{\mathcal{M}}$ by construction,
but at finite resolution the two representations differ: $\tilde{M}$
retains the non-decreasing property along outgoing null rays and
$\tilde{\mathcal{M}}$ retains the constant-in-electrovacuum property
at the discrete level, whereas $M$ and $\mathcal{M}$ do only up to
numerical error.

In spherical symmetry, only the components $E_{uu}$, $E_{ux}$,
$E_{xx}$ and $E_{\theta\theta}$ of the trace-reversed Einstein
equations $E_{ab}:= R_{ab}-8\pi(T_{ab}-(1/3)g_{ab}T)=0$ are
algebraically independent. We can write $E_{xx}=0$ as
\begin{equation}
\label{Reqn}
R_{,xx}-{G_{,x}\over G}R_{,x} +\frac{8\pi}{3}
R(\psi_{,x}^2+\chi_{,x}^2)=0,
\end{equation}
which can be solved as a second-order linear ODE for $R$, given $G$,
$\psi$ and $\chi$. More geometrically, $E_{xx}=0$ can be written as
\begin{equation}
\label{UUR}
UUR+\frac{8\pi}{3}[(U\psi)^2+(U\chi)^2]R=0,
\end{equation}
the Raychaudhuri equation for the generators of the coordinate null
cones.

We can write $E_{\theta\theta}=0$ and the real and imaginary parts of the complex
wave equation as
\begin{eqnarray}
\label{XiReqn}
(R^2\Xi R)_{,x} &=& -GR\left(1 - \frac{ c_Q^2 Q^2}{c_J R^4}\right), \\
\label{Xipsieqn}
(R^{3/2}\hat{\Xi}\psi)_{,x} &=&
  -\frac{Gqc_Q Q\chi}{2R^{3/2}}
-\tfrac{3}{2}(\Xi R)R^{1/2}\psi_{,x}, \\
\label{Xichieqn}
(R^{3/2}\hat{\Xi}\chi)_{,x} &=&
  \frac{Gqc_Q Q\psi}{2R^{3/2}}
-\tfrac{3}{2}(\Xi R)R^{1/2}\chi_{,x}.
\end{eqnarray}
$E_{ux}=0$ can be written as
\begin{eqnarray}
\label{calHeqn}
{\cal H}_{,x} &=&
  \frac{7Gc_Q^2 Q^2}{c_J R^6}
-\frac{3\bigl(G + 2(\Xi R)R_{,x}\bigr)}{R^2}
  \nonumber \\ &&
  {}+8\pi\bigl[(\hat{\Xi}\chi)\chi_{,x}
    +(\hat{\Xi}\psi)\psi_{,x}\bigr] \\
\label{calHeqnbis}
  &=& -\frac{6G\mathcal{M}}{c_M R^4}
  + \frac{10Gc_Q^2 Q^2}{c_J R^6}
  \nonumber \\ &&
  {}+8\pi\bigl[(\hat{\Xi}\chi)\chi_{,x}
    +(\hat{\Xi}\psi)\psi_{,x}\bigr],
\end{eqnarray}
where we have defined
\begin{equation}
\label{calHdef}
{\cal H}:=B_{,x}-\Xi\ln G.
\end{equation}

Eqn.~(\ref{calHeqn}) is the only field equation where the metric
function $B$ appears explicitly in the Einstein equations, rather than
implicitly through the operator $\Xi$. In this paper, we use the eG
formulation of \cite{GundlachMartel26}, where $B$ can be specified
arbitrarily, and (\ref{calHeqnbis}) is used to compute $\Xi\ln G$.

The remaining Einstein equation $E_{uu}=0$ can be written as
\begin{equation}
\label{EEuu}
\Xi\Xi R+{\cal H}\Xi R+\frac{8\pi}{3}
R(\hat\Xi\psi^2+\hat\Xi\chi^2)=0.
\end{equation}
In analogy to (\ref{UUR}), we can write it more
geometrically as
\begin{equation}
\label{XXR}
X\!XR+\frac{8\pi}{3}[(\hat X\psi)^2+(\hat X\chi)^2]R=0,
\end{equation}
where
\begin{equation}
X:=\bar G^{-1}\Xi, \qquad \hat X:=\bar G^{-1}\hat\Xi.
\end{equation}
$X$ is tangent to the affinely parameterised ingoing null geodesics (as $U$
is to the outgoing ones), and $\bar G$ is defined by
\begin{equation}
\Xi\ln \bar G=-{\cal H},
\end{equation}
In double-null gauge $B=0$, we have $\bar G=G$ and
$X=G^{-1}\partial_u$, in analogy to $U=G^{-1}\partial_x$ (the latter
holds for any choice of $B$).

Following \cite{GundlachMartel26}, when setting up initial data on an
outgoing null cone $u=0$ with a regular center, we ensure $\Xi R < 0$,
which then guarantees $\Xi R < 0$ for all $u>0$.

Following \cite{GundlachMartel26}, our primary diagnostics are the
Hawking compactness $C$, the Hawking mass $M$, the charge $Q$, and the
augmented mass $\mathcal{M}$, as defined previously. To capture
behavior on the dispersion side of the collapse threshold, we
introduce an additional curvature diagnostic, $|T|_{\text{max}}$, defined as the maximum value of $|T|$ over
the entire spacetime. Here, $T$ is the trace of the stress-energy
tensor~\eqref{stressenergytensor}, given by
\begin{eqnarray}
T &:=& -\frac{3}{2}\left(
(D_a\phi)^*D^a\phi + \frac{1}{8\pi c_J}F_{ab}F^{ab}
\right) \nonumber \\
&&= \frac{3}{G} \left({\hat{\Xi}\chi \, \chi_{,x}} + {\hat{\Xi}\psi \, \psi_{,x}}\right) + \frac{3c_Q^2 Q^2}{8 c_J \pi R^6} 
\label{diagnostic_T}
\end{eqnarray}
As discussed previously, the electromagnetic part of the stress-energy
tensor is not traceless in 4+1 spacetime dimensions. Consequently, our
diagnostic $T$ includes a term proportional to $F_{ab}F^{ab}$, or $Q^2/R^6$, which represents, in an average sense, the
electromagnetic charge energy density in 4+1 spacetime dimensions.


\section{Numerical method}
\label{sec:algorithm}


\subsection{Initial data}
\label{initialdata}


On the initial null cone $u=0$ the scalar field profiles $\psi(x)$ and
$\chi(x)$ are specified freely.  The metric is initialised in affine
gauge $G(0,x)=1$, after which~(\ref{Reqn}) is integrated for $R(0,x)$, starting from the regular-centre expansions of
Appendix~\ref{expansions}.

We consider initial data for the complex scalar field $\Phi=\psi +
i\chi$ of the form
\begin{equation}
    \Phi(0,x)=g(x)\, e^{\,i\omega f(x)}.
    \label{initial_data}
\end{equation}
Here $g(x)$ is a smooth radial profile function, $f(x)$ is a real
phase function, and $\omega$ is a frequency parameter that directly
controls the complex phase of the scalar field.

Unless specified otherwise, we use
\begin{equation}
g(x)=\mathcal{A}e^{-\frac{(x-c)^2}{\sigma^2}},
\label{Gaussian}
\end{equation}
where $\mathcal{A}$ is the amplitude, $c$ is the center of the
Gaussian pulse, and $\sigma$ determines its width, and
\begin{equation}
f(x)=x-c,
\label{linear}
\end{equation}
where $c$ is the same as the center of the Gaussian pulse
above. We also use the alternative, power-law, profile
\begin{equation}
g(x)=\mathcal{A}\left[1+\frac{(x-c)^2}{\sigma^2}\right]^{-1},
\label{pow}
\end{equation}
or a profile given by the sum of two Gaussians (\ref{Gaussian}), with
amplitudes ${\cal A}_1$ and ${\cal A}_2$, etc.


\subsection{Evolution}
\label{gaugechoice}


We then evolve to $u>0$ using the eG formulation of
\cite{GundlachMartel26}. Because this formulation does not require
$R_{,x}>0$, we can evolve through apparent and event horizons. As
already mentioned, in all simulations we use $c_J=c_M=c_Q=1$, and we
do not write these constants in the remainder of the paper.  For
$u>0$, the evolutions are carried out in the shifted double-null (sdn)
gauge of \cite{GundlachMartel26},
\begin{equation}
\label{sdnshift}
  B(u,x)=\frac{1}{2R_{,x}(u,0)}\left(1 - \frac{x}{x_0}\right).
\end{equation}
The first factor is the value of the shift at $x=0$ required to fix
$R=0$ at $x=0$, while the second factor makes $x=x_0$ an ingoing null
cone. For type-II critical collapse, $x_0$ is set to approximately
coincide with the past light cone of the accumulation point of
near-critical echoes, so that the computational domain is adapted to
the underlying self-similarity. In Sec.~\ref{sec:extremal},
  where self-similarity does not arise, $x_0$ is chosen just smaller
than $x_\text{max}$, namely $x_0=x_\text{max}-2.5$, so that
the outer grid boundary is approximately null. We then find the
solution on nearly the entire domain of dependence of the initial data
on the initial null cone. (The width of this buffer is
arbitrary, but must be independent of the numerical resolution in
convergence testing.)

If a black hole forms, its Hawking mass $M$ and charge $Q$ are
strictly defined only at future null infinity, but we can evaluate
them approximately on the first marginally outer-trapped surface (from
now on, FMOTS), where in this paper ``first'' always refers to
the smallest value of the retarded time $u$, and to our finite
numerical domain only, ignoring that a MOTS may have formed at smaller
$u$ outside the numerical domain. In spherical symmetry, we only
consider spherically symmetric MOTS. As there are no anti-trapped
surfaces in our solutions, any spherical MOTS is given by $C=1$. (In
the terminology of \cite{GundlachMartel26}, we always take ${C}_{\rm
  thr}=1$ in this paper.)


\subsection{Regularity at the centre}
\label{section:centre}


Because $R=0$ at the regular centre $x=0$, the hierarchy equations
develop coordinate singularities there.  Rather than imposing boundary
conditions directly, we expand each field as a power series in $x$ and
initialise the integration from the leading coefficients.  The full
4+1-dimensional expansions are given in
Appendix~\ref{expansions}; here we record the four key conditions that
follow from regularity and normalising $u$ to be the proper time at
the centre:
\begin{align}
  G(u,0) &= R_{,x}(u,0), &
  B(u,0) &= \frac{1}{2R_{,x}(u,0)}, \notag\\[4pt]
  \Xi R\big|_{x=0} &= -\tfrac{1}{2}, &
  \mathcal{H}\big|_{x=0} &= 0.
  \label{centre_conditions}
\end{align}
At each time step, the evolved variables $\psi$, $\chi$ and $G$ are
least-squares fitted to linear (or quadratic) polynomials over the
innermost grid points.  The expansion coefficients of $R$, $Q$, $A$,
$\Xi R$, $\hat{\Xi}\psi$, $\hat{\Xi}\chi$, and $\mathcal{H}$ are then
set from the fitted scalar and metric coefficients following
Appendix~\ref{expansions}, and the discrete integration starts from
the outermost expansion point.


\subsection{Discretisation}
\label{sec:numericalscheme}


The time evolution and spatial discretisation
follow~\cite{GundlachMartel26} essentially without modification. We
use the eG formulation in sdn gauge~\eqref{sdnshift}, advancing the
fields $\psi$, $\chi$ and $\ln G$ in retarded time $u$ with a
second-order Runge-Kutta method (method of lines), while the hierarchy
equations are solved in $x$ at each Runge-Kutta substep.

The time step and singularity excision scheme are likewise taken
directly from~\cite{GundlachMartel26}, by default the timestep $\Delta
u_4$ with the CFL parameter $C_2=0.1$ and slowdown parameter
$C_3=5\times10^{-4}$ or, where explicitly stated, the constant
timestep $\Delta u_0:=C_0\Delta x$.

Midpoint values of ratios such as $Q/R^3$ or $M/R^4$ are formed by
averaging the ratio at adjacent grid points, not by averaging
numerator and denominator separately.


\subsection{Integration near the centre}
\label{section:powerscaling}


In 4+1 dimensions $Q$, $M$, $\tilde{M}$, $\mathcal{M}$ and
$\tilde{\mathcal{M}}$ all scale as $R^4\sim x^4$ near the regular centre
(see Appendix~\ref{expansions}), so their source terms in the hierarchy
equations behave as $R^3$.  A naive midpoint rule is insufficiently
accurate for these integrals near $x=0$.  We instead use the corrected rule similar to what~\cite{GundlachMartel26} have used as
\begin{equation}
\label{myfac4p1}
\int_{x_{i-1}}^{x_i} f R^3\,dx
\;\simeq\;
  w(\epsilon)\;f_{i-1/2}(R_{i-1/2})^3\,\Delta x,
\end{equation}
where $\epsilon:=1/i$ and
\begin{equation}
\label{myfac4p1weight}
w(\epsilon) :=
  \frac{1-\tfrac{3}{2}\epsilon+\epsilon^2-\tfrac{1}{4}\epsilon^3}
       {1-\tfrac{3}{2}\epsilon+\tfrac{3}{4}\epsilon^2-\tfrac{1}{8}\epsilon^3}.
\end{equation}
The weight $w(\epsilon)$ equals unity at $\epsilon=0$ (large $i$),
recovering the plain midpoint rule, and (\ref{myfac4p1}) is
exact when $R/x$ and $f$ are both constant. 


\subsection{Convergence testing}
\label{section:convergence}


Our convergence testing methodology and results
follow~\cite{GundlachMartel26} entirely. For convergence testing, we need a fixed ratio between $\Delta x$
and $\Delta u$, where $\Delta x=h$ and $\Delta u=C_0h$, as
$h\to0$. We use $C_0=0.1$ for our testing, with the functions
\eqref{Gaussian} and \eqref{linear} below, with $\mathcal{A} =
0.15$, $c=1.5$, $\sigma= 0.5$, and $\omega=2$, and set $q =
1/(2\sqrt{\pi}) \simeq 0.2821$. We evolve with $N_x=100\cdot 2^n$ grid points for $n =
0,\ldots,8$ with $x_{\text{max}}=5$, and compare consecutive
resolutions up to the last output time before apparent horizon
formation. We find the same two-group convergence behaviour as
in~\cite{GundlachMartel26}: variables such as $R$, $Q$, $A$, $M$ and
$\tilde{M}$ show clean pointwise second-order convergence including at
the centre, while variables such as $\psi$, $\chi$, $\ln G$ and
$\Xi\psi$ converge pointwise at second order everywhere except within
a few grid spacings of $x=0$, where convergence still holds in any
integral norm.  The independent residual of the Raychaudhuri equation
\eqref{UUR} and the difference $M - \tilde{M}$ both converge to zero
at second order.


\section{Type-II critical collapse}
\label{sec:typeII}


\subsection{Derivation of the scaling laws}
\label{sec:typeII_recap}


At the threshold between dispersion and black hole formation, we see
type-II critical collapse. This is governed by a universal
self-similar intermediate attractor, the critical
solution~\cite{critreview}. Generic 1-parameter families of initial
data fine-tuned to this threshold lose memory of their initial
conditions and are funneled toward this solution, which is discretely
self-similar (DSS). In coordinates $x^\mu$ adapted to the DSS
symmetry, the metric takes the form
\begin{equation}
    g_{\mu\nu}=e^{-2\tau}\,\tilde{g}_{\mu\nu},
    \label{metric_DSS}
\end{equation}
where $\tilde{g}_{\mu\nu}$ is periodic in $\tau$ with period $\Delta$,
the echoing period. For a scalar field $\psi$ to possess a
stress-energy tensor compatible with this metric, it must exhibit the
same periodicity, $\psi(\tau + \Delta)=\psi(\tau)$. For the analysis
of near-critical solutions we use the null-coordinate similarity
variables
\begin{equation}
    \tau := -\ln(u_* - u), \qquad \hat{R} := \frac{R}{u_* - u},
    \label{tauandR}
\end{equation}
where $u_*$ is the retarded time of the echo accumulation point.

We begin by reviewing the derivation for the mass and curvature
scaling laws for a real, uncharged, scalar field, as given
in~\cite{critreview}. There is then no length scale in the problem.
Let $Z(\hat{R}, \tau)$ denote a set of scale-invariant variables of
the problem. Sufficiently close to the threshold of collapse, there is
then a spacetime region where the solution admits a perturbative
decomposition
\begin{equation}
    Z(\hat{R},\tau) \simeq Z_*(\hat{R},\tau) + \sum_{i=1}^{\infty} C_i(p)\,e^{\lambda_i \tau}\,Z_i(\hat{R},\tau),
    \label{perturbation_full}
\end{equation}
where $Z_*(\hat{R},\tau)$ is the critical solution (periodic in $\tau$
with period $\Delta$) and the $C_i$ depend on the initial data with
parameter $p$. By definition of the critical solution, there is
exactly one mode $\lambda_0$ with positive real part; as $\tau \to
\infty$ all other perturbations vanish. Retaining only this growing
mode and linearising around the critical parameter value $p_*$, nearby
solutions can be approximated in the approximately self-similar region
as
\begin{equation}
    \left. Z(\hat R,\tau) \right|_{\tau \to \infty} \simeq Z_*(\hat
    R,\tau) + \frac{dC_0}{dp}(p - p_*)\,e^{\lambda_0 \tau}\,Z_0(\hat
    R,\tau),
    \label{perturbation}
\end{equation}
where $\lambda_0 > 0$ is the Lyapunov exponent of the unique unstable
growing mode.  The system closely tracks the critical solution until
this mode grows to order unity at the departure scale $\tau_*$, after
which nonlinear collapse proceeds to black hole formation. The overall
physical length scale of the configuration at $\tau_*$ is set entirely
by the factor
\begin{equation}
    e^{-\tau_*} \propto (p - p_*)^{1/\lambda_0}.
    \label{length_scale}
\end{equation}

The mass scaling law follows from dimensional analysis. In a
$D$-dimensional spacetime, mass carries dimensions of
$[\text{length}]^{D-3}$ in $c=G_N=1$ units. Since the field
  equations contain no intrinsic scale, the sole available length
scale at departure is $e^{-\tau_*}$, giving
\begin{equation}
    M \sim \left(e^{-\tau_*}\right)^{D-3} \sim (p-p_*)^{\gamma_{M}}, 
\label{mass_scaling}
\end{equation}
where
\begin{equation}
\gamma_M=(D-3)\gamma, 
\end{equation}
and we have defined
\begin{equation}
\gamma := {1\over\lambda_0}.
\end{equation}

Since the Ricci scalar has dimension $[\text{length}]^{-2}$ (in any
spacetime dimension $D$),
\begin{equation}
|R|_{\text{max}} \propto e^{2\tau_*} \propto (p_* - p)^{-2/\lambda_0}
= (p_* - p)^{-2\gamma}, \label{Rmax_scaling}
\end{equation}
on the dispersal side \cite{critreview}.

Following \cite{GundlachMartin96}, consider now the charged complex
scalar field system. All solutions with purely real scalar field and
$Q=0$, and in particular the critical solution (the 4+1 equivalent of
the Choptuik solution for the 3+1 scalar field) are still solutions of
this system.  In the charged scalar field system the critical solution
actually extends to a 1-parameter family of solutions, with the
complex $\phi$ equal to $e^{i\alpha}$ times the real scalar field of
the critical solution, for any constant $\alpha$. What we have said
about perturbing around the purely real critical solution is also true
for the phase-rotated one.

Now perturb about the critical solution, and assume for simplicity
that it has been phase-rotated so that its scalar field is real. To
leading order in the magnitude of the imaginary part $\chi$ of the
scalar field, $\chi$ obeys a linear real wave equation on the critical
spacetime, and $Q$ is of the same order, while $\chi$ and $Q$ affect
the metric, and $Q$ affects $\psi$ and $\chi$, only at quadratic
order. Hence we can generalise (\ref{perturbation}) to include
additional linear perturbation modes $Z_i$ in which only $\chi$ and
$Q$ are perturbed, but not $\psi$ or the
metric. \cite{GundlachMartin96} demonstrates numerically that all
perturbation modes of $\chi$ (and hence the corresponding linear
perturbation $Q$) decay, that is have ${\rm Re}\,\lambda_i<0$.

This perturbative analysis suggests that the critical solution is
still an attractor in at least some region of phase space, up to a
constant phase $\alpha$. Our numerical experiments in 4+1 are
consistent with what was previously found numerically in 3+1: by
fine-tuning {\em any} 1-parameter family of charged scalar field
initial data to the threshold of collapse we again find a discretely
self-similar, real critical solution, up to a complex rotation
$e^{i\alpha}$, for some family-dependent constant
  $\alpha$. This solution is qualitatively similar to the Choptuik
solution in 3+1.

The charge scaling follows again from dimensional analysis. The
spacetime variable $Q/M$ is odd, and hence to leading order linear, in
both $\chi$ and $q$. $Q/M$ and $\chi$ are dimensionless, but $q$ has
dimension $(\text{length})^{-1}$. Hence, $e^{-\tau}q$ is
again dimensionless.\footnote{Our
$\tau$ has the opposite sign from that of \cite{GundlachMartin96}.} We then have the dimensionless relation
\begin{equation}
{Q\over M}\sim e^{-\tau}q\,\chi. 
\end{equation}
Denote the least negative $\lambda_i$ associated with $\chi$ by
$\mu$. At late times in the expansion (\ref{perturbation}), $\chi$
then decays as $e^{\mu\tau}$. Hence at the end of the similar phase,
at $\tau\simeq \tau_*$, we have
\begin{equation}
    \frac{Q}{M} \sim e^{\mu \tau_*}\,e^{-\tau_*}q.
    \label{charge_mass_ratio}
\end{equation}
Taking into account the mass scaling, the charge scales as
\begin{equation}
    Q \sim e^{\mu \tau_*}\,e^{-\tau_*}q\,\left(e^{-\tau_*}\right)^{D-3} 
\sim q\,e^{-(D - 2 - \mu)\tau_*},
    \label{Q_tau_scaling}
\end{equation}
and hence
\begin{equation}
Q\sim q\,(p-p_*)^{\delta_Q},
\end{equation}
with 
\begin{equation}
\delta_Q=(D-2-\mu)\gamma. \label{delta_Q}
\end{equation}


\subsection{Uncharged (real or complex) scalar field}
\label{sec:typeII_extraction}



\subsubsection{Curvature, mass critical exponent and echoing period}
\label{sec:exponents}


In terms of the scale-invariant logarithmic time $\tau$, the critical
scalar field echoes with a fundamental period $\Delta$, undergoing a
strict parity inversion at the half-cycle, $\psi(x, \tau + \Delta/2) =
-\psi(x, \tau)$. Because the associated stress-energy tensor is
strictly quadratic in the scalar field gradients
(\ref{stressenergytensor}), it is invariant under this
inversion. Hence the metric components must repeat at twice the
frequency of the scalar field i.e. $g_{\mu\nu}(x, \tau + \Delta/2) =
g_{\mu\nu}(x, \tau)$.

Because the critical solution is discretely self-similar, the exact
scaling relation for the Hawking mass of the FMOTS is
\begin{equation}
\label{mass_fine_struc}
    \ln M=A + \gamma_M \ln(p - p_*) + \Psi_M(\gamma \ln(p - p_*) + B),
\end{equation}
where $\gamma_M =2\gamma$ in $D=5$, see (\ref{mass_scaling}), and
$\Psi_M$ denotes a periodic fine structure on top of the dominant
power-law scaling \cite{critreview}. $A$ and $B$ are constants that
depend on the 1-parameter-family of initial data.

Conversely, in the subcritical regime ($p < p_*$), the maximum
spacetime curvature reached before dispersion scales as
\begin{equation}
\label{curv_fine_struc}
    -\frac{1}{2} \ln |R|_\text{max}=
A + \gamma \ln(p_* - p) + \Psi_R(\gamma \ln(p_* - p)+ B),
\end{equation}
with $\Psi_R$ denoting the curvature fine structure. Rather than using
the Ricci scalar directly, we adopt the curvature diagnostic defined
in Eq.~(\ref{diagnostic_T}). Thus we have
\begin{equation}
\label{curv_fine_struc_T}
    -\frac{1}{2} \ln |T|_\text{max}=
A + \gamma \ln(p_* - p) + \Psi_T(\gamma \ln(p_* - p)+ B)
\end{equation}
The two are trivially related by $R=-\frac{16\pi}{3} T$.  For each
family of initial data, we fix $A$ so that $\Psi_T$ oscillates with
approximately a zero mean (Fig.~\ref{fig:curv_fine_structure}). We do
not fix $B$ a priori, but set it to zero for one family of initial
data, and determine it so that fine structures for different families
align.

From the arguments above, $\Psi_M$ and $\Psi_T$ are each periodic in
their own argument with period $\Delta/2$. Consequently, the echoing
period $\Delta$ can be extracted from the fine structure: plotting
$\Psi_M$ or $\Psi_T$ against $\ln|p-p_*|$ yields a periodic signal
with period $P=(\Delta/2)/\gamma$, from which the spacetime
echoing period follows as
\begin{equation}
    \Delta=2\gamma P.
    \label{echoingperiod}
\end{equation}
To determine the mass and curvature scaling exponents, we consider a
real scalar field by setting $\omega=0$ in~(\ref{initial_data}). We
use Gaussian initial data with fixed $c=1.5$, $\sigma= 0.5$, and a
reference amplitude $\mathcal{A}=0.05$. We then fine-tune ${\mathcal
  A}$ to the threshold of collapse by multiplying it with factor $p$,
and bisection in $p$. We have fixed the reference amplitude ${\mathcal
  A}$ so that $p_*\simeq 1$. We will later adopt a similar procedure
for fine-tuning any one parameter of the initial data. A real scalar
field remains real under time evolution regardless of $q$, and so the
value of $q$ is then irrelevant. The outer boundary is placed at
$x_\text{max}=5$, with $N_x=350$ grid points. We take $x_0 =
2.912950$ and find $p_* \simeq 1.139$.

\begin{figure}[h]
\includegraphics[width=\linewidth]{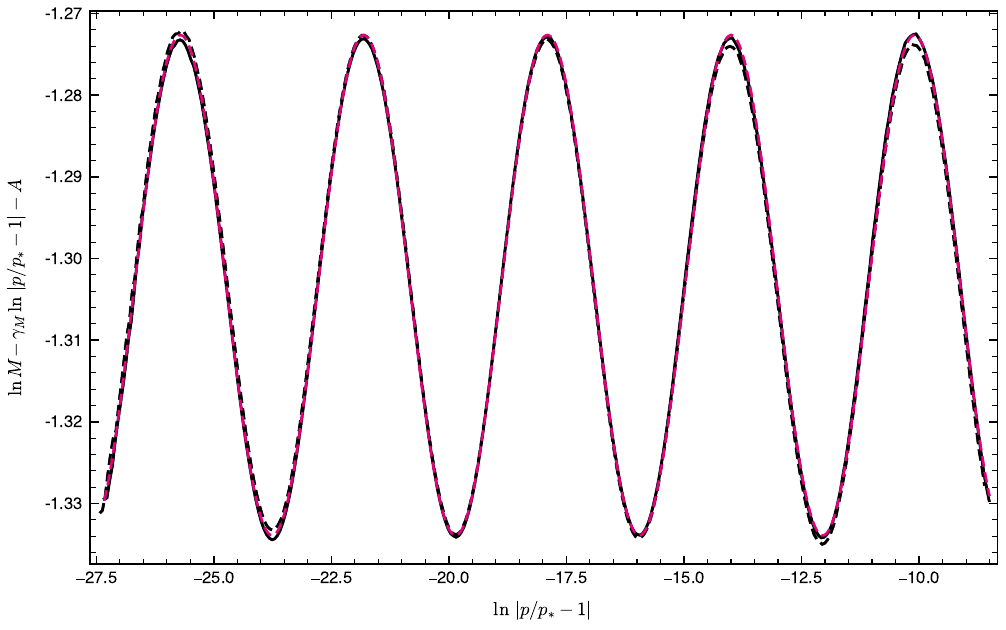}
\caption{The FMOTS mass fine structure $\Psi_M=\ln M - \gamma_M
  \ln|p/p_* - 1| - A$ against $\ln| p/p_{*}-1|$. The solid black curve
  shows $\Psi_M$ for the single-Gaussian family of real initial data
  (with $A\simeq0.828$ fixed using Fig.~\ref{fig:curv_fine_structure},
  and $B=0$). The dashed black curve is for the two-Gaussian complex
  family (with $A=0.933$ and $B/\gamma=-0.097$, chosen such that the
  curve overlaps with the other one). The dashed magenta curve shows
  the corresponding best-fit sinusoid. The three curves are almost
  identical.}
\label{fig:mass_fine_structure_sin_fit}
\end{figure}

\begin{figure}[h]
\includegraphics[width=\linewidth]{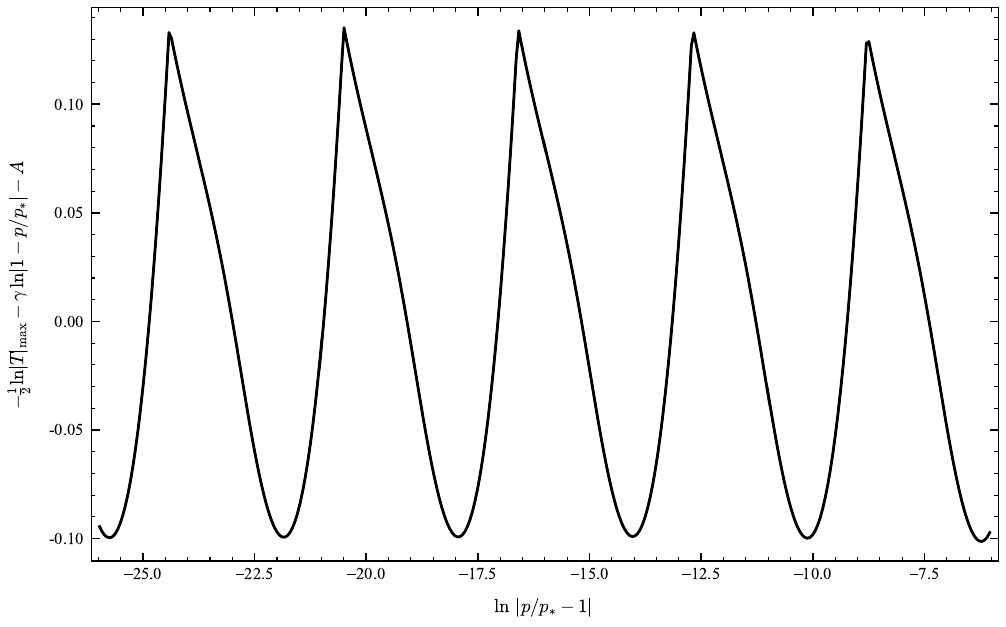}
\caption{Maximum curvature fine structure $\Psi_T =
  -\tfrac{1}{2}\ln|T|_\text{max} - \gamma\ln|p/p_*-1| - A$ against
  $\ln\lvert p/p_{*}-1\rvert$ for the family of real, single-Gaussian
  initial data on the dispersion side, with $A \simeq 0.828$.}
\label{fig:curv_fine_structure}
\end{figure}

We plot the mass fine structure $\Psi_M$ for the FMOTS, selecting the
value of $\gamma_M$ for which $\Psi_M$ plotted against $\ln|p/p_*-1|$
shows no secular drift. Our best estimate is $\gamma_M=0.8265$, see
Fig.~\ref{fig:mass_fine_structure_sin_fit}. We note that the
(slicing-dependent) mass fine structure is well approximated by a sine
function, just as in 3+1 dimensions in null slicing
\cite{GundlachMartel26}.

To test universality, we have also considered initial data where
$g(x)$ is the sum of two Gaussians, each of the form (\ref{Gaussian}),
with parameters $c_1=1.5$, $\sigma_1=0.5$, $\mathcal{A}_1=0.05$
for the inner pulse and $c_2=4.5$, $\sigma_2=0.5$, $\mathcal{A}_2 =
0.025$ for the outer pulse. We also set $\omega=10^{-3}$
in~\eqref{initial_data} rather than $\omega=0$, so the initial data
are nearly, but not exactly, real, with the phase function as defined
in \eqref{linear} with centre at $c_1$, but we still set $q=0$. The
outer boundary is moved out to $x_\text{max}=10$, with $N_x=700$
grid points, keeping the grid spacing fixed relative to the
single-pulse evolutions above. We set $x_0=5.867950$ and
find $p_* \simeq 0.492$.

We compute the mass fine structure $\Psi_M$ for the FMOTS in this
two-Gaussian family in the same way as above and also plot it
on Fig.~\ref{fig:mass_fine_structure_sin_fit} (dashed black
curve). The two curves are essentially the same.

The echoing period $\Delta$ is obtained by fitting a sinusoidal
function to the mass fine structure $\Psi_M$
(Fig~\ref{fig:mass_fine_structure_sin_fit}). The fitted oscillation
period is $P \simeq 3.901$, and from~(\ref{echoingperiod}) with
$2\gamma=\gamma_M\simeq0.8265$, this gives
\begin{equation}
    \Delta=2\gamma P \simeq 3.224.
\end{equation}

The curvature scaling exponent $\gamma$ is obtained as
\begin{equation}
    \gamma=\frac{\gamma_{M}}{2} \simeq \frac{0.8265}{2} \simeq 0.41325 
\end{equation}
We plot $\Psi_{T}$ on the dispersion side in
Fig.~\ref{fig:curv_fine_structure}. As pointed out in
\cite{GundlachMartel26}, we expect $\Psi_M$ to be universal for a
given time slicing, here null cones, and we expect $\Psi_T$ to be
fully universal, across families of initial data.

\subsubsection{Estimation of charge scaling exponent from imaginary
  scalar field perturbation}


To find $\mu$ directly, rather than from $\delta_Q$, we
consider a complex scalar field with vanishing coupling $q=0$,
initialized such that the imaginary part $\chi$ is small
compared to the real part $\psi$.  This places the system within the
small imaginary perturbation regime analyzed in
\cite{GundlachMartin96}.

During evolution, nonlinearity mixes $\psi$ and $\chi$.  We resolve
this by applying a post-evolution phase rotation parameterized by an
angle $\alpha$, where $\alpha$ is independent of $u$ and $x$, but
depends on $p$:
\begin{equation}
    \psi_{\text{rot}} := \cos\alpha\,\psi + \sin\alpha\,\chi, \quad
    \chi_{\text{rot}} := \cos\alpha\,\chi - \sin\alpha\,\psi.
    \label{phase_rotation}
\end{equation}
We choose $\alpha$ such that $\psi_{\text{rot}}$ aligns
with the real critical solution and $\chi_{\text{rot}}$ decays at
large $\tau$.

For the initial data, we use a single Gaussian with
$\mathcal{A}=0.05$, $c=1.5$, and $\sigma= 0.5$, and $\omega
 =10^{-3}$, which keeps the imaginary component $\chi$ small enough
to act as a linear perturbation to the real field $\psi$. We use the
same boundary $x_\text{max}=5$ and $N_x=350$ grid points. We
again set $x_0=2.912950$ and find $p_* \simeq 1.139$, which
is very close to the value obtained for a real scalar field with the
same parameters. To analyze the dynamics near the intermediate attractor,
we analyze the last dispersing solution we find right before the
bisection algorithm stops.

\begin{figure}[h]
\includegraphics[width=\linewidth]{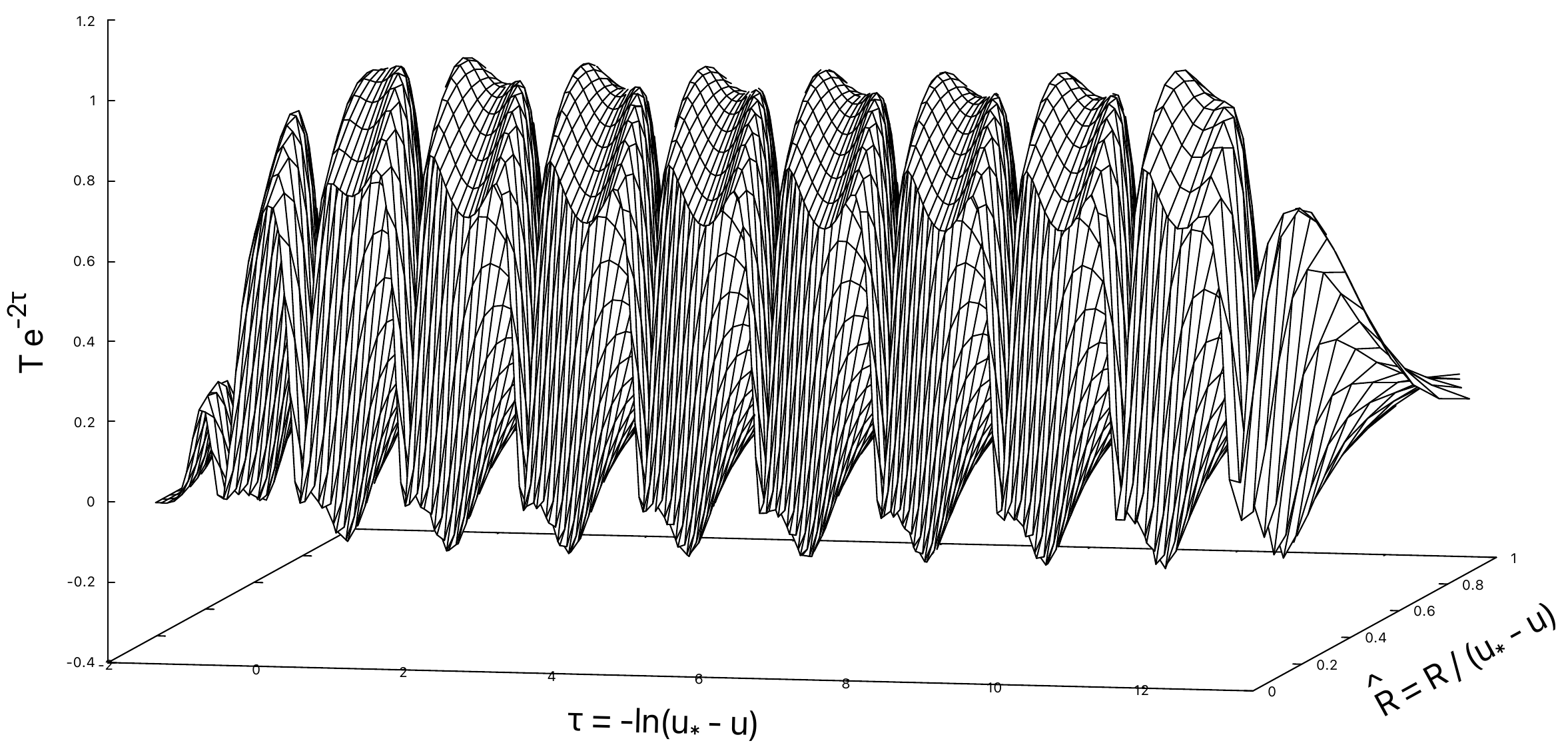}
\caption{Scale-invariant curvature diagnostic $Te^{-2\tau}$
  (\ref{diagnostic_T}) for the almost real scalar field initial data,
  against the similarity coordinates $(\tau,\hat R)$.}
\label{fig:T_scalar}
\end{figure}

For plotting, we use the self-similarity adapted coordinates
$\tau$ and $\hat R$ as defined in~(\ref{tauandR}). For the optimal
choice of $x_0$ in sdn gauge, our numerical coordinate $x$ is also
adapted to self-similarity, but $\hat{R}$ does not rely on a choice of
$x_0$ and so is more robust. To find $\tau$, we must determine $u_*$,
the retarded time corresponding to the echo accumulation point. We
first approximate $u_*$ as the earliest time $u$ at which the
compactness reaches the dispersion threshold of $0.01$. We then refine
it by selecting the value that produces the clearest periodic behavior
(with almost equal peaks) in the scale invariant curvature scalar
diagnostic $Te^{-2\tau}$ before dispersion occurs. This procedure
yields $u_*=3.92527419$; the resulting periodic curvature diagnostic
is shown in Fig.~\ref{fig:T_scalar}.

Following the $U(1)$ gauge rotation procedure described above, we find
the rotation angle to be $\alpha=0.00077$.

\begin{figure}[h]
\includegraphics[width=\linewidth]{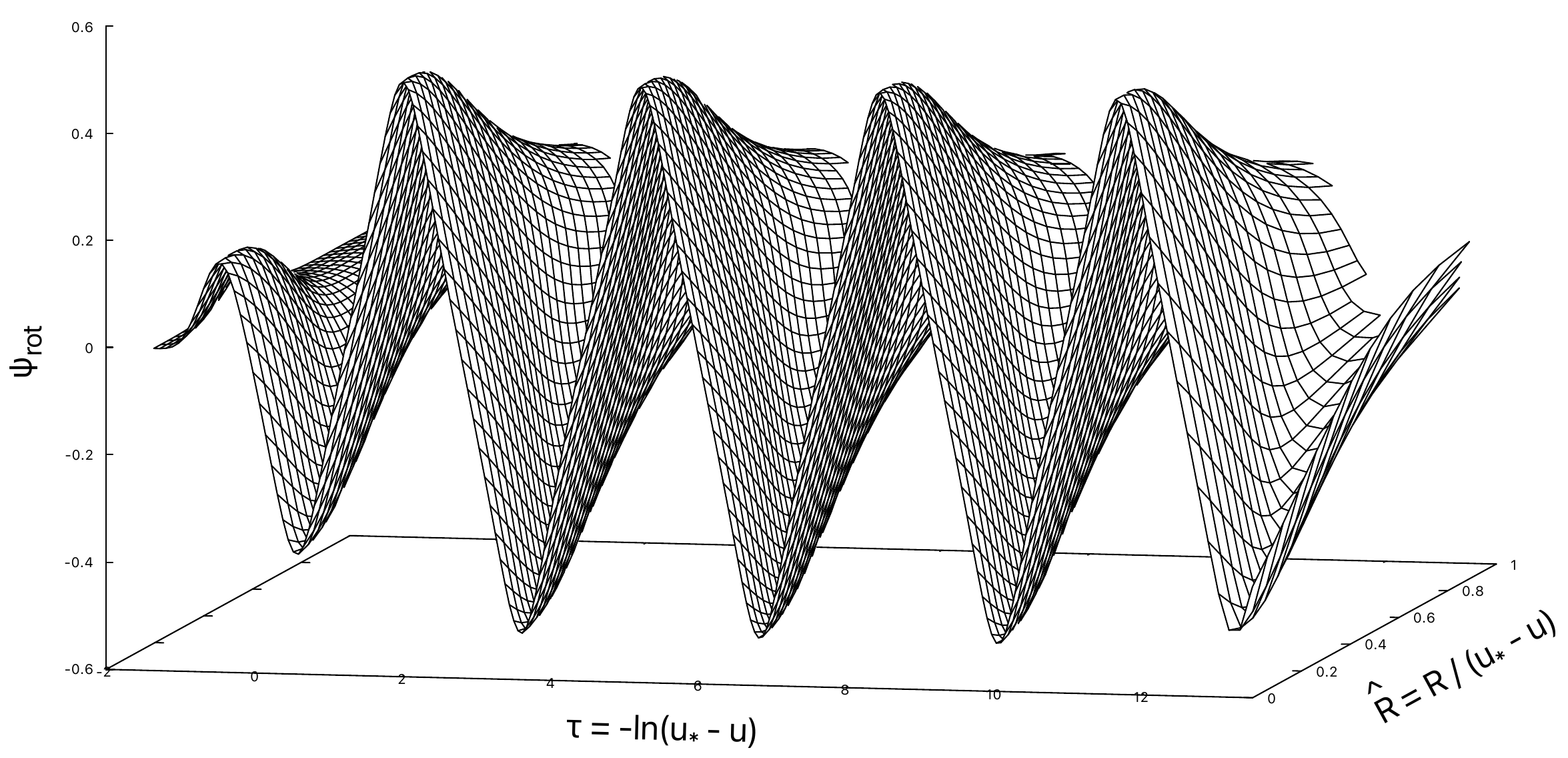}
\caption{Phase-rotated real part of the scalar field
  $\psi_{\text{rot}}(\tau, \hat R)$, evaluated at a rotation angle of
  $\alpha=0.00077$.}
\label{fig:phase_rotated_fields}
\end{figure}

The resulting phase-rotated field $\psi_{\text{rot}}$ is shown in
Fig.~\ref{fig:phase_rotated_fields}.

\begin{figure}[h]
\includegraphics[width=\linewidth]{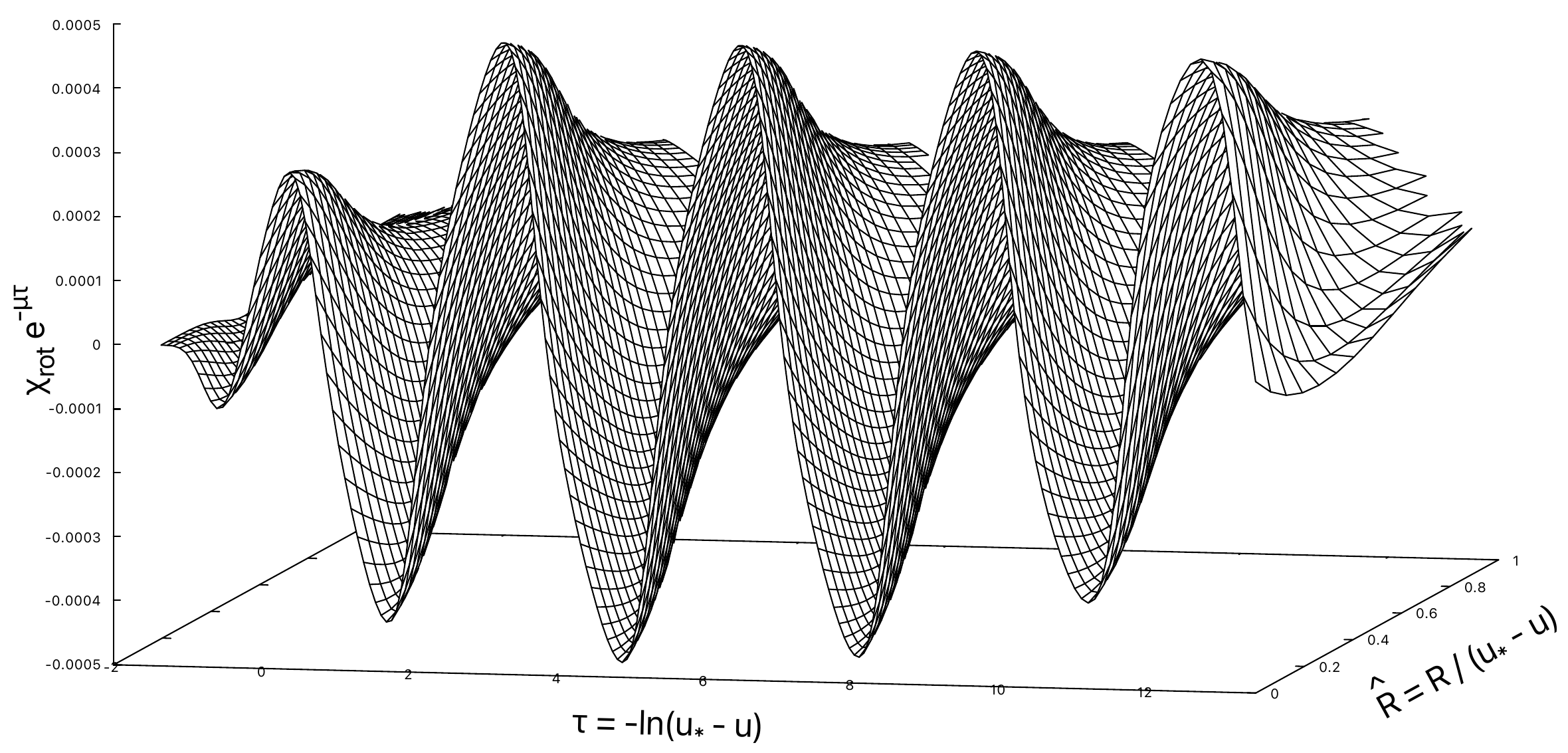}
\caption{Decay-canceled phase rotated imaginary perturbation,
  $\chi_{\text{rot}} e^{-\mu\tau}(\tau, \hat R)$, evaluated at the
  central critical exponent estimate of $\mu=-0.448$.}
\label{fig:decay_canceled_field}
\end{figure}

To extract the decay rate $\mu<0$, we look for the value that makes
the scaled imaginary field $\chi_{\text{rot}} e^{-\mu\tau}$ most
strictly periodic (with almost equal peaks) within the self-similar
regime. We find that this periodicity holds well across the narrow
window $\mu \in [-0.450, -0.446]$, from which we adopt the midpoint
$\mu=-0.448$ as our best estimate. The resulting decay-canceled
perturbation is plotted in Fig.~\ref{fig:decay_canceled_field}.

Finally, we calculate the charge scaling exponent $\delta_Q$ via
Eq.~(\ref{delta_Q}). Substituting our central estimate $\mu=-0.448$
alongside the curvature scaling exponent $\gamma=0.41325$ yields
$\delta_Q \simeq 1.425$.

\begin{table*}[htbp]
\caption{Initial-data families used in the $\mu$-universality
  comparison.  All families share amplitude $\mathcal{A}=0.05$, centre
  $c=1.5$, width $\sigma=0.5$ and coupling constant $q=0$. We use grid
  parameters $x_\text{max}=5$ and $N_x=350$. $\alpha$ and $\Delta\tau$
  are the complex phase shift and self-similarity phase shift used in
  Figs.~\ref{fig:universality2_psirot}-\ref{fig:universality2_chargedensity}.
  The colours in brackets are those used in the plots. The
  family-dependent quantity $\Delta\tau$ is related to the constant
  $B$ in the fine structure functions $\Psi_M$ and $\Psi_T$. $S$ is
  the amplitude scaling for $\chi_{\text{rot}}e^{-\mu\tau}$. $|S|$ is
  related to $A_Q$ in the definition of $\Psi_Q$. $\bar
  Qe^{-(1+\mu)\tau}/{R}^4$ is also scaled with $S$, to test the
  hypothesis that $\bar Q$ is proportional to $\chi$ for small
  $\chi$. Note: Family~1 (black) is the same as used in
  Figs.~\ref{fig:T_scalar}, \ref{fig:phase_rotated_fields},
  \ref{fig:decay_canceled_field}, \ref{fig:Qbar}, and
  \ref{fig:Qbar_avg_density}.}
\label{tab:id_families}
\renewcommand{\arraystretch}{1.3}
\setlength{\tabcolsep}{8pt}
\begin{tabular}{l||l|l|l|l|l|l|l|l|l}
  \text{Family} & $g(x)$ & $f(x)$ & $\omega$ & $x_0$ & $p_*$ & $\alpha$ & $u_*$ & $\Delta\tau$ & $S$ \\
\hline
  $1$ (black) &
    $\text{Gau}$ &
    $\text{lin}$ &
    $10^{-3}$ &
    $2.912950$ &
    $1.139$ &
    $0.00077$ &
    $3.92527419$ & 
    $0$ & $1$ \\
  $2$ (blue) &
    $\text{dGaudx}$ &
    $\text{lin}$ &
    $10^{-3}$ &
    $2.839175$ &
    $0.537$ &
    $0.000835$ &
    $3.95359690$ & 
     $1.13$ & $0.9772$
    \\
  $3$ (red) &
    $\text{pow}$ &
    $\text{Gau}$ &
    $10^{-3}$ &
    $2.554275$ &
    $1.492$ &
    $0.000054+\pi$ &
    $3.52329380$ & 
     $0.93$ & $-6.3232$
    \\ 
  $4$ (purple) &
    $\text{pow}$ &
    $\text{Gau}$ &
    $2$ &
    $2.556275$ &
    $1.482$ &
    $0.11+\pi$ &
    $3.52685185$ & 
     $0.93$ & $-0.0032$
  \\
\end{tabular}
\end{table*}

\begin{figure}[h]
\includegraphics[width=\linewidth]{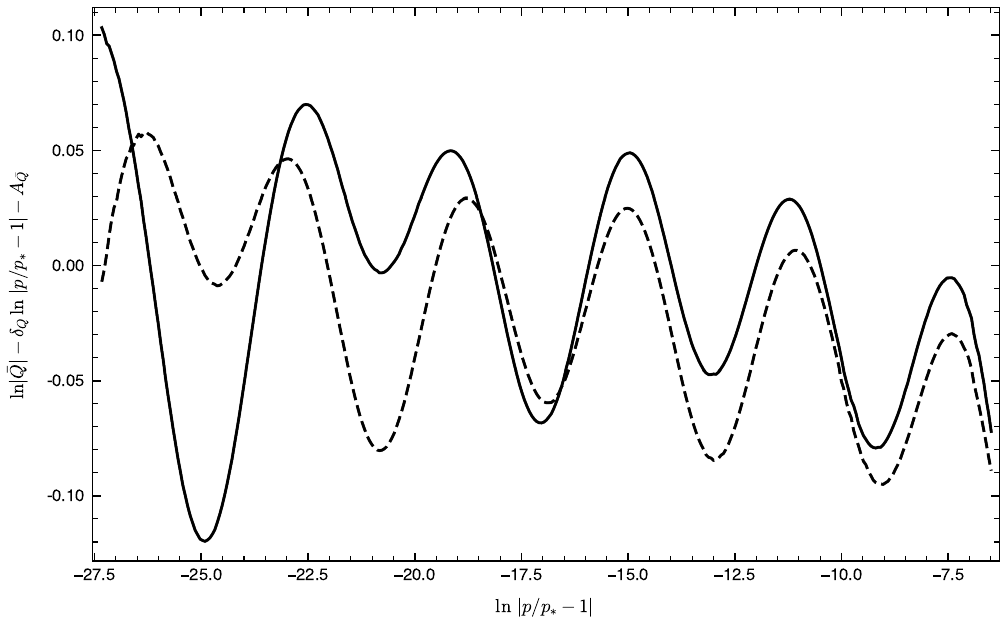}
\caption{Pseudo-charge fine structure $\Psi_{\bar Q} = \ln|\bar{Q}| -
  \delta_Q \ln|p/p_* - 1| - A_{\bar Q}$ plotted against $\ln|p/p_* -
  1|$ for the family of initial data with a small imaginary scalar
  field perturbation with $\delta_Q=1.425$. The solid curve shows
  $\Psi_{\bar Q}$ for the single-Gaussian family with $A_{\bar Q}
  \simeq -7.162$, and the dashed curve represents the two-Gaussian
  family with $A_{\bar Q} \simeq -6.960$. The two curves oscillate
  with almost the same period. Note the horizontal shift $B$ has
  already been fixed from the mass and curvature fine-structures.}
\label{fig:charge_fine_structure}
\end{figure}

Since the coupling constant $q$ is set to zero throughout this
analysis, no dynamical electromagnetic charge is present in the
evolved fields. To construct a $U(1)$-invariant measure of the
imaginary perturbation, we can nevertheless define a pseudo-charge
$\bar{Q}$ by evaluating Eq.~\eqref{Qeqn} with a nominal value of $q=1$
on the fields $\psi$ and $\chi$ obtained from the $q=0$ evolution of
Eqs.~\eqref{Xipsieqn}--\eqref{Xichieqn}. When $q$ and hence
$Q$ are small, so that the backreaction of the electromagnetic field
on the metric and scalar field can be neglected, we have $Q\simeq
q\bar Q$, with equality as $q\to 0$. Hence we can use the
scaling fine structure of the pseudo-charge $\bar Q$ to estimate the
(true) charge scaling fine structure, which is defined similar to
\eqref{mass_fine_struc} as
\begin{equation}
\label{charge_fine_struc}
    \Psi_Q=\ln|Q| - \delta_Q \ln|p/p_* - 1| - A_Q
\end{equation}
Analogous to the mass scaling, we now analyse the pseudo-charge
scaling and plot its fine structure, see
Fig.~\ref{fig:charge_fine_structure}. Here, we fix the value of
$A_{Q}$ for a given family such that the mean of the fine structure is
approximately zero over the high fine-tuning half of the data: the
lower 50\% of the $\ln|p/p_* - 1|$ range spanned by the data. The
values chosen for $A_Q$ are somewhat arbitrary, as the oscillations do
not have constant amplitude. Note $A_{Q}$, which encodes the
dependence of the most slowly decaying imaginary perturbation mode on
$p$ in a given family, is independent from $A$, which encodes the
dependence of the growing real perturbation mode on $p$ in that
family.

We plot $\Psi_{\bar Q}$ for the single and double complex Gaussian
initial data. The agreement between the two curves in
Fig.~\ref{fig:charge_fine_structure} is not nearly as good as for the
mass fine-structure in
Fig.~\ref{fig:mass_fine_structure_sin_fit}. Recall that here $q=Q=0$
and so this is simply the behaviour of a complex massless scalar
field, with $\psi$ and $\chi$ not coupled directly but only through
gravity. It is possible that the lack of universality is due to the
contribution of linear perturbation modes other than the least damped
one to $\chi$. Nonlinearities appear unlikely as an explanation, as
$|\chi_\text{rot}|$ is smaller than $|\psi_\text{rot}|$ by a factor of
$10^{-3}$ already at $\tau=0$, and by a factor of $10^{-5}$ at
$\tau=10$. If either explanation holds, we would expect universality
to be restored at higher fine-tuning than we can achieve in
double-precision numerics.

\begin{figure}[h]
\includegraphics[width=\linewidth]{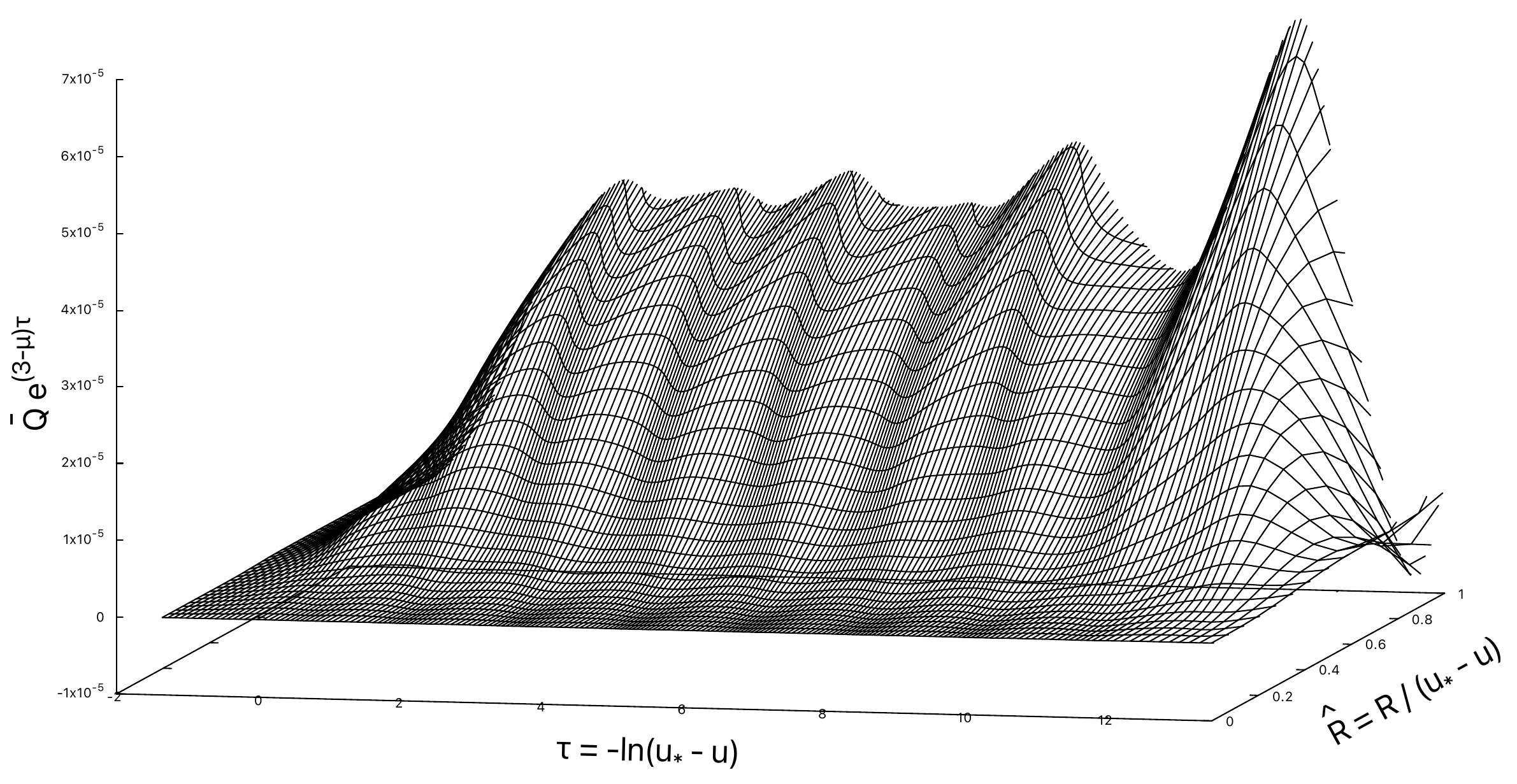}
\caption{Rescaled pseudo-charge $\bar{Q}\,e^{(3-\mu)\tau}(\tau, \hat
  R)$ for the single Gaussian initial data with small $\omega$.}
\label{fig:Qbar}
\end{figure}

\begin{figure}[h]
\includegraphics[width=\linewidth]{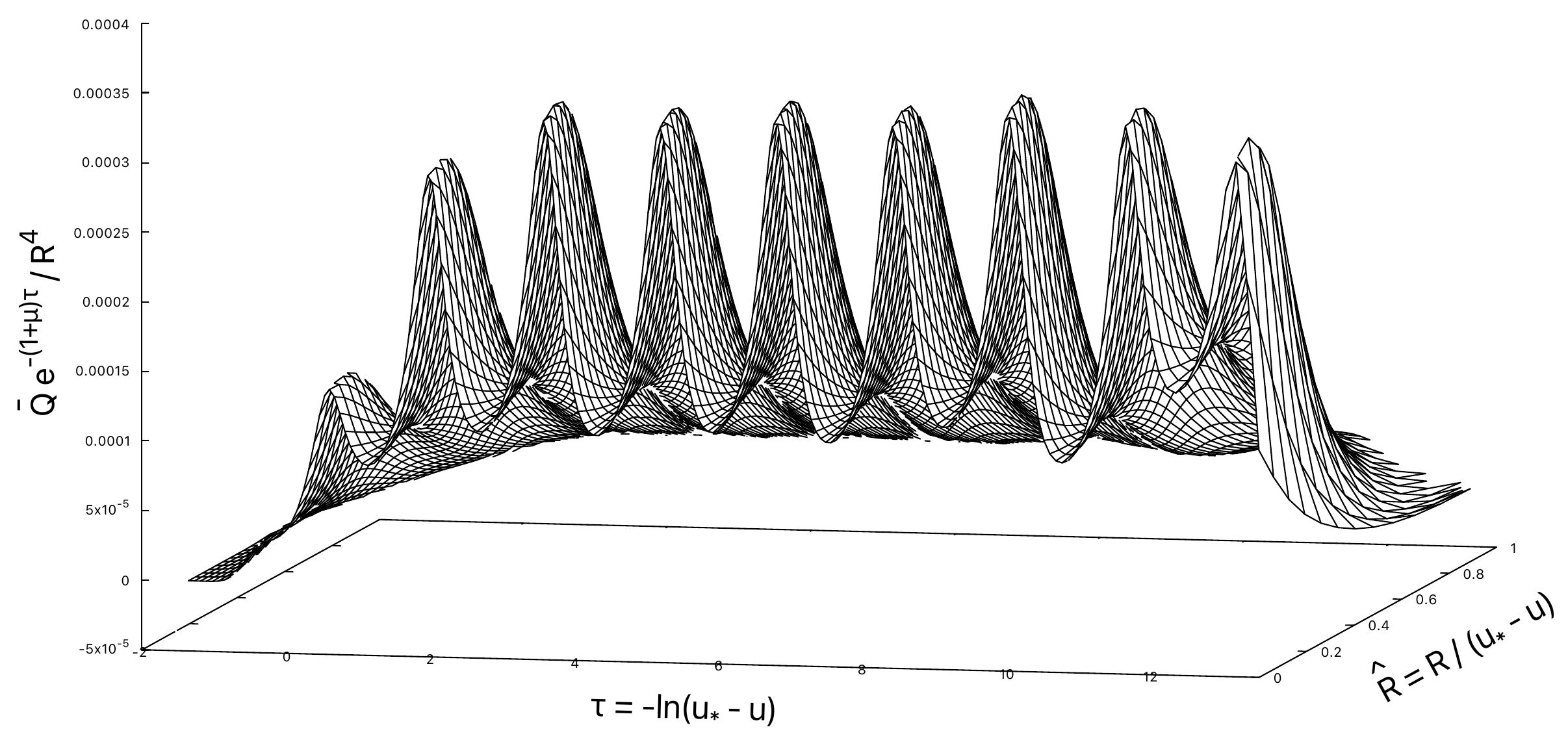}
\caption{As Fig.~\ref{fig:Qbar}, but showing the rescaled average
  pseudo-charge density $(\bar{Q}e^{-(1+\mu)\tau}/R^4)(\tau, \hat
  R)$.}
\label{fig:Qbar_avg_density}
\end{figure}

To examine the fields in the self-similar regime themselves, we plot
the pseudo-charge $\bar{Q}$ and the associated average charge density
against $(\tau,\hat{R})$.  Figure~\ref{fig:Qbar} shows
$\bar{Q}\,e^{(3-\mu)\tau}$, the natural scale-invariant measure of the
pseudo-charge. Fig.~\ref{fig:Qbar_avg_density} shows the corresponding
rescaled average charge density
$\bar{Q}e^{-(1+\mu)\tau}/{R}^4$. Obviously, these are just different
plots of the same data, but $\bar Q$ vanishes at $R=0$, while $\bar
Q/R^4$ is finite there, thus giving the behaviour near the origin more
prominence: it appears that the periodicity is clearer near the origin
than near the light cone.

\subsubsection{Universality in uncharged collapse}


To demonstrate the universality of the (real) critical solution and
its decaying imaginary perturbation, we investigate four distinct
families of initial data of the form~(\ref{initial_data}). These
families are summarized in Table~\ref{tab:id_families}. There, we
denote the initial data \eqref{linear} by ``lin'', \eqref{Gaussian} by
``Gau'' and \eqref{pow} by ``pow''. Furthermore, we denote $d\cdot
g'(x)$, where $g(x)$ is a Gaussian, by dGaudx. (We have scaled it by
the width $d$ of the Gaussian so that it remains dimensionless).

\begin{figure}[h]
\includegraphics[width=\linewidth]{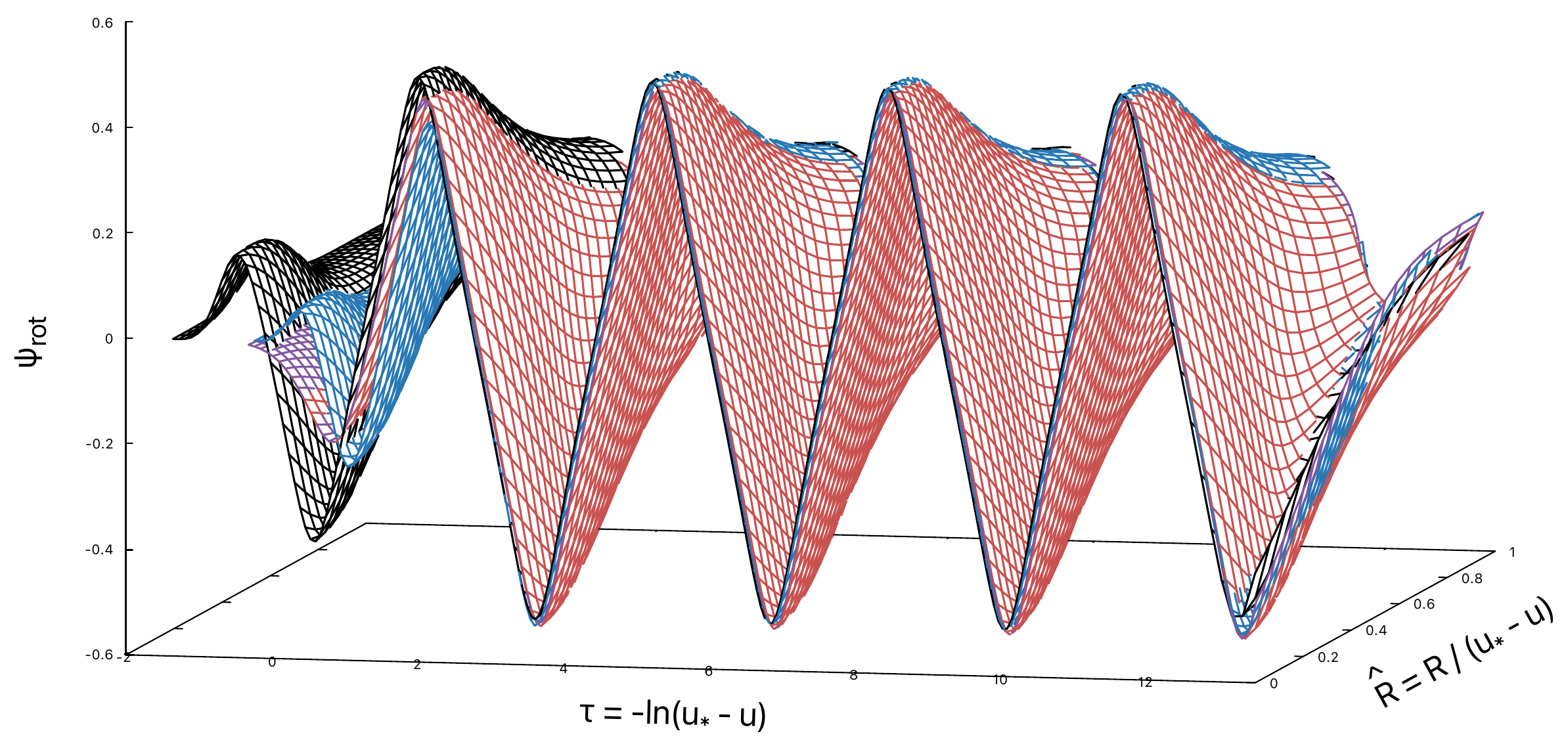}
\caption{$\psi_{\text{rot}}(\tau, \hat R)$, as in
  Fig.~\ref{fig:phase_rotated_fields} but with the four families
  mentioned in Table~\ref{tab:id_families}.}
\label{fig:universality2_psirot}
\end{figure}

\begin{figure}[h]
\includegraphics[width=\linewidth]{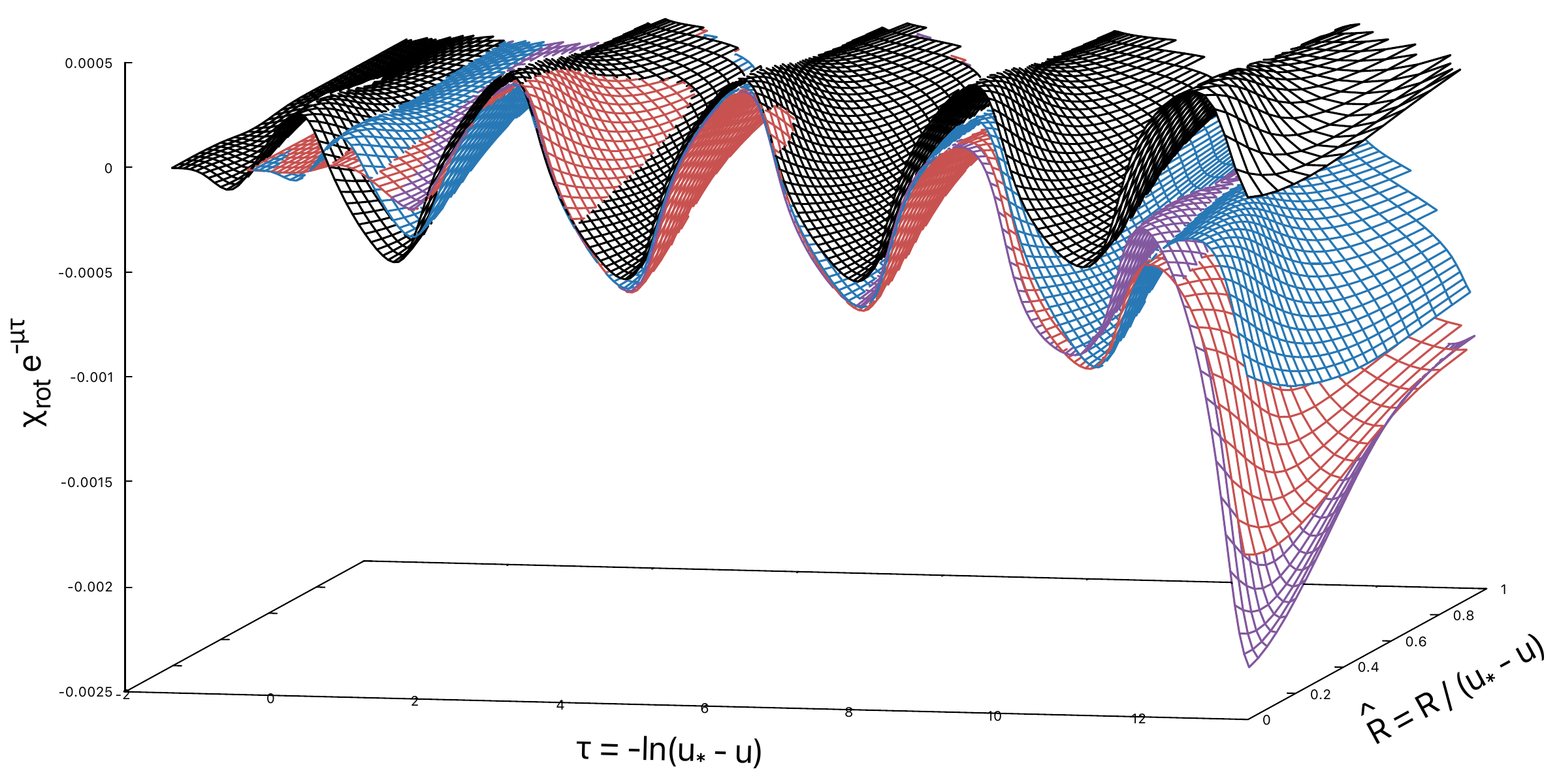}
\caption{$S\chi_{\text{rot}}e^{-\mu\tau}(\tau, \hat R)$, as in
  Fig.~\ref{fig:decay_canceled_field} but with the four families
  mentioned in Table~\ref{tab:id_families}.}
\label{fig:universality2_chirot}
\end{figure}

\begin{figure}[h]
\includegraphics[width=\linewidth]{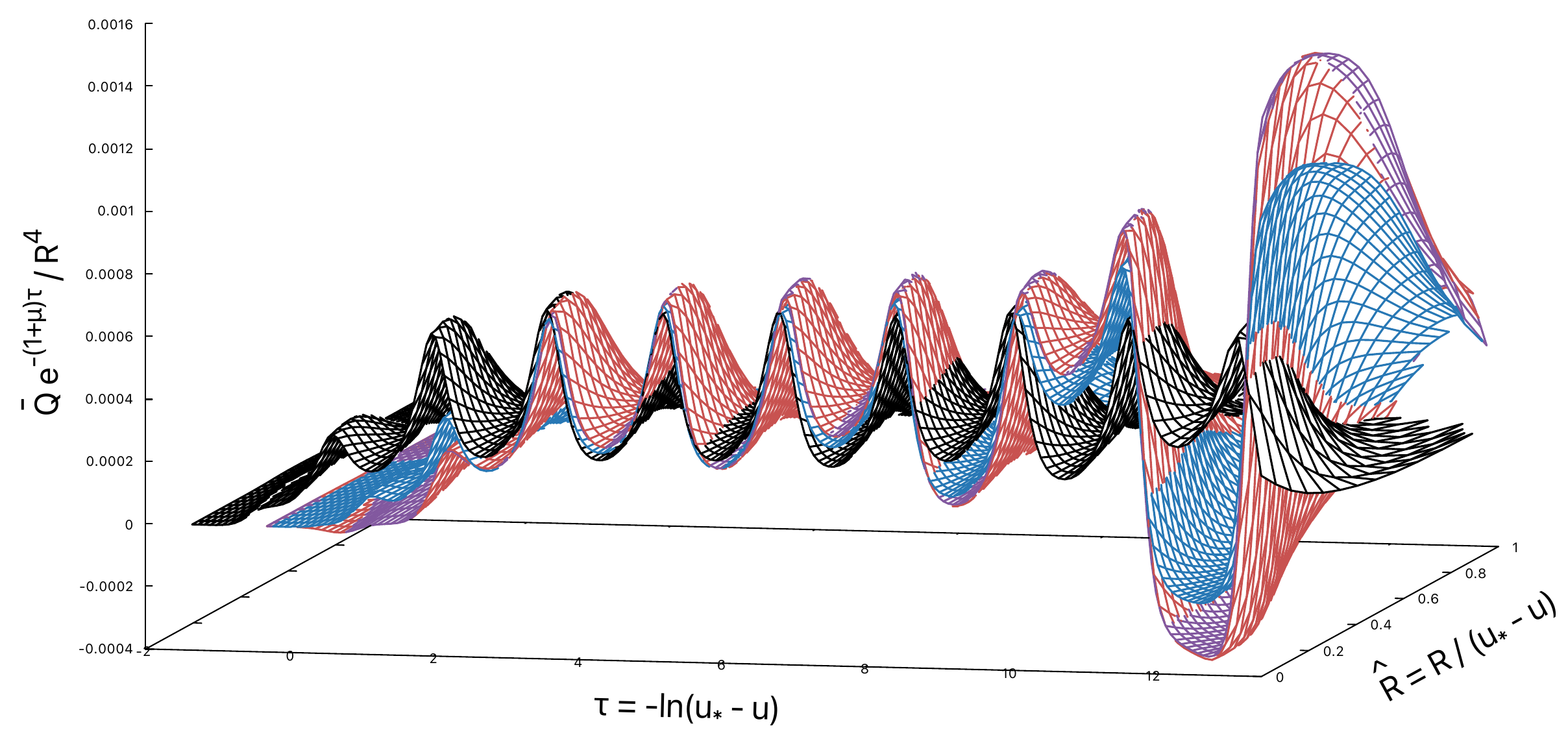}
\caption{$S(\bar Q e^{-(1+\mu)\tau}/R^4)(\tau, \hat R)$, as in
  Fig.~\ref{fig:Qbar_avg_density} but with the four families mentioned
  in Table~\ref{tab:id_families}.}
\label{fig:universality2_chargedensity}
\end{figure}

For each family we fine-tune to the collapse threshold $p_*$ and
extract the near-critical sub-threshold solution. As mentioned above,
the value $u_*$ of the echo-accumulation retarded time, and the
complex phase $\alpha$, are determined independently for each family
by considering the last dispersing solution: $u_*$ is chosen so that
$Te^{-2\tau}$ shows the clearest periodic behaviour with nearly equal
peaks, as done in Fig.~\ref{fig:T_scalar} for Family~1, while $\alpha$
is chosen so that $\psi_{\text{rot}}$ and $\chi_{\text{rot}}
e^{-\mu\tau}$ show clear periodic behaviour, with $\chi_{\text{rot}}$
decaying at large $\tau$. The $\tau$ shifts $\Delta\tau$ are measured
with respect to Family~1, from the plot of $\psi_{\text{rot}}$. The
results are displayed as functions of $(\tau,\hat{R})$ in
Figs.~\ref{fig:universality2_psirot}, \ref{fig:universality2_chirot},
and \ref{fig:universality2_chargedensity}. As already in the previous
figures, we plot for $-2\le\tau\le13.5$ and $0\le\hat R\le 1$.

Fig.~\ref{fig:universality2_psirot} displays $\psi_{\text{rot}}$ for all
four families.  The amplitude-matched profiles are superimposed
without any vertical offset. The plots coincide. This confirms that,
after the transient determined by initial conditions is absorbed into
the shift, every family converges to the same background critical solution.

Fig.~\ref{fig:universality2_chirot} displays
$Se^{-\mu\tau}\chi_{\text{rot}}$, using the reference exponent
$\mu\simeq-0.448$ computed from Family~1. After adjusting the
family-dependent constant $S$, the plots coincide. This confirms that
the damping exponent $\mu$ is universal, and that
$e^{-\mu\tau}\chi_\text{rot}$ has a universal periodic structure. From
the plot, we see that $e^{-\mu\tau}\chi_{\text{rot}}$ changes sign
after a $\tau$ translation of $\Delta/2$, similar to
$\psi_{\text{rot}}$.

Fig.~\ref{fig:universality2_chargedensity} displays the scaled average
pseudo-charge density $S\bar Qe^{-(1+\mu)\tau}/{R}^4$. This confirms the universality of
the rescaled $\chi_\text{rot}$ in a gauge-invariant way. Since
  both $\psi_\text{rot}$ and $e^{-\mu\tau}\chi_{\text{rot}}$ change
  sign with a $\tau$ translation of $\Delta/2$, $\bar{Q}$, which is
  linear in both, is periodic with period $\Delta/2$.

\subsection{Charged scalar field}
\label{subsection:sphericalscalarcollapse}


\subsubsection{Fine structure analysis for charged collapse}


We now turn our attention to the genuinely charged spherical scalar
field. Our results are qualitatively similar to those presented
in~\cite{GundlachMartel26}, who also considered Gaussian initial
data. With $q=1/\sqrt{4\pi}\simeq 0.282$, they found that, on the
collapse side, the black hole charge always had the same sign as the
initial total charge, but with $q=10/\sqrt{4\pi} \simeq 2.821$ the
black hole charge changed sign, approximately periodically in
$\ln(p-p_*)$ as the collapse threshold is approached, while still
showing an overall power-law scaling. In other words, $\exp\Psi_Q$
changed sign approximately periodically.

In our $4+1$ dimensional simulations, we set single Gaussian initial
data with the same single Gaussian envelope as defined in
\eqref{Gaussian}, with amplitude $\mathcal{A}=0.05$, centre $c =
1.5$, and width $\sigma= 0.5$. The complex phase function is
\eqref{linear}, but now with frequency $\omega=2$, and we initially set
$q=10/\sqrt{4\pi} \simeq 2.821$. We set $x_0=2.756250$ and
$x_\text{max}=5$, $N_x=350$, and find $p_* \simeq 0.722$. For the
near-critical solution at $p \simeq p_*$, the initial mass and charge
evaluated on the initial slice $u=0$ are $M\simeq0.107$ and
$Q\simeq0.069$. At this value of $q$, the black hole charge still
always has the same sign. 

\begin{figure}[h]
\includegraphics[width=\linewidth]{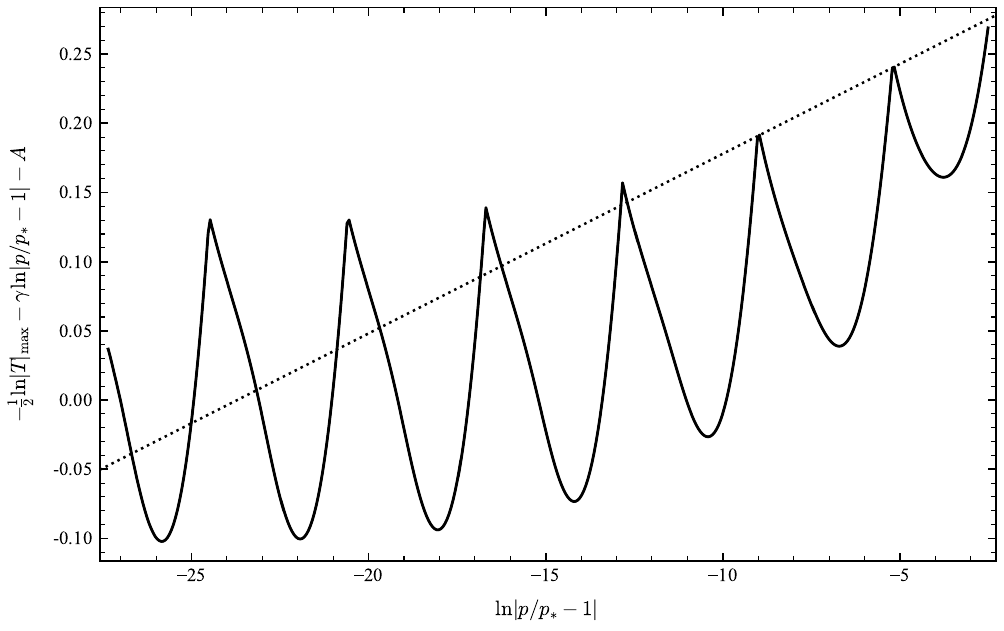}
\caption{Maximum curvature fine-structure for the family of initial
  data of Sec.~\ref{subsection:sphericalscalarcollapse}, with
  $q\simeq2.821$: we plot $\Psi_T=-\frac{1}{2} \ln |T|_{\text{max}} -
  \gamma \ln |p/p_{*} - 1| - A$ against $\ln|p/p_{*} - 1|$ with $A
  \simeq 0.566$. Following \cite{GundlachMartel26}, we fitted the thin
  dotted line by eye to the first two peaks on the right. This yields
  a slope of $\Delta\gamma \simeq 0.0130$, indicating that the
  curvature critical exponent is slightly larger at low fine-tuning.}
\label{fig:maximum_curvature_fine_structure_q10o2sqrtpi}
\end{figure}

\begin{figure}[h]
\includegraphics[width=\linewidth]{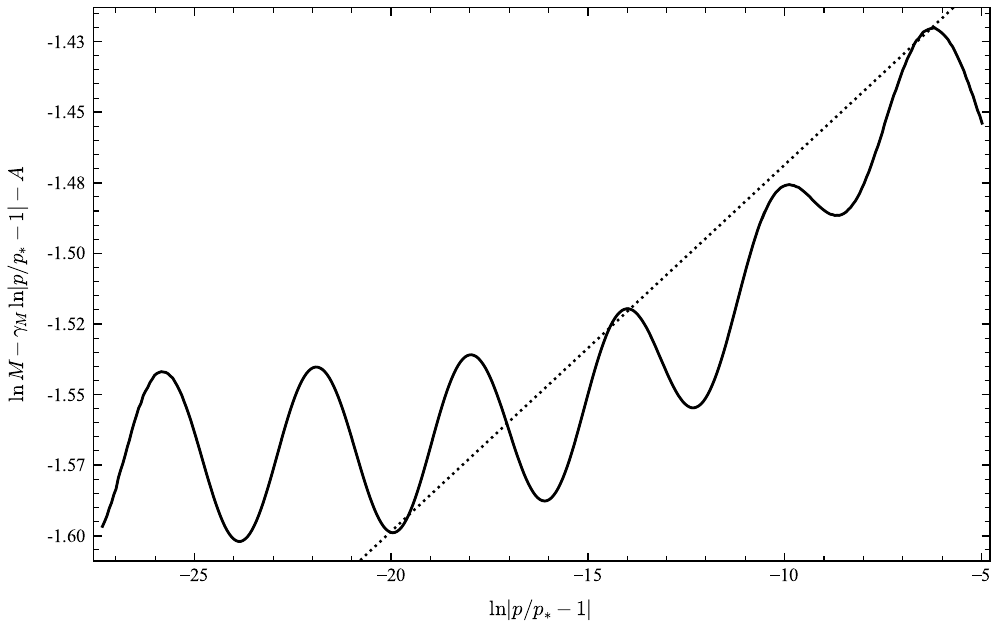}
\caption{FMOTS mass fine-structure for the same family of initial data
  and parameters as
  in~Fig.\ref{fig:maximum_curvature_fine_structure_q10o2sqrtpi}. The
  thin dotted line has the slope deduced from the curvature
  fine-structure of
  Fig.~\ref{fig:maximum_curvature_fine_structure_q10o2sqrtpi}.}
\label{fig:bh_mass_fine_structure_q10}
\end{figure}

\begin{figure}[h]
\includegraphics[width=\linewidth]{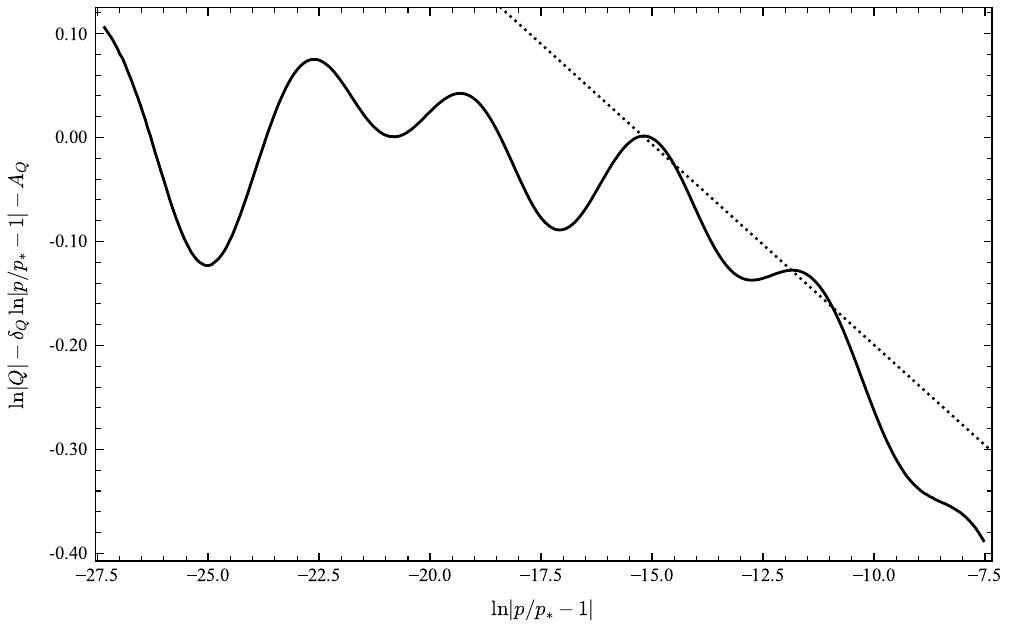}
\caption{FMOTS charge fine-structure for the same family of initial
  data as in
  Fig.~\ref{fig:maximum_curvature_fine_structure_q10o2sqrtpi}.  We
  plot $\Psi_Q=\ln |Q| - \delta_Q \ln|p/p_{*} - 1| - A_Q$ against
  $\ln|p/p_{*} - 1|$ with $A_Q\simeq1.022$. Following
  \cite{GundlachMartel26}, we have fitted the thin dotted line by eye
  to the first two peaks on the right. This yields a slope of
  $\Delta\delta_Q \simeq -0.0386$, indicating that the charge critical
  exponent is smaller at low fine-tuning.}
\label{fig:bh_charge_fine_structure_q10}
\end{figure}

The fine-structure of critical scaling on the dispersal side of the
threshold of collapse is shown in
Fig.~\ref{fig:maximum_curvature_fine_structure_q10o2sqrtpi} for the
maximum curvature diagnostic $|T|_\text{max}$, defined in
\eqref{curv_fine_struc_T}. On the collapse side,
Figs.~\ref{fig:bh_mass_fine_structure_q10} and
\ref{fig:bh_charge_fine_structure_q10} show the fine-structure for
$M$, defined in \eqref{mass_fine_struc} and $Q$, defined in
\eqref{charge_fine_struc}, of the FMOTS. The family-dependent offsets
$A$ and $A_{Q}$ for
Fig.~\ref{fig:maximum_curvature_fine_structure_q10o2sqrtpi} and
Fig.~\ref{fig:bh_charge_fine_structure_q10} respectively, have been
fitted so the mean of the fine structures is roughly zero at high
fine-tuning. We use the values $\gamma \simeq 0.41325$ and $\delta_Q =
1.425$ that we found for the uncharged scalar field.

At high fine-tuning, the fine structures are periodic, thus confirming
the values of $\gamma_M=2\gamma$ and $\delta_Q$ we extracted from real
initial data. The fine structure for $|T|_\text{max}$ in
Fig.~\ref{fig:maximum_curvature_fine_structure_q10o2sqrtpi} has the
same amplitude of about $0.23$ and shape as in
Fig.~\ref{fig:curv_fine_structure}, where we showed the fine structure
of $|T|_\text{max}$ for a 1-parameter family of real, spherically
symmetric initial data. At low fine-tuning, the curve has the same
periodic structure but rises slightly with $\ln(1 - p/p_{*})$.  The
same observation applies to the periodic fine structure of $M$, which
has a peak to peak amplitude $\simeq 0.06$ at high fine tuning from
Fig.~\ref{fig:bh_mass_fine_structure_q10}, which is compatible with
the observation from Fig.~\ref{fig:mass_fine_structure_sin_fit}. We
find that the fine structure in $\ln{|Q|}$ also appears to be
continuous, with a peak to peak amplitude approximately in the range
$0.2$ to $0.225$ at high fine-tuning, see
Fig.~\ref{fig:bh_charge_fine_structure_q10}.

Heuristically, these deviations from the expected scaling laws can be
described as modifications of the critical exponents
$\gamma_M=2\gamma$ and $\delta_Q$.  To quantify these, we have fitted
the thin dotted lines in
Figs.~\ref{fig:maximum_curvature_fine_structure_q10o2sqrtpi} and
\ref{fig:bh_charge_fine_structure_q10} by eye, obtaining slopes
$\Delta\gamma \simeq 0.0130$ and $\Delta\delta_Q \simeq -0.0386$,
respectively.  For the mass fine structure of
Fig.~\ref{fig:bh_mass_fine_structure_q10} we do not fit the slope
independently; instead we draw the dotted line with slope
$\Delta\gamma_M=\Delta\gamma \simeq 0.0130$.

\begin{figure}[h]
\includegraphics[width=\linewidth]{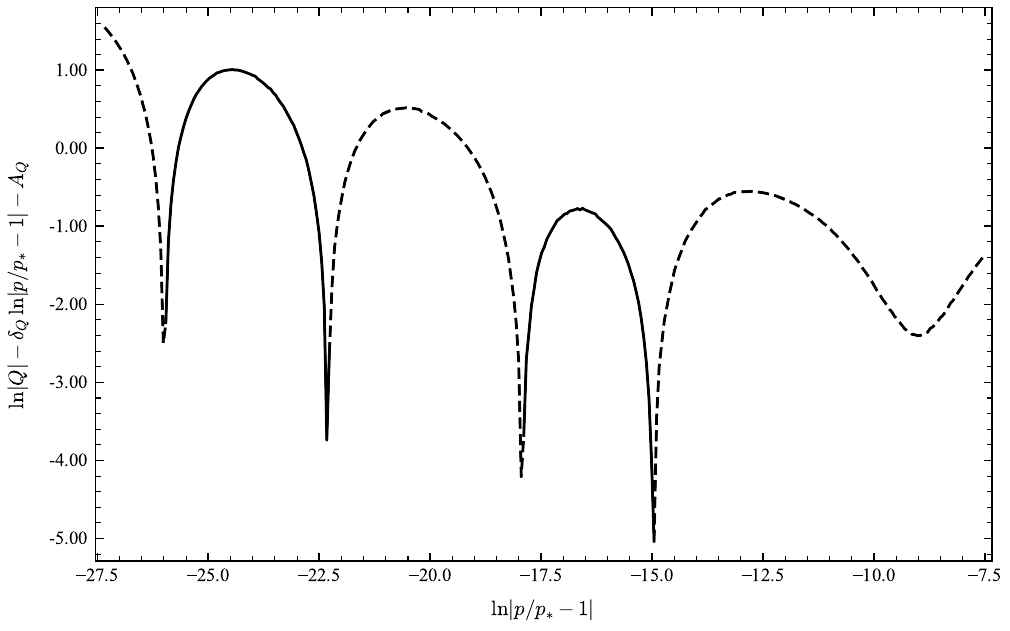}
\caption{FMOTS charge fine-structure for a different family of complex
  Gaussian initial data. The initial scalar field profile is identical
  to that of Sec.~\ref{subsection:sphericalscalarcollapse}, but with
  $q\simeq7.052$ which is $2.5$ times larger than in the family shown
  in Fig.~\ref{fig:bh_charge_fine_structure_q10}. We set $A_Q \simeq
  -0.041$ for the plot, leaving everything else as in
  Fig.~\ref{fig:bh_charge_fine_structure_q10}. The charge is positive
  where the curve is solid and negative where it is dashed.}
\label{fig:charge_fine_structure_q25}
\end{figure}

With the same initial data, but now with $q=25/\sqrt{4\pi} \simeq
7.052$, we find $p_* \simeq 1.007$. We plot its charge fine structure
in Fig.~\ref{fig:charge_fine_structure_q25}. The charge scaling
exponent $\delta_Q$ is consistent with the value found before, but the
charge now changes sign approximately periodically in $\ln|p/p_* -
1|$. The mass scaling, by contrast, remains unchanged from the
$q\simeq2.821$ family, and we therefore do not show it here. This
change of sign of $Q$ at large $q$ is similar to that observed in
\cite{GundlachMartel26}, and we do not explore this non-universality
further here.

\subsubsection{Universality in charged collapse}


When the field equations are expressed in similarity coordinates
adapted to the self-similar solution, $q$ and $e^{-\tau}$ always
appear together in the field equations
\cite{GundlachMartin96}. Therefore, on increasingly small scales, as
$\tau\to\infty$, the value of $q$ becomes irrelevant, and the fine
structures should become identical for different values of $q$ in the
high fine-tuning regime.

\begin{figure}[h]
\includegraphics[width=\linewidth]{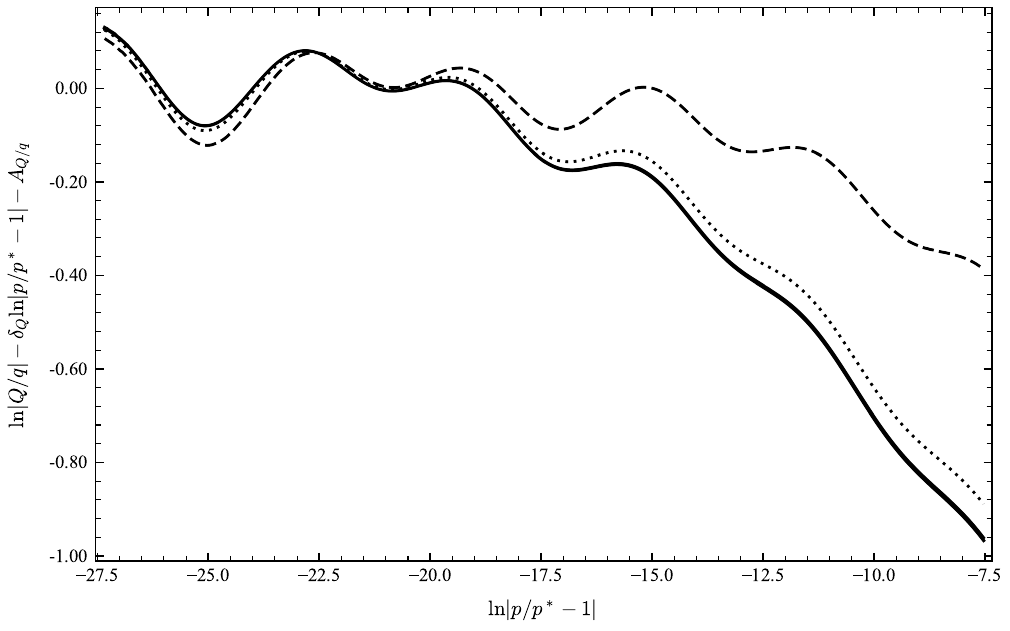}
\caption{FMOTS normalized charge fine-structure combination $\ln|Q/q|
  - \delta_Q\ln|p/p_*-1| - A_{Q/q}$ plotted against $\ln|p/p_*-1|$ for
  six values of the coupling constant, using the Gaussian initial-data
  family of Sec.~\ref{subsection:sphericalscalarcollapse} with
  $\delta_Q = 1.425$. Dashed: $q=10/(2\sqrt{\pi})\simeq2.821$,
  $A_{Q/q} \simeq -0.015$; dotted: $q=1$, $A_{Q/q} \simeq 0.823$; solid: $q
= 1/(2\sqrt{\pi})$, $A_{Q/q} \simeq 0.928$; $q=0.1$, $A_{Q/q} \simeq
0.936$; $q=0.01$, $A_{Q/q} \simeq 0.937$; and $q = 0.001$, $A_{Q/q}
\simeq 0.937$.  All six curves are superposed without horizontal
shifts. The four lowest-coupling curves (solid) overlap closely,
confirming $q$-universality.}
\label{fig:universality_q}
\end{figure}

In Figure~\ref{fig:universality_q}, we plot the normalized charge
$(Q/q)$ fine structures for the family of initial data introduced in
Sec.~\ref{subsection:sphericalscalarcollapse}, evaluated across a
range of coupling constants: $q\in\{2.821, 1, 0.282, 0.1, 0.01,
0.001\}$. This supports the universality with respect to $q$ predicted
in \cite{GundlachMartin96}. For the two largest values of $q$,
however, this universality breaks down. See also
Fig.~\ref{fig:charge_fine_structure_q25} for the even larger
$q\simeq7.052$.

\begin{figure}[h]
\includegraphics[width=\linewidth]{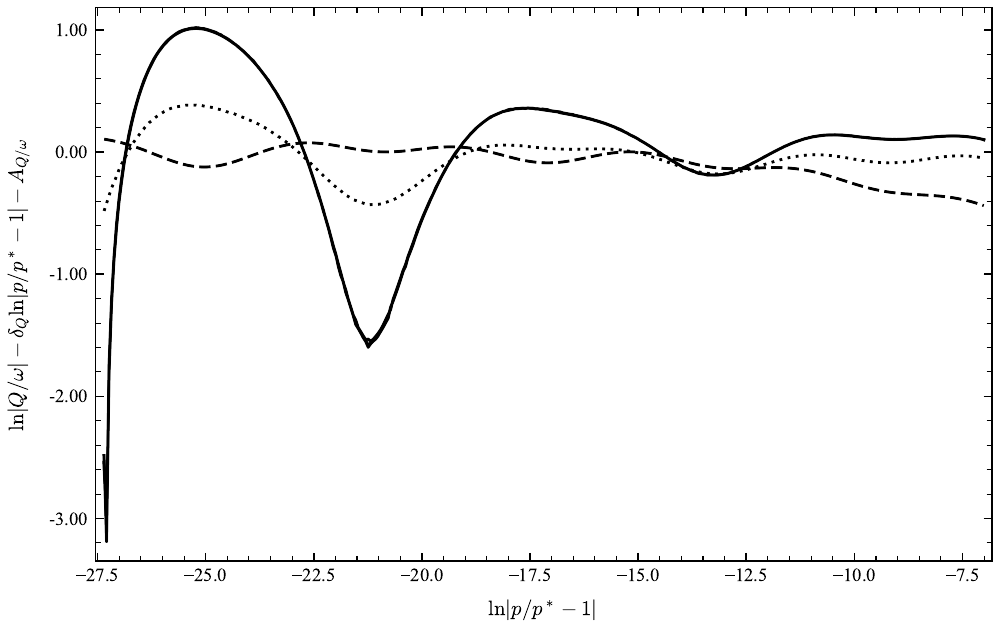}
\caption{FMOTS normalized charge fine-structure combination
  $\ln|Q/\omega| - \delta_Q\ln|p/p_*-1| - A_{Q/\omega}$ plotted
  against $\ln|p/p_*-1|$ for fixed $q\simeq 2.821$ and five values of
  $\omega$, using the Gaussian initial-data family of
  Sec.~\ref{subsection:sphericalscalarcollapse} with $\delta_Q =
  1.425$. Dashed: $\omega=2$ (same curve as in
  Fig.~\ref{fig:universality_q}), $A_{Q/\omega} \simeq 0.326$; dotted:
  $\omega=1$, $A_{Q/\omega} \simeq -0.082$; solid: $\omega=0.1$,
  $A_{Q/\omega} \simeq -0.329$; $\omega=0.01$, $A_{Q/\omega} \simeq
  -0.331$; and $\omega=0.001$, $A_{Q/\omega} \simeq -0.331$. The
  curves for the three lowest values of $\omega$ all overlap.}
\label{fig:universality_omega}
\end{figure}

However, in the derivation of the charge scaling in
\cite{GundlachMartin96}, reviewed above in
Sec.~\ref{sec:typeII_recap}, the assumption is not that $q$ is small,
but that $\chi$ is small relative to $\psi$ (after a suitable constant
phase rotation). This assumption can be violated even with $q=0$, that
is for a complex but uncharged scalar field, and we therefore return
to $q=0$. To control the size of $\chi$ instead, we vary the
parameter $\omega$ in the real data. (We accept that this changes not
only the initial amplitude of $\chi$, but also its shape.)  In
Fig.~\ref{fig:universality_omega} we plot the normalized charge
$(Q/\omega)$ fine structures for the family of initial data introduced
in Sec.~\ref{subsection:sphericalscalarcollapse}, evaluated across a
range of parameter values: $\omega \in \{2, 1, 0.1, 0.01, 0.001\}$,
while keeping the coupling constant fixed at $q\simeq 2.821$. Much
like the previous analysis, the plot demonstrates a strict overlap of
the black hole charge fine structures for $\omega \le 0.1$, thereby
confirming universality within the low-$\omega$ regime. In contrast,
at the two largest values of $\omega$, we observe significant
deviations, as before at large $q$.

\section{Extremal collapse}
\label{sec:extremal}


\subsection{The formation of an extremal black hole in collapse}


Having established that a single, suitably tuned pulse collapses to
form a sub-extremal black hole, we now turn our attention to the
formation of extremal black holes. Gelles and
Pretorius~\cite{GellesPretorius26} set $Q_0\gtrsim\mathcal{M}_0$ on
the bottom corner of their null rectangle setup, put a small scalar
field on the right edge of the null rectangle, and fine-tuned $Q_0$ to
form an exactly extremal horizon. We attempted to imitate this in
genuine collapse by initial data with two widely separated Gaussians:
the first creates the corner data $Q_0$ and $\mathcal{M}_0$, and the
second is the equivalent to the initial data on the right boundary of
the null rectangle. However, the null rectangle of
\cite{GellesPretorius26}, with zero or small scalar field on its left
boundary, cannot be embedded into genuine collapse, and conversely our
data would create large scalar field data on the left edge of an
embedded null rectangle.
  
To imitate $Q_0\gtrsim\mathcal{M}_0$, we explore initial data
satisfying the super-extremal condition $Q_{,x} > \mathcal{M}_{,x}$ on
the initial null slice $u=0$, which implies $Q/\mathcal{M}>1$. From
equations~\eqref{Qeqn} and \eqref{calMxexpr} for a complex Gaussian of
width $\sigma$ and frequency $\omega$, this condition roughly requires
$\sigma > 2/q$, with the constraint sharpest at the optimal frequency
$\omega=q/2$. The two-Gaussian parameters in
Table~\ref{tab:initial_params_sweep} satisfy this: with $q=0.8$ the
constraint gives $\sigma > 2.5$, which both widths $\sigma_1 =
\sigma_2=4.5$ satisfy comfortably, and the frequencies $\omega_1 =
\omega_2=0.4$ coincide with the optimal value $q/2=0.4$. 

Through numerical experimentation with this two-Gaussian family, we
found a configuration showing clear evidence of extremal black hole
formation, with $1-(Q/\mathcal{M})_{\text{FMOTS}}\sim 10^{-5}$ already
at resolution $h=0.15$, where $(Q/\mathcal{M})_{\text{FMOTS}}$ is the
value of $Q/\mathcal{M}$ at the FMOTS, see the first row of
Table~\ref{tab:perturbed_study}. We can push
$(Q/\mathcal{M})_{\text{FMOTS}}$ as closely to one as we wish, but we
require increasing resolution and, at each resolution, a bisection
search in $p$ to machine precision for this, see
Table~\ref{tab:resolution_convergence}. (The notation $\hat p_1$ for
the threshold value of $p$ on a finite numerical domain will be
motivated below.)

Crucially, extremality does not occur at the Type-II critical
threshold but well inside the collapse threshold. We expect that
extremality can be achieved with a single Gaussian. However, the
two-Gaussian family gives us a larger parameter space, eight
independent Gaussian parameters together with the coupling constant
$q$, in which we can probe the codimension-1 nature of the extremal
threshold. 

\begin{table}[h]
\renewcommand{\arraystretch}{1.3}
\setlength{\tabcolsep}{8pt}
\caption{Initial parameters for the two-Gaussian initial data that
  result in $(Q/{\cal M})_{\text{FMOTS}}\simeq 1$.}
\vspace{10pt}
\label{tab:initial_params_sweep}
\begin{tabular}{l||c|c}
    & Gaussian~1 & Gaussian~2 \\
  \hline
  \quad $\mathcal{A}_i$               & 0.008   & 0.036485 \\
  \quad $c_i$               & 24.5    & 64.5     \\
  \quad $\sigma_i$          & 4.5     & 4.5      \\
  \quad $\omega_i$          & 0.4     & 0.4      \\
\hline
  \quad $q$ & \multicolumn{2}{c}{0.8} \\
\end{tabular}
\end{table}

\begin{table}[h]
\caption{The 1-parameter families we have tuned to extremality. The
  values of all parameters are as in
  Table~\ref{tab:initial_params_sweep}, except for the parameter $X$
  that is being varied with $p$ scaling $X$. We set
  $x_{\text{max}}=120$ and $x_0=117.5$ with $h:=\Delta x=0.15.$ In
  the first family, ${\cal A}_2$ is tuned. We tune the other 8
  parameters, one by one, for two values of ${\cal A}_2$ that lie on
  either side of the extremal value $\mathcal{A}_{2,\text{crit}}=\hat
  p_1\mathcal{A}_2=0.036408$ of the first family. This means that the
  resulting additional 16 families all intersect the extremality
  hypersurface at different points, as well as in different
  directions. In each family, $p$ is evaluated at a fixed small
  increment $\delta p=6\times10^{-4},4.8\times10^{-4}$ for
  $\mathcal{A}_2=0.0365,0.0362$, respectively. Comment 1: Scanning
  over $\sigma_1$ did not reach extremality for
  $\mathcal{A}_2=0.0362$, so for this family we used
  $\mathcal{A}_2=0.0363$. Comment 2: $\sigma_2$ yields two distinct
  extremal solutions at each of the two values of $\mathcal{A}_2$.}
\label{tab:perturbed_study}
\renewcommand{\arraystretch}{1.3}
\setlength{\tabcolsep}{7pt}
\begin{tabular}{l||l|l|l|l}
  $X$ & $X_{\text{init}}$ & $\mathcal{A}_2$ & $\hat p_{1}$ &
  $\!1\!-\!\left(\frac{Q}{\mathcal{M}}\right)_{\text{\tiny FMOTS}}$ \\
  \hline
${\cal A}_2$ & \multicolumn{2}{c|}{0.036485} & 0.9979 & $3.801\times10^{-5}$ \\
  \hline
  \multirow{2}{*}{$\mathcal{A}_1$}
    & 0.008 & 0.0365 & 0.9036 & $1.86\times10^{-6}$ \\
    & 0.008 & 0.0362 & 1.1697 & $5.22\times10^{-6}$ \\
  \hline
  \multirow{2}{*}{$c_1$}
    & 24.5 & 0.0365 & 0.7955 & $3.78\times10^{-6}$ \\
    & 24.5 & 0.0362 & 1.4617 & $1.62\times10^{-6}$ \\
  \hline
  \multirow{2}{*}{$c_2$}
    & 64.5 & 0.0365 & 0.9949 & $4.97\times10^{-5}$ \\
    & 64.5 & 0.0362 & 1.0118 & $1.73\times10^{-5}$ \\
  \hline
  \multirow{2}{*}{$\omega_1$}
    & 0.4 & 0.0365 & 0.8700 & $2.66\times10^{-6}$ \\
    & 0.4 & 0.0362 & 1.2158 & $4.83\times10^{-6}$ \\
  \hline
  \multirow{2}{*}{$\omega_2$}
    & 0.4 & 0.0365 & 0.9979 & $4.35\times10^{-5}$ \\
    & 0.4 & 0.0362 & 1.0050 & $8.87\times10^{-5}$ \\
  \hline
  \multirow{2}{*}{$q$}
    & 0.8 & 0.0365 & 1.0027 & $7.31\times10^{-5}$ \\
    & 0.8 & 0.0362 & 0.9935 & $7.22\times10^{-5}$ \\
  \hline
  \multirow{2}{*}{$\sigma_1$}
    & 4.5 & 0.0365 & 0.1996 & $2.71\times10^{-5}$ \\
    & 4.5 & 0.0363(!) & 1.3556 & $1.41\times10^{-6}$ \\
  \hline
  \multirow{4}{*}{$\sigma_2$}
    & 4.5 & 0.0365 & 1.0201 & $1.45\times10^{-5}$ \\
    & 4.5 & 0.0365 & 1.1180 & $6.02\times10^{-6}$ \\
    & 4.5 & 0.0362 & 0.9753 & $1.18\times10^{-5}$ \\
    & 4.5 & 0.0362 & 1.1971 & $2.16\times10^{-5}$ \\
\end{tabular}
\end{table}

\begin{table}[h]
\renewcommand{\arraystretch}{1.3}
\setlength{\tabcolsep}{8pt}
\caption{Convergence of $1-(Q/\mathcal{M})_{\text{FMOTS}}$ with
  resolution $h=\Delta x$, but at fixed $x_\text{max}=120$ and
  $x_0=117.5$, for the initial data considered in
  Table~\ref{tab:initial_params_sweep}, finetuned by scaling
  $\mathcal{A}_2$ with $p$.  In contrast to
  Table~\ref{tab:perturbed_study}, and only for this Table, $\hat p_1$
  is now obtained by bisection down to machine precision. Only five
  digits of $\hat p_1$ are shown, to indicate how it depends on
  $h$. We use a fixed time step $\Delta u=C_0 \Delta x$ with $C_0 =
  0.1$.}
\label{tab:resolution_convergence}
\begin{tabular}{l||l|l}
  Resolution $h$ & $\hat p_1$ & $1-(Q/\mathcal{M})_{\text{FMOTS}}$ \\
  \hline
  $0.3$      & 1.00962  & $3.157\times10^{-5}$  \\
  $0.15$     & 0.99771 & $7.853\times10^{-7}$  \\
  $0.075$    & 0.99587 & $2.848\times10^{-8}$  \\
  $0.0375$   & 0.99557 & $2.231\times10^{-9}$  \\
  $0.01875$  & 0.99551 & $2.267\times10^{-10}$ \\
  $0.009375$ & 0.99550 & $1.167\times10^{-11}$ \\
\end{tabular}
\end{table}


\subsection{Evidence for codimension-1 behaviour}


The fact that we have found initial data that form an extremal black
hole at all suggests that such initial data are codimension-1 in the
space of initial data. To check the codimension-1 hypothesis in our
8-dimensional space of initial data, we probe it by 17 1-parameter
families intersecting the extremality surface at 17 different points
and in 9 different directions. (Two of the 17 families intersect the
extremality surface twice, thus giving us 19 intersection in total,
see already Table~\ref{tab:perturbed_study}).

In each family, we see that $(Q/\mathcal{M})_{\text{FMOTS}}$ rises to
$1$ smoothly with nonvanishing derivative (that is, approximately
linearly) then jumps to a value clearly below one, varying again
smoothly on the other side of the jump. The derivative also jumps. We
refer to the (family-dependent) {\em location} of the jump as
$p_1$. We refer to the one-sided {\em limit} from the extremal side as
$p\to p_{1e}$, and the one-sided {\em limit} from the subextremal side
as $p\to p_{1s}$.

The numerical value of $p_1$ that we obtain depends not only on the
family but on our numerical method and all numerical parameters, in
particular the numerical resolution $h:=\Delta x$ and the size
$x_\text{max}$ of the numerical domain. For given numerical
parameters, we define a numerical approximation $\hat p_1$ to the true
extremal threshold $p_1$ by
\begin{equation} 
\lim_{p\to \hat p_{1e}}(Q/{\cal M})_\text{FMOTS}=1.
\label{hatp1def}
\end{equation}
In practice we can find $\hat p_1$, for given resolution and
$x_\text{max}$, to machine precision by bisection to the jump in
$(Q/{\cal M})_\text{FMOTS}$. We have done this, however, only
in Table~\ref{tab:resolution_convergence}, in order to demonstrate
convergence with $h$.

In a first family, we keep all parameters at the values where we first
found a very-near extremal black hole (from now on the ``base
values''), and vary ${\mathcal A_2}$ by multiplying it with a tuning
parameter $p$. For $2\times 8=16$ further families, we ``detune'' one
arbitrarily chosen parameter, namely ${\mathcal A_2}$, from its
extremal value: we either increase the amplitude of the second
Gaussian to $\mathcal{A}_2=0.0365$, which is on the extremal side, or
decrease it to $\mathcal{A}_2=0.0362$, which is on the subextremal
side. These correspond to $p_{\mathcal{A}_2}-1\simeq 0.00041$ and
$p_{\mathcal{A}_2}-1\simeq -0.00781$, and result in
$(Q/\mathcal{M})_{\text{FMOTS}}\simeq0.99937$ and
$(Q/\mathcal{M})_{\text{FMOTS}}\simeq0.99608$, respectively. Then we
vary each one of the other 7 parameters of the initial data, and also
$q$, in turn, one at a time, by multiplying each by $p$. We found that
we could re-tune to extremality along each of these 16
families. Moreover, we see the same kind of jump in
$(Q/\mathcal{M})_{\text{FMOTS}}$ as in the first family. (In one
family, there are two such jumps, in opposite directions.)  The scans
of $(Q/\mathcal{M})_{\text{FMOTS}}(p)$ for all 17 families are shown
in Fig.~\ref{fig:allfamilies} at the end of the paper.

\begin{figure}[h]
\includegraphics[width=\linewidth]{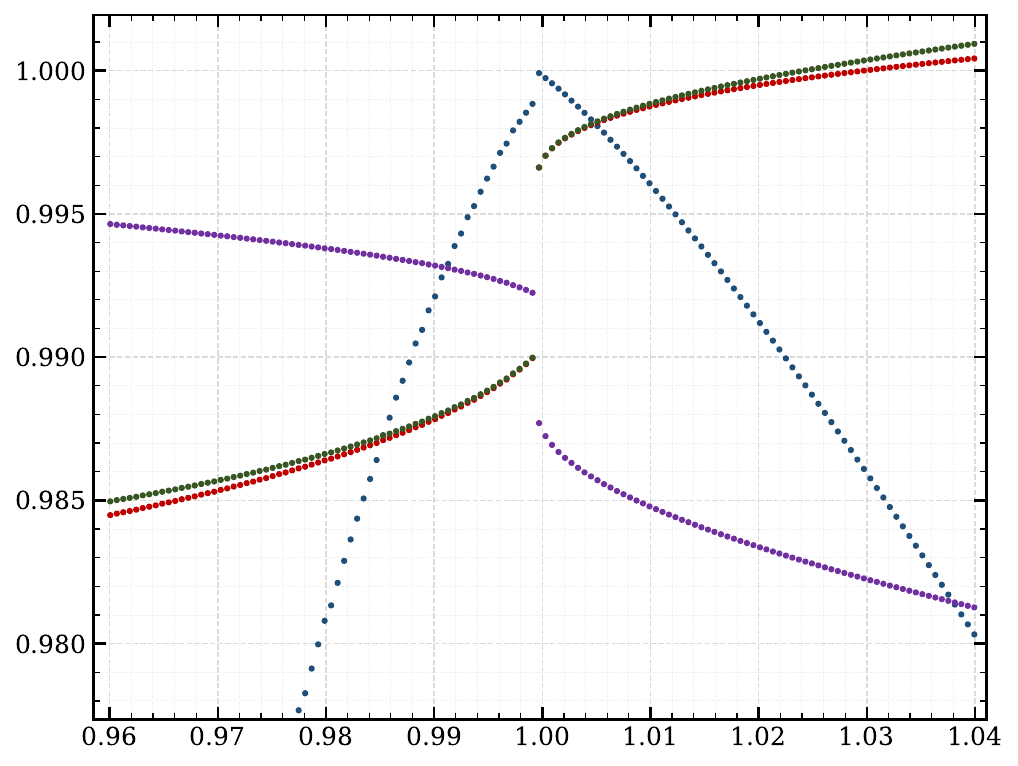}
\caption{FMOTS charge $Q$ (red), mass $\mathcal{M}$ (green),
  charge-to-mass ratio $Q/\mathcal{M}$ (blue), and retarded time
  $u_\text{FMOTS}$ of the FMOTS (purple), plotted against $p$.  For
  plotting, $Q$ and ${\cal M}$ are scaled to $\tilde{y}=s\, y + y_0$,
  with $s=10^{-5}$ and $y_0=0.975$, while $u_{\text{FMOTS}}$ is scaled
  with $s=4\times10^{-4}$ and $y_0=0.975$; the ratio $Q/\mathcal{M}$
  is plotted without rescaling.  The initial data are from the
  1-parameter family in which $p$ scales $\omega_2$, with
  $\mathcal{A}_2=0.0365$ and all remaining parameters held fixed at
  the reference values in Table~\ref{tab:initial_params_sweep}. $Q$,
  $\mathcal{M}$, and $Q/\mathcal{M}$ are all evaluated at
  $(u_\text{FMOTS},x_\text{max})$. At $p=\hat p_1 \simeq 0.998$, $Q$,
  $\mathcal{M}$ and $Q/\mathcal{M}$ jump down as $p$ crosses from the
  extremal side (here, $p>\hat p_1$) to the subextremal side (here,
  $p<\hat p_1$), while $u_\text{FMOTS}$ jumps up.}
\label{fig:jump_plot}
\end{figure}

In all 17 families, both ${\cal M}_{\text{FMOTS}}$ and ${
  Q}_{\text{FMOTS}}$ jump, as well as
$(Q/\mathcal{M})_{\text{FMOTS}}$, as does the location
$u_\text{FMOTS}$ of the first trapped sphere on the numerical
domain. This is shown for one of these families in
Fig.~\ref{fig:jump_plot}. If any sphere $(u,x)$ is (marginally)
  trapped, then from the Raychaudhuri equation any sphere at the same
  $u$ and larger $x$ is also (at least marginally)
  trapped. Therefore, the true FMOTS on a finite domain must always
be at $(u_{\text{FMOTS}}, x_{\max})$. Because of numerical error and
the shallowness of the $C=1$ contour, the first occurrence of a
  grid point with $C\ge 1$ may not be at $x_\text{max}$. We have used
that point as our FMOTS anyway, except in
Fig.~\ref{fig:jump_plot}, where we have determined the FMOTS as the
first $u$ where $C\ge 1$ at $x_\text{max}$. This makes almost no difference to the determination of the mass and charge of the FMOTS.

\subsection{Theoretical understanding}


The obvious interpretation of these numerical results is that $Q/{\cal
  M}=1$ is achieved on a codimension-1 hypersurface in a small open
region of our 8-dimensional phase space, and therefore probably on a
codimension-1 hypersurface in a small open region of the full,
infinite-dimensional phase space (in the topology of a suitable
function norm).

This interpretation is in the spirit of extremal critical collapse, as
proved for the Einstein-Maxwell-{\em uncharged} scalar field system in
a null rectangle setting in Thm.~II of
\cite{AngelopoulosKehleUnger26}, and conjectured there for the
Einstein-Maxwell-{\em charged} scalar field system in genuine
collapse: a $C^1$ function $\sigma$ is (teleologically) defined on the
space of initial data, such that data $|\sigma|\le 1$ evolve into a
black hole with $Q/{\cal M}=\sigma$ and data with $|\sigma|>1$
disperse. However, on what we have called the subextremal side of the
jump we do not observe dispersion, but a sub-extremal black hole:
\begin{equation}
\lim_{p\to \hat p_{1e}}{|Q|\over{\cal M}}=1, \qquad 
\lim_{p\to \hat p_{1s}}{|Q|\over{\cal M}}<1.
\end{equation}
The key observation is that when $p=\hat p_1$ is crossed from the
extremal to the subextremal side, not only does $(|Q|/{\cal
  M})_{\text{FMOTS}}$ jump down but also $u_\text{FMOTS}$, its
retarded time, jumps up:
\begin{equation}
\begin{split}
u_{\text{FMOTSe}}
&:= \lim_{p\to\hat p_{1e}}u_\text{FMOTS} \\
&< u_{\text{FMOTSs}}
:= \lim_{p\to\hat p_{1s}}u_\text{FMOTS}.
\end{split}
\end{equation}
This type of jump is described in Thm.~3 of \cite{KehleUnger24},
although for the Einstein-Maxwell-charged {\em Vlasov} system, and for
the event horizon, rather than the FMOTS.

\begin{figure}[h]
\includegraphics[width=\linewidth]{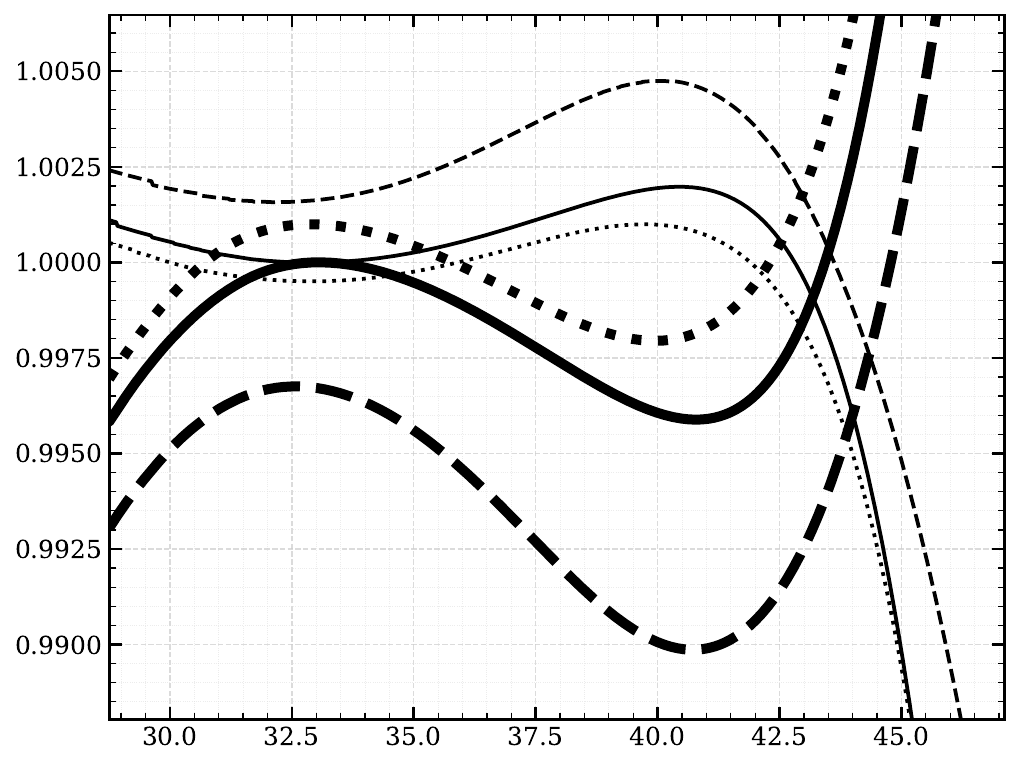}
\caption{Maximum compactness $\max_x C(u,x)$, and charge-to-mass
  $(Q/\mathcal{M})$ ratio at the point of maximum compactness, plotted
  against $u$. The initial data are from the 1-parameter family where
  $p$ scales $c_1$, with $\mathcal{A}_2=0.0365$ and other parameters
  fixed to Table~\ref{tab:initial_params_sweep} reference
  values. Thick curves denote $\max_x C$ and thin curves
  $Q/\mathcal{M}$. Dotted lines denote a value of $p$ on the extremal
  side, solid lines $p\to \hat p_{1e}$, and dashed $p\to \hat p_{1s}$.}
\label{fig:maxC_QoM_p1}
\end{figure}

As $p\to \hat p_{1e}$, the trapped surface at $u_{\text{FMOTSs}}$ is
already there, but the {\em first} trapped surface is the one at
$u_{\text{FMOTSe}}$. As $p\to \hat p_{1s}$, at $u_{\text{FMOTSe}}$
there is a surface that is almost but not quite trapped, so that now
the first trapped surface occurs at $u_{\text{FMOTSs}}$.  We can see
this by plotting $\max_xC(u,x)$ against $u$ for the two solutions just
either side of $\hat p_1$. This is shown in
Fig.~\ref{fig:maxC_QoM_p1}.

\begin{figure}[h]
\includegraphics[scale=0.35]{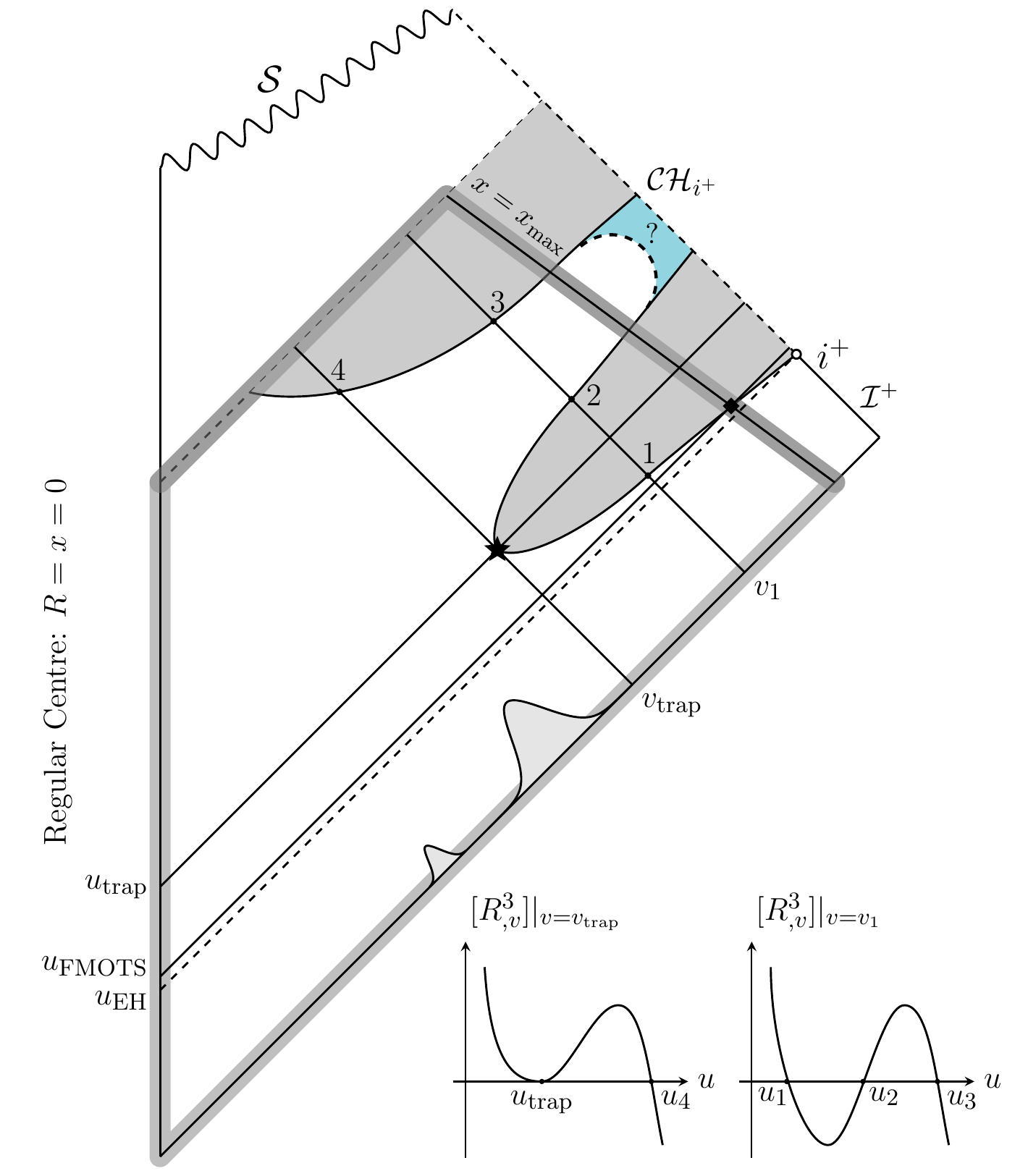}
\caption{Theoretical spacetime diagram in double-null coordinates
  $(u,v)$ with angular coordinates suppressed, illustrating black hole
  formation on the extremal side from a superposition of two-Gaussian
  initial data Table~\ref{tab:initial_params_sweep}. The fat boundary
  marks our finite numerical domain. The inset shows $R^3_{,v}$ along
  $v=v_1$ and $v=v_\text{trap}$, where the retarded times $u_i$
  correspond to points 1 to 4 marked in the main diagram. The star
  marks the extremal 3-sphere at the tip of the cigar. The diamond
  marks the FMOTS on our numerical domain, where we measure the black
  hole charge $|Q|$ and mass $\mathcal{M}$. Due to the finite
  $x_\text{max}$ of our domain, the true event horizon
  $\mathcal{H}^{+}$ (dashed line $u_{\text{EH}}$) lies behind
  $u_\text{FMOTS}$, though the FMOTS would asymptote to it as
  $x_\text{max}\to\infty$.}
\label{fig:extremal_side_spacetime}
\end{figure}

Fig.~\ref{fig:extremal_side_spacetime} gives a schematic spacetime
picture (with lightcones at 45 degrees) of a solution with the
  value of $p$ lying a small finite distance to the extremal side of
$\hat p_1$. This shows two distinct trapped regions. The boundary of
the earlier trapped region is almost extremal. From its shape, we call
this region the cigar. We call the point on its boundary with the
smallest value of $v$ the tip of the cigar. Following
\cite{GellesPretorius26}, we denote its coordinate values by
$(u_\text{trap},v_\text{trap})$.

An argument first given in \cite{MurReaTan13} shows that the 3-sphere
at the tip of the cigar is not only marginally trapped but also
extremal. In double-null coordinates, the wave equation for the area
radius $R$, Eq.~(\ref{XiReqn}), can be written as
\begin{equation}
\left(R^2R_{,v}\right)_{,u}=-GR\left(1-{Q^2\over R^4}\right).
\label{R3uv}
\end{equation}
We have $R_{,u}<0$, $R>0$ and $G>0$ everywhere, so that both sides of
the equation are negative if and only if $|Q|<R^2$. At the tip of the
cigar we must have $R_{,v}=0$ (it is marginally trapped), but also
$(R_{,v})_{,u}=0$ (because $R_{,v}>0$ at both infinitesimally larger
and smaller $u$). From (\ref{R3uv}) we then have $|Q|=R^2$, and from
this, $C=1$ and \eqref{calMdef}, we have $|Q|=\cal M$.

\begin{figure*}[htbp]
\includegraphics[width=\linewidth]{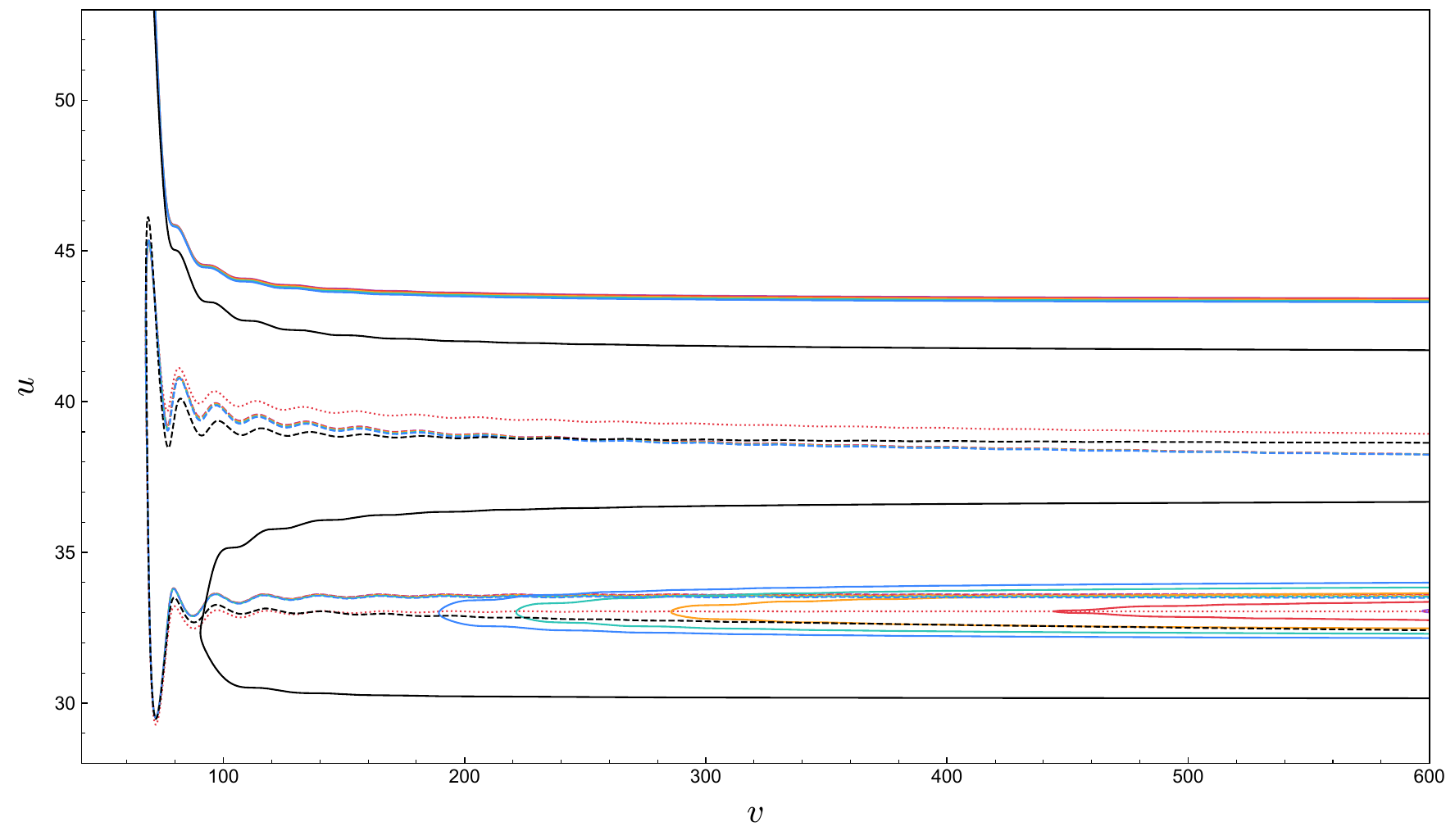}
\caption{Locations of the first and second trapped region as the
  initial parameter $p$ is tuned toward the extremal threshold $\hat
  p_1\simeq0.9487$. For these simulations, $p$ scales $c_1$, with
  $\mathcal{A}_2=0.0365$. The trapped regions are bounded by solid
  contours of compactness $C=1$. The tiny purple contour just at the
  right edge of the numerical domain corresponds to $p=\hat p_1\simeq
  0.9487$, red corresponds to $p\simeq0.9498$, and so on, up to black,
  which corresponds to $p\simeq1.059$. We also plot dashed contours
  for the ratio $Q/R^2=1$; however, we find that for the near-extremal
  $p \to \hat p_1$ case ($p\simeq0.9498$), it is the contour $Q/R^2 =
  0.9969$ (dotted red line) that passes exactly through the tip of the
  trapped region. We believe this is due to numerical error, as we do
  not explicitly use Eq.~\eqref{XiReqn} to evolve $R$. For $\hat
  p_1\le p\lesssim 1.1588$, the two trapped regions remain disjoint
  within our finite numerical domain. (They may be joined at larger
  $v$.)}
\label{fig:contour_comparison}
\end{figure*}

Fig.~\ref{fig:contour_comparison} shows the locations of the first and
second trapped region, as $p$ is tuned toward $\hat p_1$ from the
extremal side, for five values of the parameter. The trapped regions
are bounded by the compactness contour $C=1$, and the figure also
displays the locus of $|Q|/R^2=1$ whose intersection with $C=1$
marks the tip of the first trapped region. This figure, rotated
anticlockwise by 45 degrees, provides a numerical counterpart of the
schematic picture of Fig.~\ref{fig:extremal_side_spacetime}.

The cigar-shaped first trapped region recedes rapidly to the Cauchy
horizon until only the tip of the cigar is left. We will give
numerical evidence that $v_\text{trap}\to\infty$ as $p\to p_{1e}$. (By
contrast, the second trapped region depends continuously on $p$ as
$p\to p_{1e}$.) The event horizon then intersects $\mathcal{I}^+$ at
that point, but that point is extremal, and so, in a handwaving sense,
the event horizon becomes extremal as $v\to\infty$ in the true
extremal limit. The receding cigar mechanism is described explicitly
in \cite{GellesPretorius26}, in particular its Fig.~2, and is
presumably implicit in \cite{MurReaTan13} and
\cite{AngelopoulosKehleUnger26}. Fig.~\ref{fig:contour_comparison}
demonstrates that the receding cigar mechanism operates at the
extremal threshold in {\em genuine} collapse of a {\em charged} scalar
field in {\em 4+1} dimensions as well, and we expect that a proof of
this can also be given.

\begin{figure*}[htbp]
\includegraphics[width=\linewidth]{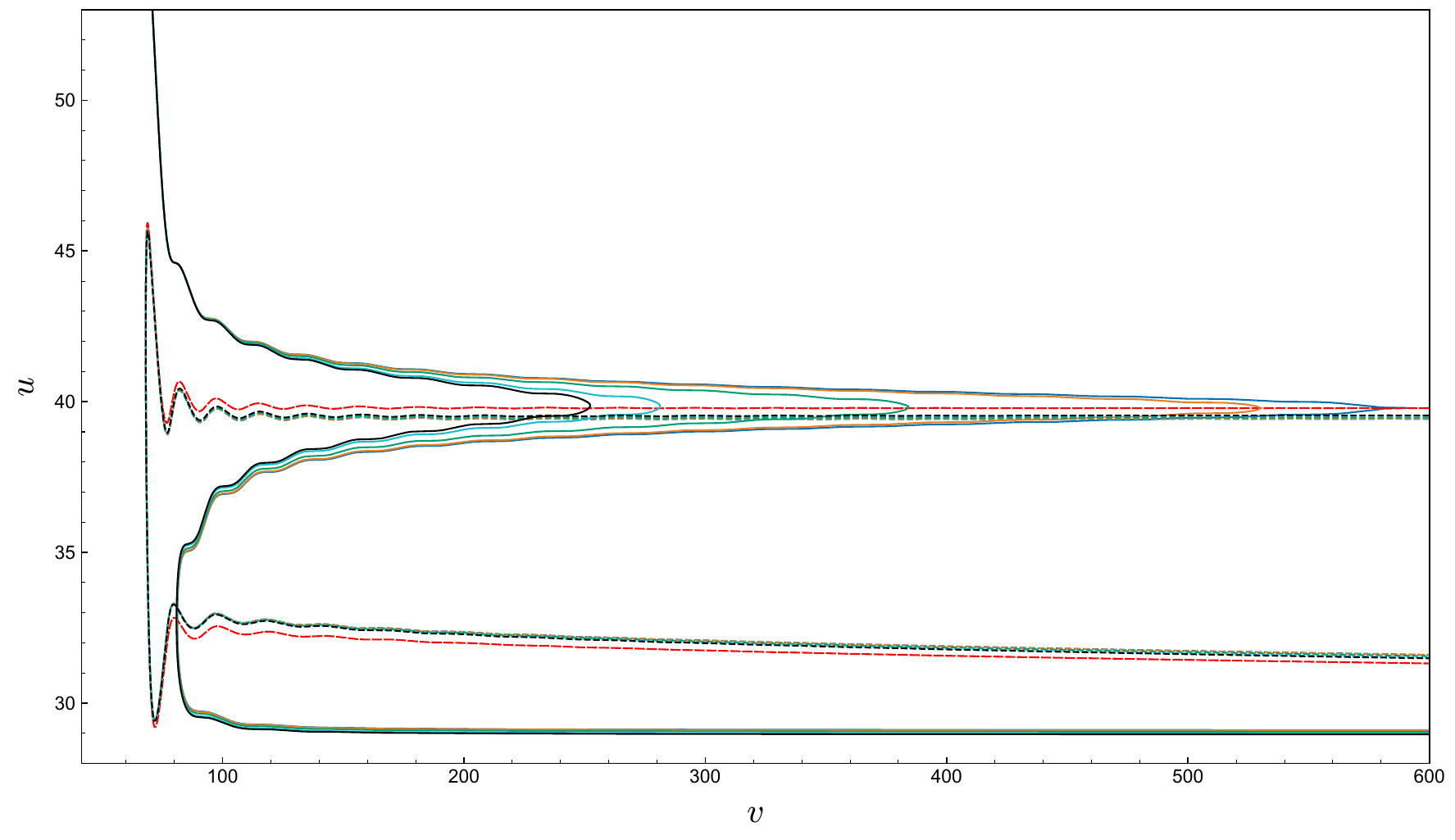}
\caption{For $p\gtrsim1.159$, and up to the value of $p$ at which
  trapped surfaces are already present on the initial null slice, what
  were two trapped regions in Fig.~\ref{fig:contour_comparison} are
  now joined already on our finite numerical domain. The recess
  between the two lobes of this single trapped region becomes
  shallower (moves to smaller $v$) as $p$ moves away from $\hat
  p_1$. Note that now the cigar moves little with $p$, and instead the
  recess does. The trapped region is bounded by contours of
  compactness $C=1$ (solid lines); the blue contour corresponds to
  $p\simeq1.159$ and the black contour to $p\simeq1.175$. The dashed
  lines are again contours of $Q/R^2=1$; except that the red dashed
  line is the contour $Q/R^2=0.9970788$ of the solution with
  $p\simeq1.159$, for which this contour passes exactly through the
  tip of the recess. As in Fig.~\ref{fig:contour_comparison}, we
  attribute this deviation from $Q/R^2=1$ to numerical error, since we
  do not explicitly use Eq.~\eqref{XiReqn} to evolve $R$.}
\label{fig:contour_comparison_joint_trapped_region}
\end{figure*}

Fig.~\ref{fig:contour_comparison_joint_trapped_region} is equivalent
to Fig.~\ref{fig:contour_comparison}, but shows values of $p$ further
away from $\hat p_1$ on the extremal side. We see that then the two
trapped regions are actually connected already on our
numerical domain, so that there is only a single trapped region with a
recess between them. Even further from $\hat p_1$, even this recess
disappears. On the other hand, on our finite numerical domain we
cannot see what happens as $p\to p_{1e}$, so we do not know if there
is always a single trapped region (connected outside our numerical
domain) or if there really are two trapped regions that remain
separate all the way to the Cauchy horizon. Therefore we have sketched
the bridge between the two trapped regions in
Fig.~\ref{fig:extremal_side_spacetime} in blue with a question
mark. Even if there is a single trapped region, $|Q|/\cal M$ of the
black hole could still be extremal, namely if the apparent horizon
meets the event horizon tangentially to the Cauchy horizon. We note in
passing that the tip of the recess in the single trapped region that
we see in Fig.~\ref{fig:contour_comparison_joint_trapped_region} far
from $\hat p_1$ is extremal for the same reason that the tip of the
cigar is.

\cite{GellesPretorius26} have made a key quantitative observation
about the receding cigar in the Einstein-Maxwell-charged scalar field
system (not already made in \cite{AngelopoulosKehleUnger26}): in the
near-extremal regime, $v_{\text{trap}}$ diverges as
\begin{equation}
\label{vtrapscaling}    
v_{\text{trap}} \sim |p - p_1|^{-1/2}
\end{equation}
as $p \to p_1$. In the gauge of \cite{GellesPretorius26}, the metric
function $G$ is approximately constant, and so the affine parameter
$\lambda=\int G\,dx$ along the outgoing null geodesics is
approximately linear in the null coordinate $v$, but in contrast to
$v$ itself, it is also gauge-invariant under changes of the coordinate
$x$. In our evolutions $G$ is far from constant near the horizon, and
so our $v$ is not even approximately a linear function of that of
\cite{GellesPretorius26}. We will therefore attempt to derive and
verify \eqref{vtrapscaling} in terms of $\lambda$ instead of $v$.

\cite{GellesPretorius26} justify (\ref{vtrapscaling}) by identifying
their $v_\text{trap}$ with the ``decay time'' for a real scalar field
evolving along a near-extremal RN horizon derived in
\cite{MurReaTan13}. We propose a more explicit, and apparently
different, derivation in
Appendix~\ref{appendix:lambdatrapderivation}. This links the following
three observations:

\begin{itemize}

\item At $p=p_1$, on the event horizon, $e^{-i\alpha}\phi$
  becomes essentially real, and its real part decays as 
\begin{equation}
\label{psirotdecay}
\psi_\text{rot}:={\rm Re}(e^{-i\alpha}\phi)\sim (\lambda-\lambda_0)^{-1/2}.
\end{equation}
(The imaginary part $\chi_\text{rot}$ decays faster while
oscillating).

\item Also at $p=p_1$, on the event horizon, the expansion of the
  horizon generators decays as
\begin{equation}
\label{Vdecay}
V:=R_{,\lambda}={R_{,x}\over G} \sim (\lambda-\lambda_0)^{-2}.
\end{equation}

\item Finally, as $p\to p_{1e}$, the $v$-location of the tip of the
  cigar, expressed in terms of the affine parameter $\lambda$ along
  outgoing null rays, diverges as
\begin{equation}
\label{lambdatrapscalingbis}
\lambda_\text{trap}-\lambda_0 \sim |p-p_1|^{-1/2}.
\end{equation}

\end{itemize}

Introducing a family-dependent parameter $\lambda_0$ in these scaling
laws is necessary because we fix $\lambda=0$ at the regular centre,
which is not relevant for the local receding-cigar mechanism. By
contrast \cite{GellesPretorius26} fix $v=0$ at the left side of their
null rectangle, which is much closer to the receding-cigar region.

In addition, for the Einstein-Maxwell-{\em real} scalar field system
on a null rectangle, \cite{AngelopoulosKehleUnger26} have proved a
number of scaling laws at the threshold of horizon formation, as it is
approached from the collapse side. Three of these are algebraically
independent, and can be taken to be
\begin{eqnarray}
\label{uEHscaling}
u_1-u_{\text{EH}} &\sim& |p-p_1|^{1/2}, \\
\label{REHscaling}
R_{\text{EH}}-R_1 &\sim& |p-p_1|^{1/2}, \\
\label{QoMscaling}
1-(|Q|/{\cal M})_{\text{EH}} &\sim& |p-p_1|.
\end{eqnarray}
Scaling laws for $Q$, ${\cal M}$, the surface gravity and the event
horizon temperature also given in \cite{AngelopoulosKehleUnger26}
follow from (\ref{uEHscaling}-\ref{QoMscaling}) and formulas for RN
black holes, but (\ref{psirotdecay}-\ref{lambdatrapscalingbis}) do not
(as far as we know). We expect the scaling laws
(\ref{uEHscaling}-\ref{QoMscaling}) to carry over to our extremal
threshold $p\to p_{1e}$ (not the collapse threshold $p\to p_*$) in
genuine collapse in the Einstein-Maxwell-{\em charged} scalar field
system, but we make no attempt here at deriving them.


\subsection{Numerical tests of the extremal scaling laws}
\label{sec:numextrscaling}


For numerical simulations we use the same two-Gaussian initial data as
specified in Table~\ref{tab:initial_params_sweep}, but with the outer
boundary extended to $x_\text{max}=600$ with $x_0=597.5$, so that
we can follow the cigar further as it recedes to the Cauchy
horizon. We tune $c_1$, with $\mathcal{A}_2=0.0365$. All evolutions
use resolution $h=0.15$.

Before we proceed, we need to consider the dependence of our results
on the fact that we only have a finite domain, bounded in our method
by $x\le x_\text{max}$, not one that reaches to ${\cal I}^+$ and the
Cauchy horizon.  We ignore, for this discussion, the effects of
numerical error, in particular from finite grid resolution $h$. For
definiteness, we also assume that the initial data are defined on
$u=0$ out to ${\cal I}^+$, with finite mass and charge, even if we
construct a numerical solution only out to $x_\text{max}$.

On a finite domain, we cannot compute the true black hole mass, charge
and location for given $p$, but we can estimate them as the mass,
charge and location of the FMOTS, which (up to discretisation error)
will always occur at the outer boundary of the finite domain. On an
infinite domain, it would occur where the event horizon meets ${\cal
  I}^+$ (and the Cauchy horizon).

Above, we have defined an approximation $\hat p_1$ to the true
extremal threshold through $(Q/{\cal M})_\text{FMOTS}=1$ at $p=\hat
p_1$. Recall that the tip of the cigar is extremal while the rest of
its boundary is subextremal. Hence the limit above occurs as the cigar
is about to disappear from the numerical domain, at which point the
FMOTS is the tip of the cigar. This explains why we can fine-tune to
extremality of the FMOTS even on a finite domain. By contrast, the
event horizon touches the underside of the trapped region (the cigar)
only where it terminates at ${\cal I}^+$, and so
$u_\text{EH}<u_{\text{trap}}$. The final black hole mass and charge
are defined only at that point, and are sub-extremal. Therefore, $\hat
p_1$ is only an approximation to the true $p_1$, although $\hat p_1$
will approach the true $p_1$ as the domain becomes larger.

By contrast, the location $(u_\text{trap},\lambda_\text{trap})$ of the
tip of the cigar depends on $p$ but not on the domain, as long as the
tip of the cigar is still on the domain. We define alternative proxies
$\bar p_1$ and $\bar u_1$ for $p_1$ and $u_1$ by obtaining the best
fit to a new scaling law
\begin{equation}
\bar u_1-u_\text{trap}(p)\sim (p-\bar p_1)^{1\over 2},
\label{utrapu1barp1}
\end{equation}
and, subsequently, $\lambda_0$ as a best fit to 
in\begin{equation}
\lambda_\text{trap}(p)-\lambda_0\sim (p-\bar p_1)^{-1/2}.
\label{lambdatrapu1barp1lambda0}
\end{equation}
$\bar p_1$, $\bar u_1$ and $\lambda_0$ depend weakly on the domain,
but only because by extending the domain we add more data points,
closer to extremality, to the set of data points to which we fit the
power laws. This suggests to that $\hat p_1$ depends more strongly on
the domain than $\bar p_1$ does, but at finite resolution $h$ there is
no clear numerical evidence for this, see
Table~\ref{tab:adaptive_timestep} for some experimentation.

\begin{table*}[htpb]
\caption{Comparison of evolutions of the same 1-parameter family of
  initial data, with different $x_\text{max}$ and $x_0$, but the same
  $h=0.15$, using the adaptive timestep. The first line corresponds to
  the results presented here in Sec.~\ref{sec:numextrscaling}. $\bar
  p_1$ seems to depend only on $x_0$. By contrast, $\hat p_1-\bar p_1$
  seems to depend almost only on $x_\text{max}$, going down as
  $x_\text{max}$ goes up. $\bar u_1$ fitted to (\ref{utrapu1barp1})
  and $\lambda_0$ fitted to (\ref{Vdecay}) seem to be completely
  independent of $x_\text{max}$ and $x_0$.}
\label{tab:adaptive_timestep}
\renewcommand{\arraystretch}{1.3}
\setlength{\tabcolsep}{8pt}
\begin{tabular}{r|r||r|r|r|r|r|r|r|r}
\thead{$x_{\text{max}}$} &
\thead{$x_0$} &
\thead{$\hat p_1$} &
\thead{$\bar p_1$} &
\thead{$\hat p_1-\bar p_1$} &
\thead{$u_{\text{FMOTS}}$\\at $p=\hat p_1$} &
\thead{$u_{\text{trap}}$\\at $p=\hat p_1$} &
\thead{$u_{\text{falloff}}$\\at $p=\hat p_1$} &
\thead{$\bar u_1$\\in
  (\ref{utrapu1barp1})} &
\thead{$\lambda_0$\\ in (\ref{Vdecay})} \\
\hline
$600$  & $597.5$  & $0.9487$ & $0.9455$ & $0.0032$ & $33.0160$ &
$33.0452$ & $32.5786$ & $33.115$ & $57$ \\
$1200$ & $597.5$  & $0.9472$ & $0.9455$ & $0.0017$ & $33.0430$ & $33.0719$ & $32.7250$ & $33.115$ & $57$ \\
$1200$ & $1197.5$ & $0.9739$ & $0.9721$ & $0.0018$ & $33.0392$ & $33.0688$ & $32.7434$ & $33.115$ & $57$ \\
$2400$ & $597.5$ & $0.9467$ & $0.9455$ & $0.0012$ & $32.9491$  & $33.0747$ & $32.8138$ & $33.115$ & $57$ \\
$2400$ & $2397.5$ & $0.9865$ & $0.9855$ & $0.0010$ & $33.0308$ & $33.0904$ & $32.8521$ & $33.115$ & $57$ \\
\end{tabular}
\end{table*}

For the evolutions we use the following two schemes:

\paragraph*{Adaptive time step method:}

\begin{figure}[h]
\includegraphics[width=\linewidth]{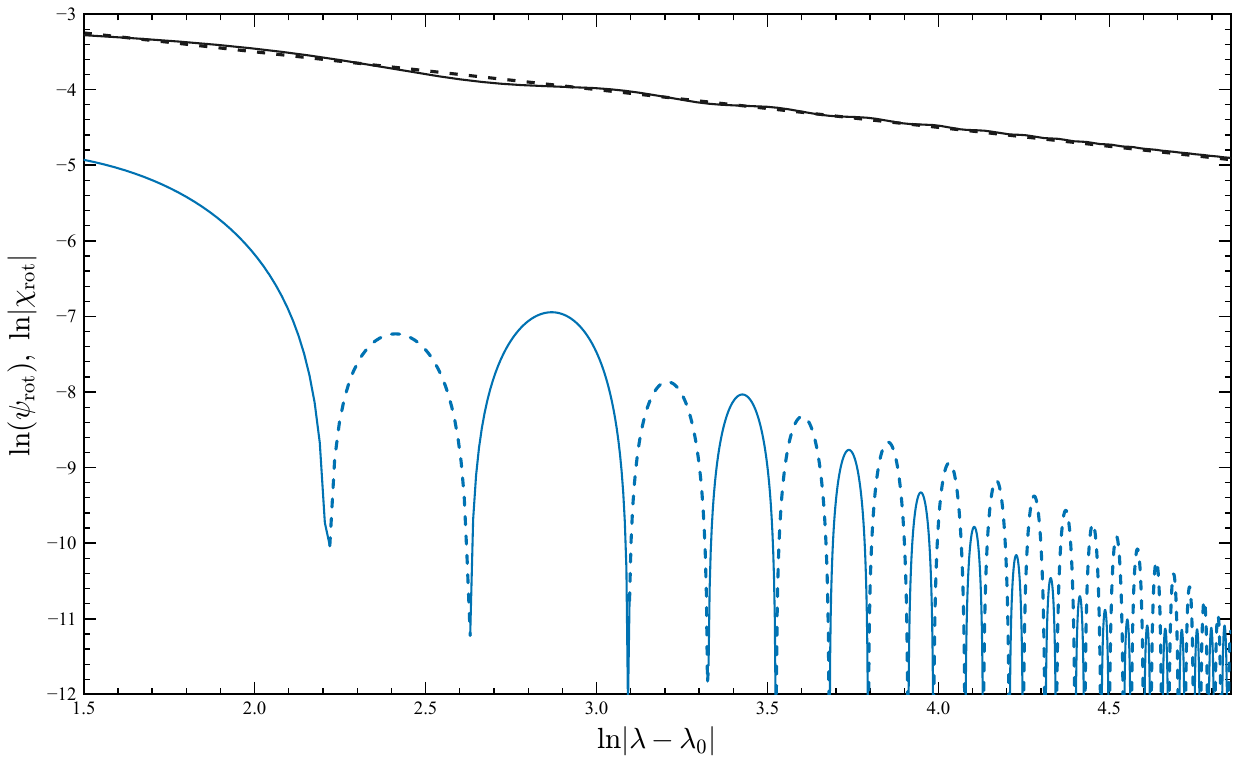}
\caption{Scaling of $\psi_{\rm rot}(u_\text{falloff},\lambda)$ (black)
  and $\chi_{\rm rot}(u_\text{falloff},\lambda)$ (blue, dashed where
  negative) on a proxy for the event horizon of an extremal black
  hole, as a function of $\lambda - \lambda_0$.  The initial data is
  the complex double Gaussian with parameters listed in
  Table~\ref{tab:initial_params_sweep}, with $\mathcal{A}_2=0.0365$
  and $p$ scaling $c_1$. We have fine-tuned to the extremal threshold
  $\hat p_1\simeq 0.9487$. We evolve with $x_\text{max}=600$ and
  $x_0=597.5$ at resolution $h=0.15$. We evaluate at
  $u_\text{falloff}= 32.5786$. We have fitted $\lambda_0=57$
  and $\alpha =0.2424$. The dashed black reference
  line is $\ln\psi_{\rm rot}=-0.5\ln(\lambda - \lambda_0) - 2.5$.}
\label{fig:psi_rot_chi_rot_power_scaling}
\end{figure}

\begin{figure}[h]
\includegraphics[width=\linewidth]{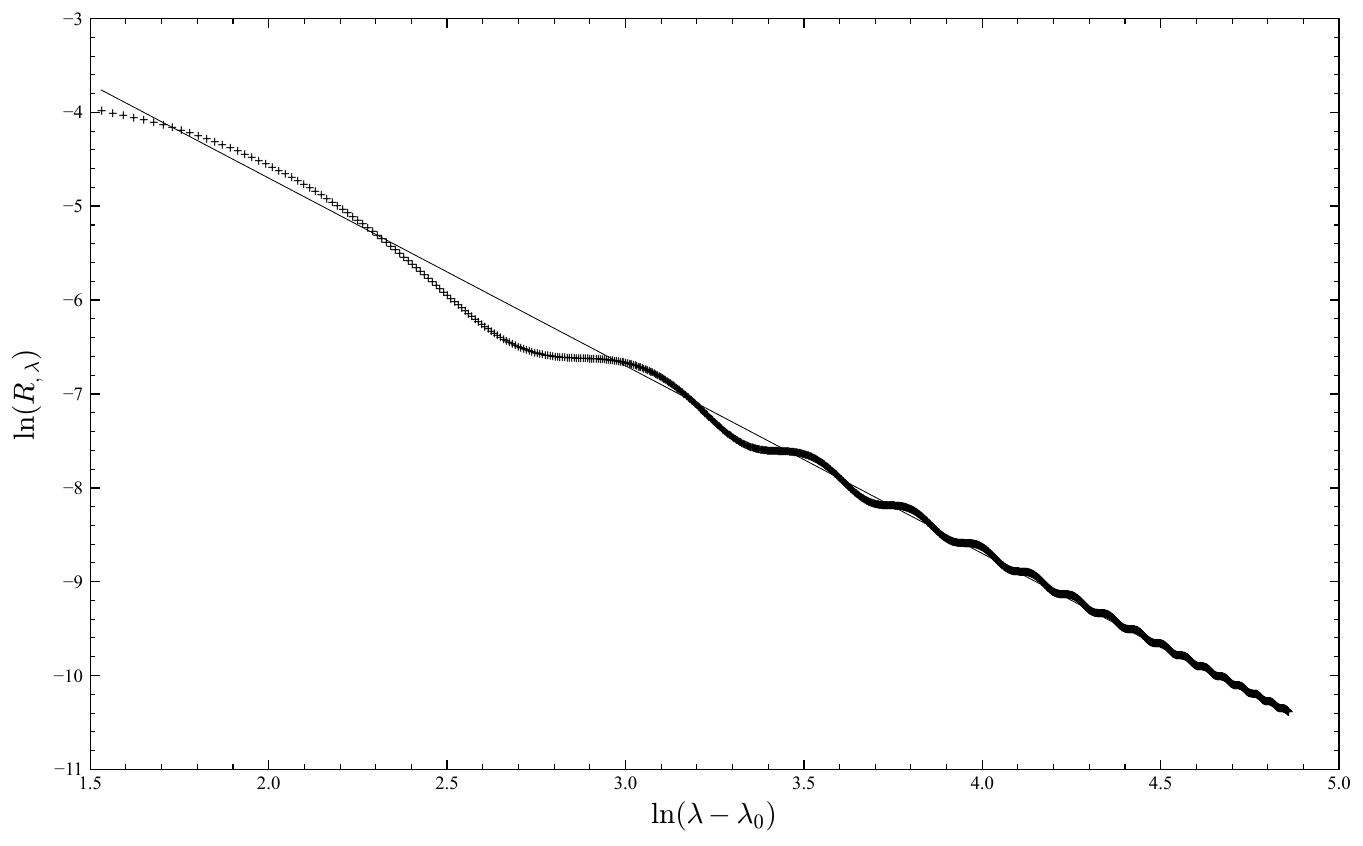}
\caption{Scaling of $V(u,\lambda)=R_{,\lambda}=R_{,x}/G$ on the proxy
  extremal event horizon, as a function of $\lambda - \lambda_0$. The
  initial data, $\lambda_0$ and $u$ are as in
  Fig~\ref{fig:psi_rot_chi_rot_power_scaling}. The reference line is
  $\ln V=-2\ln(\lambda - \lambda_0) - 0.7$.}
\label{fig:V_power_scaling}
\end{figure}

\begin{figure}[h]
\includegraphics[width=\linewidth]{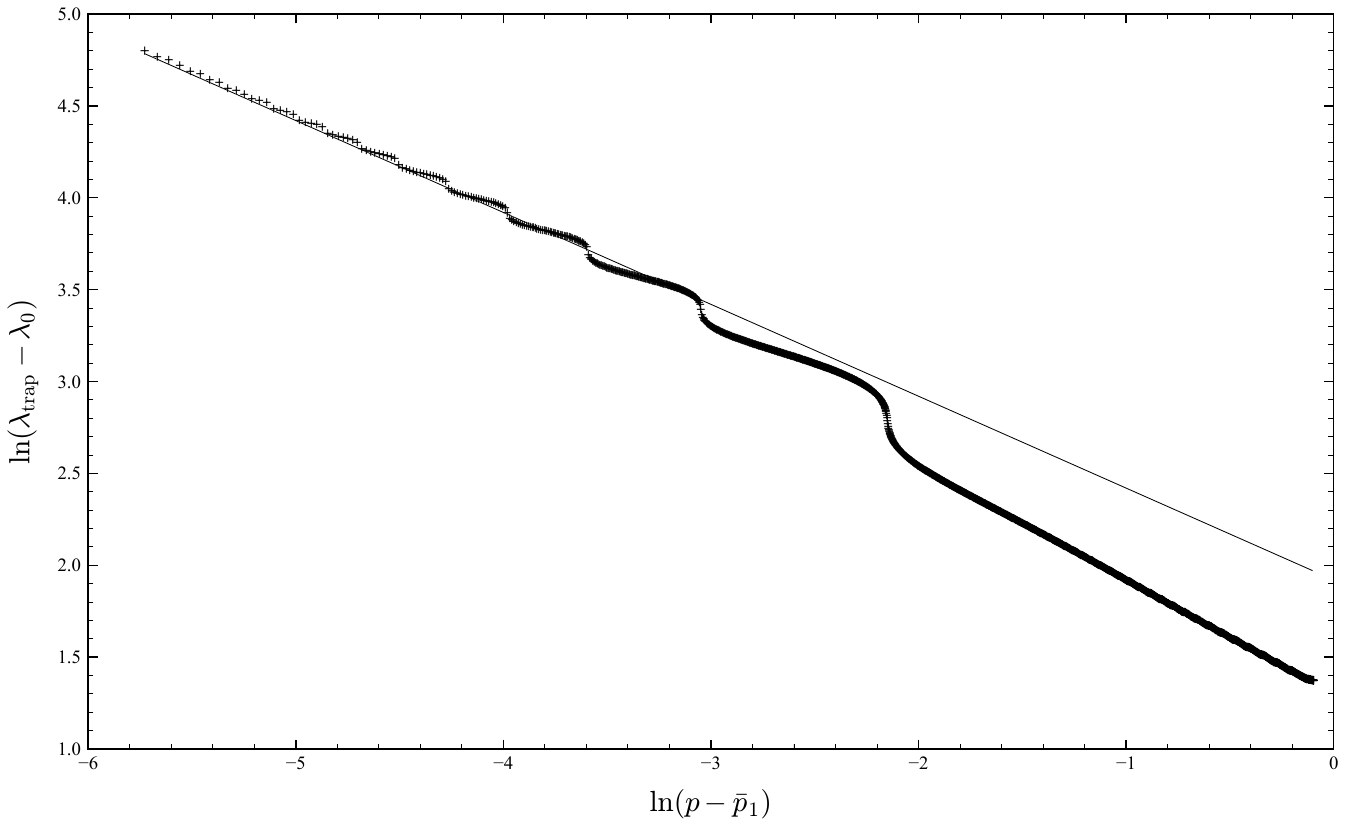}
\includegraphics[width=\linewidth]{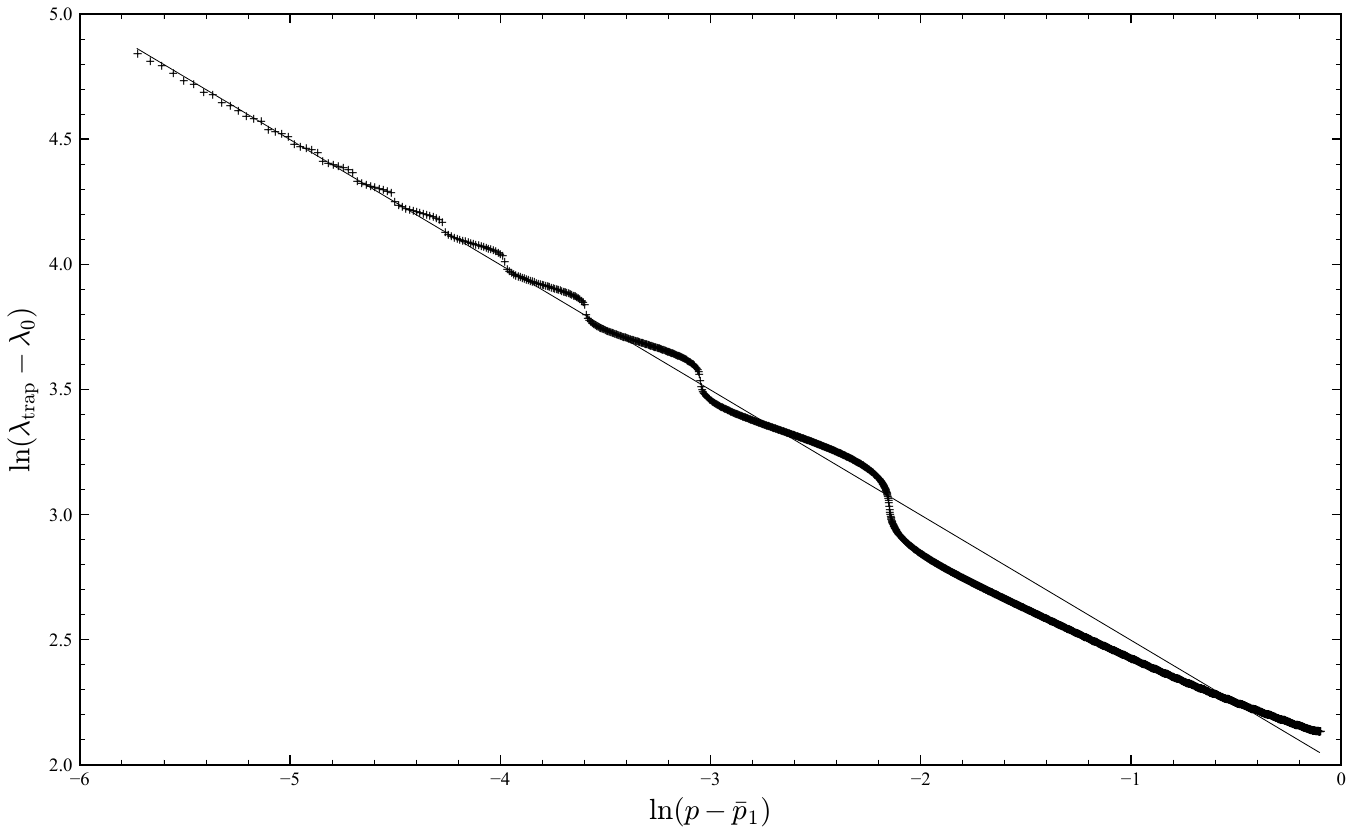}
\caption{Scaling of $\lambda_{\text{trap}}$ as a function of $p - \bar
  p_1$, with $\bar p_1=0.9455$. The 1-parameter family of initial data
  is the same as in Figs.~\ref{fig:psi_rot_chi_rot_power_scaling} and
  \ref{fig:V_power_scaling}, but here we plot against $p$, whereas
  there the specific data with $p=\hat p_1\simeq 0.9487$ were
  used. Top panel: $\lambda_0=57$ (as fitted in
  Figs.~\ref{fig:psi_rot_chi_rot_power_scaling} and
  \ref{fig:V_power_scaling}), with the reference line
  $\ln(\lambda_{\text{trap}} - \lambda_0)=-\tfrac{1}{2}\ln(p - \bar
  p_1) + 1.92$.  Bottom panel: $\lambda_0=52.5$ (fitted only to these
  data), with the reference line $\ln(\lambda_{\text{trap}} -
  \lambda_0)=-\tfrac{1}{2}\ln(p - \bar p_1) + 1.998$.  Note that the
  fit in the top panel is not as clean as in the bottom panel, with
  the disagreement at low fine-tuning.}
\label{fig:lambda_trap_scaling}
\end{figure}

This is described in Sec.~\ref{sec:numericalscheme}, and we use it for
Figs.~\ref{fig:contour_comparison}-\ref{fig:lambda_trap_scaling} and
\ref{fig:r+_scaling_0.9455}-\ref{fig:oneminusQbyM_log}. A scan in $p$
similar to Table~\ref{tab:perturbed_study}, followed by a second scan
with a finer $\delta p$, gives $\hat{p}_1 \simeq 0.9487$. This value
could be refined with higher resolution in $u$ and $x$, see
Table~\ref{tab:resolution_convergence}, but at the resolution $h=0.15$
used here, bisection alone does not improve $\hat{p}_1$ further
relative to the fine scan.  At this $\hat{p}_1$, $1
-(Q/\mathcal{M})_{\text{FMOTS}} \simeq 6\times10^{-6}$. (By
comparison, for $h=0.15$ and $x_{\text{max}}=120$, bisection
refinement in Table~\ref{tab:resolution_convergence} significantly
improves on Table~\ref{tab:perturbed_study}, since the latter used
only a single linear scan with a coarser $\delta p$).

\paragraph*{Fixed time step:}

\begin{figure}[h]
\includegraphics[width=\linewidth]{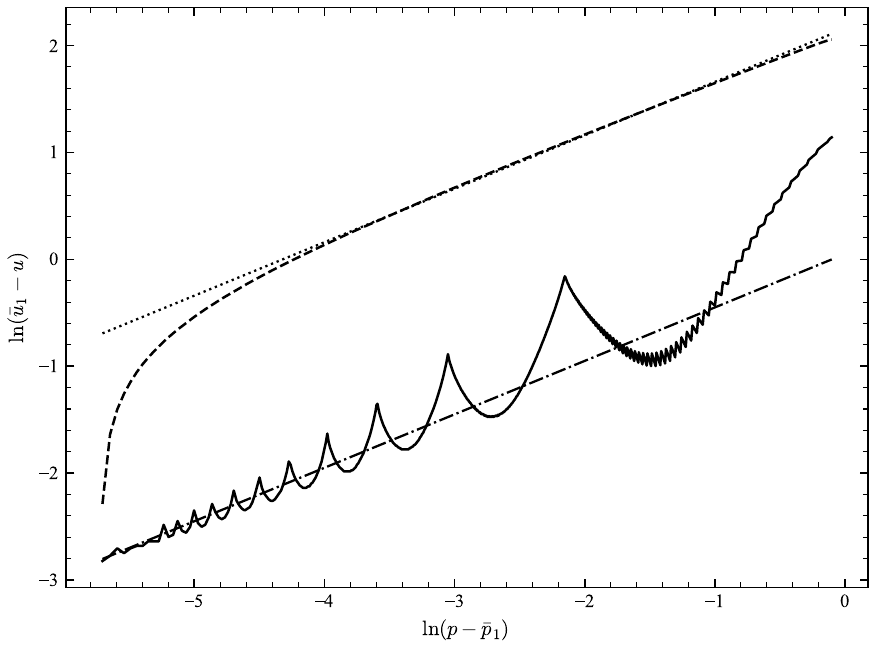}
\caption{Scaling of $u_{\text{FMOTS}}$ and $u_{\text{trap}}$ using
  fixed time step evolutions. Dashed: $\ln(\bar u_1 -
  u_\text{FMOTS})$; solid: $\ln(\bar u_1 - u_{\text{trap}})$, both
  versus \ $\ln|p-\bar p_1|$, with $\bar u_1=33.115$, $\bar p_1
  =0.9455$ and $\hat p_1 \simeq 0.9488$. The two curves would coincide
  at $p=\hat{p}_1$ if $\hat{p}_1$ were determined at infinite
  resolution in $u$ and $x$. The reference lines have slope $-1/2$.}
\label{fig:u_scaling_0.9455}
\end{figure}

\begin{figure}[h]
\includegraphics[width=\linewidth]{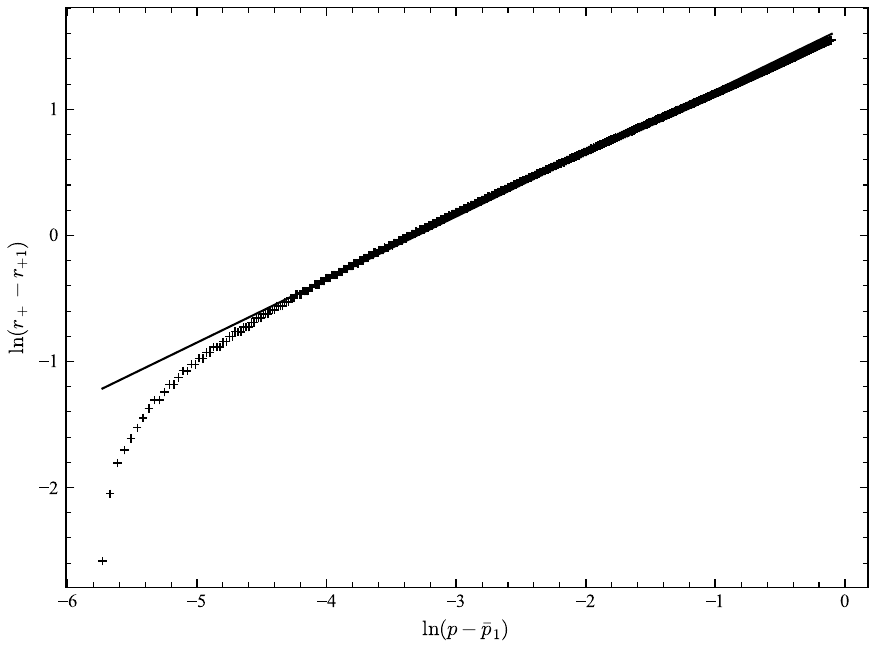}
\caption{Plot of $\ln(r_{+}-r_{+1})$ against $\ln(p - \bar{p}_1)$ with
  $r_{+1}=45.93$, with $\bar{p}_1=0.9455$. Here $r_+$ is computed from
  the $Q_\text{FMOTS}$ and ${\cal M}_\text{FMOTS}$ using the RN
  formula \eqref{appendix:r+-}. The family of initial data is as in
  Fig.~\ref{fig:lambda_trap_scaling}. The reference line has slope
  $-1/2$.}
\label{fig:r+_scaling_0.9455}
\end{figure}

\begin{figure}[h]
\includegraphics[width=\linewidth]{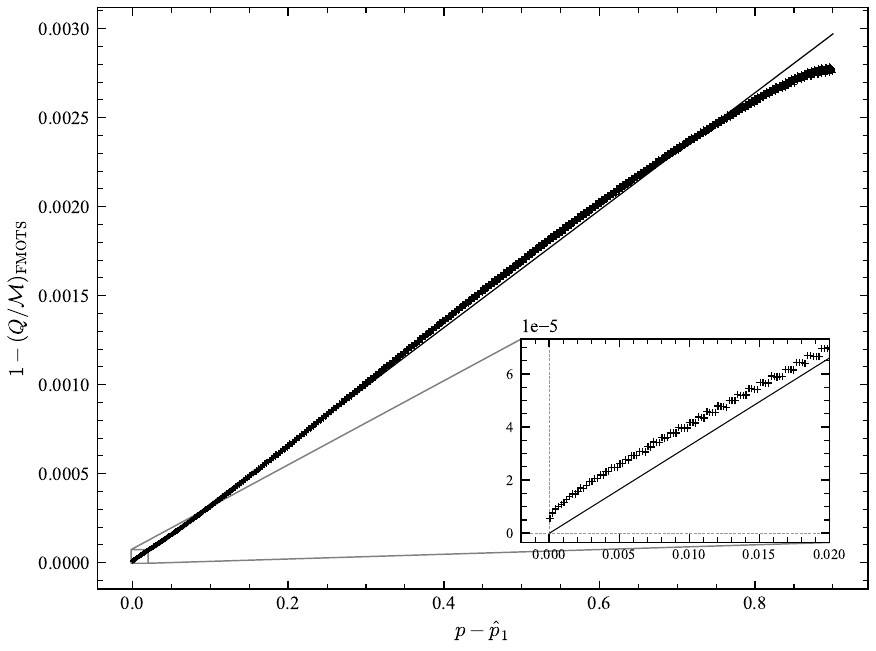}
\caption{Linear plot of $1-(Q/\mathcal{M})_\text{FMOTS}$ against $(p -
  \hat{p}_1)$ overlaid with the linear reference $1 -
  (Q/\mathcal{M})_\text{FMOTS}=0.0033(p - \hat{p}_1)$ (solid black
  line). The family of initial data is as in
  Fig.~\ref{fig:lambda_trap_scaling}.}
\label{fig:oneminusQbyM_0.9487_lin}
\end{figure}

\begin{figure}[h]
\includegraphics[width=\linewidth]{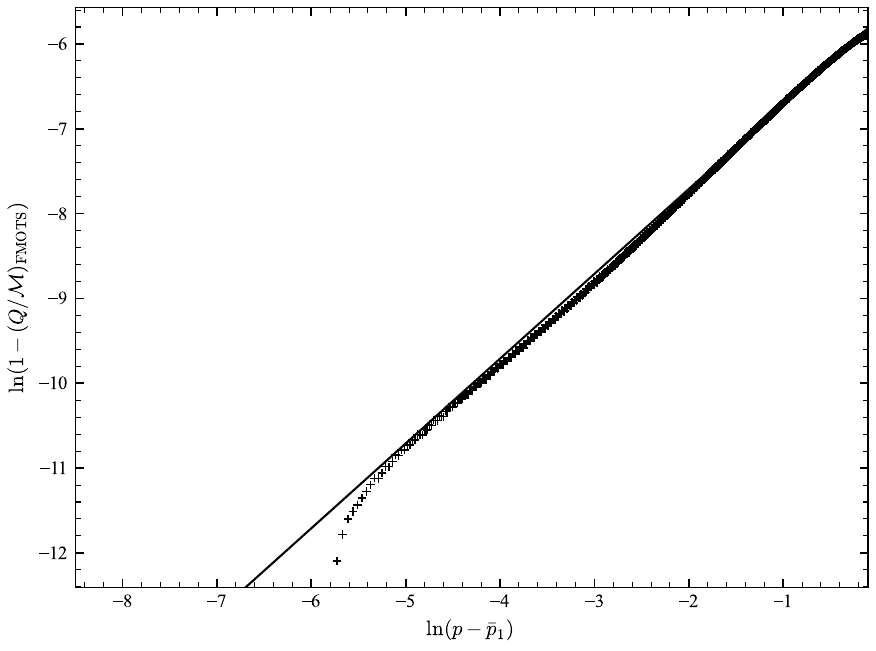}
\includegraphics[width=\linewidth]{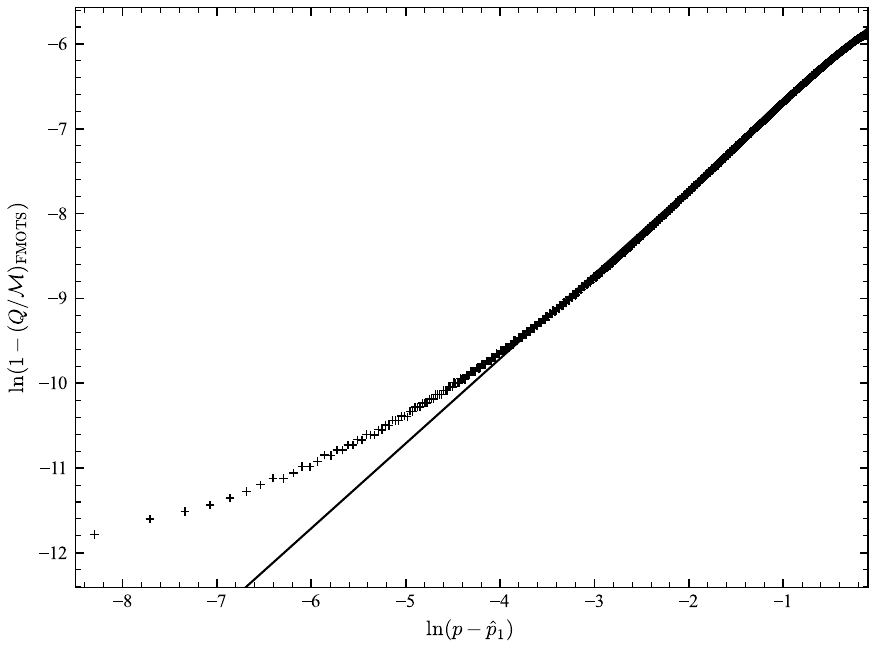}
\caption{Top panel: plot of $\ln(1-(Q/\mathcal{M})_\text{FMOTS})$
  against $\ln(p - \bar{p}_1)$, with $\bar{p}_1=0.9455$. The curve
  must bend down because (numerical error aside)
  $1-(Q/\mathcal{M})_{\text{FMOTS}}=0$ at $p=\hat p_1$. Bottom panel:
  plot against $\ln(p - \hat{p}_1)$, with $\hat{p}_1 \simeq
  0.9487$. The plot appears to switch to a lower power as $p\to\hat
  p_1$ and $1-(Q/\mathcal{M})_\text{FMOTS}\to0$. The family of initial
  data is as in Fig.~\ref{fig:lambda_trap_scaling}. The reference line
  in both plots has slope $-1$ and the same intercept. (To keep the
  same range in both plots, one point is excluded from the left side
  of the bottom panel.)}
\label{fig:oneminusQbyM_log}
\end{figure}

This means $\Delta u=C_0 \, \Delta x$, with $C_0=0.01$, and we have
used it for Fig.~\ref{fig:u_scaling_0.9455}, since it provides higher
resolution in $u$ at high fine-tuning. We obtain $\hat{p}_1 \simeq
0.9488$, using the same method as mentioned above. Using the adaptive
timestep, we get the same picture as in
Fig.~\ref{fig:u_scaling_0.9455}, just with less resolution in $u$.
$\bar p_1$ is also independent of the timestep.

We now present numerical evidence for  the scaling laws
(\ref{psirotdecay}-\ref{QoMscaling}), in this order. 

\paragraph*{Scaling \eqref{psirotdecay} of $\psi_\text{rot}(\lambda)$:}

The decay (\ref{psirotdecay}) of the real part of the scalar field on
(approximately) the event horizon, and faster decay of the imaginary
part, are illustrated in Fig.~\ref{fig:psi_rot_chi_rot_power_scaling},
where we have used $\hat p_1$ as a proxy for $p_1$, and have adjusted
$\alpha$ so that $\chi_\text{rot}$ falls off as rapidly as
possible. We extract $\psi_\text{rot}$ at the value of $u$ at which
the falloff of $V$ with $\lambda$ is cleanest,
$u=u_\text{falloff}=32.5786$. We expect that this is close to the
(unknown, on a finite domain) event horizon time $u_\text{EH}(\hat
p_1)$. Consistently with that interpretation, the FMOTS forms later,
at $u_\text{FMOTS} \simeq 33.016$.  At this $u$, we find that
$1-(Q/\mathcal{M})(u_\text{falloff},x_{\text{max}})\simeq
6\times 10^{-7}$.

From \cite{Gajic26}, we expect that the value of the dimensionless
quantity $qQ$ is relevant for the power of the falloff and/or the
oscillation of the complex part (but we have no theoretical model for
this). We observe that $qQ \simeq 1702.16$, close to constant and $\gg
1$.

\paragraph*{Scaling \eqref{Vdecay} of $V(\lambda)$:}

The decay (\ref{Vdecay}) of $V$ on the horizon is illustrated in
Fig.~\ref{fig:V_power_scaling}, for the same solution and at the same
$u=u_\text{falloff}$. (As already noted, this is actually how we fix
$u_\text{falloff}$.)

\paragraph*{Scaling \eqref{lambdatrapscalingbis} of $\lambda_\text{trap}(p)$:}

The scaling (\ref{lambdatrapscalingbis}) of $\lambda_\text{trap}$ is
illustrated in Fig.~\ref{fig:lambda_trap_scaling}. Note this plot is
derived from a 1-parameter family of solutions, not the extremal
solution in that family, and that on the horizontal axis we now have
$\ln(p-\bar p_1)$, not $u$. For this plot, we should logically use the
same value of $\lambda_0$ as already fitted in
Fig.~\ref{fig:psi_rot_chi_rot_power_scaling} and
Fig.~\ref{fig:V_power_scaling}, but this shows a deviation from the
power law at low fine-tuning (far from extremality). The lower panel
of Fig.~\ref{fig:lambda_trap_scaling} shows that a slightly different
value of $\lambda_0$ gives a better fit at low fine tuning. We have no
explanation for this, but are not concerned, as we expect the power
laws to hold increasingly well as extremality is approached with
increasing fine-tuning. From the fit we find $\bar p_1=0.9455$,
consistent with Fig.~\ref{fig:u_scaling_0.9455}. The different
timestep used in the evolutions underlying the two figures seems to
make no difference to $\bar p_1$.

\paragraph*{Scaling \eqref{uEHscaling} of $u_\text{EH}(p)$:}

As a proxy for $u_\text{EH}(p)$, which cannot be known on a finite
domain, in Fig.~\ref{fig:u_scaling_0.9455} we plot both
$u_\text{trap}(p)$ and $u_\text{FMOTS}(p)$ against $\ln(p-\bar
p_1)$. Data for $u_\text{trap}(p)$ and a best fit to
(\ref{utrapu1barp1}) are shown as the lower pair of curves. From the
fit we find $\bar p_1=0.9455$.  We believe that the oscillations
overlaying the power law are related to similar oscillations of the
scalar field. The upper pair of curves shows $\bar u_1-u_\text{FMOTS}$
and a power-law fit with the same $\bar u_1$ and power $-1/2$. The
curve follows the expected scaling initially but bends down to
intersect the curve representing $u_\text{trap}$ at $p=\hat p_1$. This
is expected because by definition $u_\text{FMOTS}=u_\text{trap}$ at
$p=\hat p_1$ (up to numerical error). We expect this break to move
left when we extend the domain, as $\hat p_1$ then approaches $\bar
p_1$.

In Fig.~\ref{fig:u_scaling_0.9455}, in contrast to the other scaling
plots, we have used the fixed timestep in order to obtain more
resolution in $u$ (at the same resolution in $x$). The reason for this
is still visible at the right end of the lower curve, which shows the
finite resolution in $u$: with the adaptive timestep, the plot is much
more jagged, but otherwise agrees with the fixed timestep. In
particular, we find the same $\bar p_1$ and $\bar u_1$ by fitting.

In this context we note that reaching the value of $\hat{p}_1$ at
which $u_{\text{trap}}$ and $u_{\text{FMOTS}}$ (at finite
$x_\text{max})$ coincide requires infinite resolution in both $u$ and
$x$ in order to locate the tip of the cigar precisely as it leaves the
numerical domain.

\paragraph*{Scaling \eqref{REHscaling} of $R_\text{EH}(p)$:} 

As a proxy for $R_\text{EH}(p)$, in Fig.~\ref{fig:r+_scaling_0.9455}
we plotthe event horizon radius $r_+({\cal M},Q)$ given by
\eqref{appendix:r+-}, using $\mathcal{M}_{\text{FMOTS}}(p)$ and
$Q_{\text{FMOTS}}(p)$. We see the expected power $1/2$ scaling, with a
similar downward bend as $p \to \hat{p}_{1e}$ and, we believe, for the
same reason: as the FMOTS approaches the tip of the cigar while both
approach the outer boundary of the finite domain, it becomes a bad
approximation to the event horizon.

\paragraph*{Scaling \eqref{QoMscaling} of $(Q/{\cal M})_\text{EH}(p)$:}

As a proxy for $(Q/{\cal M})_\text{EH}$ we use $(Q/{\cal
  M})_\text{FMOTS}$. We plot this against $p-\hat p_1$ in
Fig.~\ref{fig:oneminusQbyM_0.9487_lin}. By definition, this goes
through $(0,0)$, up to numerical error (due to finite resolution,
$1-(Q/\mathcal{M})_{\text{FMOTS}} \sim 10^{-6}$ at $\hat{p}_1\simeq
0.9487$), but in the inset we already see a deviation from strict
linearity. To investigate this more, in the upper panel of
Fig.~\ref{fig:oneminusQbyM_log} we plot $\ln(1-(Q/{\cal
  M})_\text{FMOTS})$ against $\ln(p-\bar p_1)$. Because $(Q/{\cal
  M})_\text{FMOTS}=1$ at $p=\hat p_1$, the resulting curve bends below
the best-fit straight line with slope 1, and eventually down to
$-\infty$, as $p\to\hat p_{1e}$. We expect, however, that the
deviation from a straight line happens closer to extremality on a
larger domain. (This has indeed been demonstrated in 3+1 dimensions
\cite{MartelGundlachMittal26}). In the lower panel, we plot against
$\ln(p-\hat p_1)$ instead. However, this does not give a more
convincing power law, but instead results in an upwards break
of the curve to a smaller slope (smaller power) as $p\to\hat
p_{1e}$. We have no reason to prefer this plot, and show it here just
to illustrate that fact. Experimenting with other approximations to
$p_1$ also just results in a more complicated deviation from a
straight line close to $p=\hat p_1$.

In the family of initial data and at the resolution described above,
we found $\hat p_1 \simeq 0.9487$ from locating the jump in
$(Q/\mathcal{M})_{\text{FMOTS}}$. We found $\bar p_1=0.9455$ from a
fit to the scaling laws (\ref{utrapu1barp1}) and
(\ref{lambdatrapu1barp1lambda0}), with the fit to (\ref{utrapu1barp1})
also giving $\bar u_1=33.115$. This puts $\bar p_1$ on the subextremal
side of $\hat p_1$. This is expected as we believe that, for $p$
between the two, the cigar is still present but has receded too far to
be seen on the numerical domain. We expect that both $\bar p_1$ and
$\hat p_1$ approach the (unknown) continuum value $p_1$, and hence
each other, in the limit $h\to 0$, $x_\text{max}\to\infty$. However,
we have no evidence for this double limit.


\section{Conclusions}
\label{sec:conclusions}


Using a straightforward extension of the time evolution code of
\cite{GundlachMartel26}, we have investigated the gravitational
collapse of a charged massless scalar field in spherical symmetry in
4+1 spacetime dimensions.

At the threshold of collapse, in the space of initial data (or
equivalently, the space of solutions), we have found the expected
type-II critical phenomena. This is qualitatively similar to previous
results for spherical charged scalar field collapse in 3+1 dimensions
\cite{HodPiran97,Petryk05,GundlachMartel26}, and also consistent with
previous results for spherical real scalar field collapse in 4+1
dimensions \cite{BlandEtAl05,PortoGundlach2022}. We find that at the
threshold, the FMOTS mass, a proxy for the final black hole mass,
vanishes as ${M}\sim (p-p_*)^{0.8265}$, and the black hole charge
vanishes as $Q\sim (p-p_*)^{1.425}$, so that $|Q|/{M}\to 0$ at the
threshold.

In particular, we have found no evidence, (in 4+1 dimensions), for the
stable extremal critical collapse conjecture of
\cite{KehleUnger24,AngelopoulosKehleUnger26}, which says that there is
a neighbourhood within the threshold of collapse where $|Q|/\mathcal{M}\to
1$. We note that there is also no evidence in the literature to date
for this in 3+1 dimensions, if by collapse we mean the time evolution
of initial data that are posed either on a simply connected Cauchy
surface or on an outgoing null cone with regular vertex, and without
trapped surfaces present in these initial data.

The most concrete support to date for the stable extremal critical
collapse conjecture for the {\em uncharged} scalar field are the
numerical work \cite{MurReaTan13} and the mathematical work
\cite{AngelopoulosKehleUnger26} which confirms it. Here, at the
threshold of horizon formation the horizon becomes extremal. Numerical
evidence for the {\em charged} scalar field is given in
\cite{GellesPretorius26}. This last paper also proposes an intuitive
mechanism for reaching extremality: near-extremal collapsing solutions
have a cigar-shaped trapped region whose tip is always extremal, and
which recedes to the Cauchy horizon exactly at the threshold, leaving
only an extremal horizon behind. We believe the same mechanism applies
to uncharged scalar field matter, but this is less explicit in
\cite{MurReaTan13,AngelopoulosKehleUnger26}.

However, these three papers consider a space of solutions
whose initial data are posed on the union of an ingoing and an
outgoing null cone, with $|Q|\simeq{\cal M}$ on the sphere where the
two intersect, and the initial scalar field small. Such solutions
cannot be embedded into genuine collapse spacetimes.

To our surprise, we found that extremal black holes {\em can} be
created also in genuine collapse, and we have found strong
evidence that this happens for a codimension-1 set of initial
data. However, these extremal black holes are not created at the
threshold of (genuine) collapse, but on a hypersurface inside the
collapse half of solution space. The jump across this threshold is not
from extremal black hole to dispersion, but from extremal black hole
to subextremal black hole.

Both types of critical behaviour had previously been proved for the
spherical Einstein-Maxwell-charged {\em Vlasov} system in
\cite{KehleUnger24}: the jump from extremal black hole formation to
dispersion in Thm.~1, and the jump from extremal to subextremal black
hole formation in Thm.~3: the latter is similar to what we observe.

It is not clear to us, however, how relevant those results are for the
scalar field, as Vlasov matter is very different. In particular, the
matter is supported only inside a timelike tube, surrounded by a RN
exterior, see Figs.~1 and 2 of \cite{KehleUnger24}. In extremal
collapse this exterior must then be eRN, which has no trapped or
antitrapped surfaces, and one can use the ingoing and outgoing
Raychaudhuri equation to conclude that there are neither trapped nor
antitrapped surfaces in the matter region either.

Why is the second trapped region that we observe not seen in
\cite{MurReaTan13,GellesPretorius26}?  In \cite{MurReaTan13} the
scalar field is real and uncharged, so that $Q$ is constant in the
spacetime, and not generated by the matter at all. This means that an
ingoing null cone can, with increasing $u$, enter and then leave a
trapped region $R_{,v}<0$ only once. In \cite{GellesPretorius26}, the
scalar field is charged, this argument does not apply, and so in
principle there could be two trapped regions. We believe that they do
not arise because the solutions constructed there are close to the
vacuum RN solution, in the sense that the scalar field initial data
contribute only a small part of the charge (and mass), and that these
solutions cannot be embedded into genuine collapse spacetimes.

Conversely, why do we not see extremal black holes at the threshold of
collapse? We believe the reason is the difference between genuine
collapse, where the spacetime has a regular centre, and the null
rectangle setup of
\cite{MurReaTan13,GellesPretorius26,AngelopoulosKehleUnger26}. To
extend the ingoing null cone in the setup of \cite{GellesPretorius26}
to a regular centre, the charged scalar field on the initial ingoing
null cone would need to exactly supply the mass ${\cal M}$ and charge
$Q$ on the corner sphere. We believe it would then be difficult to
avoid forming trapped surfaces at small $R$. In fact, we found our
first instance of extremal collapse precisely by trying to find
spacetimes with a regular centre into which the examples of
\cite{GellesPretorius26} could be embedded, but in doing that we also
found a second trapped region behind the cigar, so that the jump from
extremal black hole formation at the threshold is not to dispersion
but to subextremal black hole formation. Of course, extremal (genuine)
critical collapse of a charged scalar field may exist in a different
region of solution space that has not yet been explored.

We note that after our initial discovery of extremal genuine collapse
solutions described here, these have also been found to exist and to
be codimension-1 in spherical charged scalar field collapse in 3+1
\cite{MartelGundlachMittal26}.


\acknowledgments

LM was supported by an EPSRC Doctoral Training Grant to the University
of Southampton. CG would like to thank T. Baumgarte, M. Dafermos,
Z. Gelles, D. Hilditch, C. Kehle, F. Pretorius, H. Reall and R. Unger
for helpful conversations.

\bigskip

\centerline{\bf Data Availability Statement}

No data for this paper are available publicly. The code and sample
parameter files for creating the data underlying our figures are
available from the authors upon reasonable request. 


\appendix


\section{Expansions at the centre}
\label{expansions}


Since we use the eG formulation of~\cite{GundlachMartel26}, we have
the following expansions at the regular centre. For finding $R$ by
solving a system of two first-order ODEs, given $G$, we require
\begin{eqnarray}
\label{Rexpansion}
R &=& G_{(0)} x + \frac{G_{(1)}}{2} x^2
\nonumber \\ &&
{}+ \frac{1}{3} \Bigl( G_{(2)} - \tfrac{4}{3} G_{(0)} \pi
( \chi_{(1)}^2 + \psi_{(1)}^2 ) \Bigr) x^3
\nonumber \\ &&
{}+ O(x^4),  \\
V &=& 1 - \tfrac{4\pi}{3} (\chi_{(1)}^2 + \psi_{(1)}^2) x^2
\nonumber \\ &&
{}+ \frac{4\pi}{9 G_{(0)}} \Bigl[ G_{(1)}
(\chi_{(1)}^2 + \psi_{(1)}^2)
\nonumber \\ &&
{}- 8 G_{(0)} (\chi_{(1)} \chi_{(2)}
+ \psi_{(1)} \psi_{(2)}) \Bigr] x^3
\nonumber \\ &&
{}+ O(x^4),
\end{eqnarray}
where we fit $G$ via the least-squares method on the initial few grid
points to obtain $G_{(0)}$, $G_{(1)}$, and $G_{(2)}$.  Note that,
given the definition of $V:=R_{,x}/G$, to be consistent we should
truncate the expansion of $V$ at one order lower in $x$ than that of
$R$, but here we have written out all expansion coefficients that we
know in full, given a quadratic fit to the evolved quantities.

We next have
\begin{eqnarray}
{\frac{Q}{4\pi q c_J}} &=&
  \frac{R_{(1)}^3}{12 c_Q}
  (\chi_{(1)}\psi_{(0)} - \chi_{(0)}\psi_{(1)}) x^4
  \nonumber \\ &&
  {}+ \frac{R_{(1)}^2}{15 c_Q} \Bigl[
  3R_{(2)} (\chi_{(1)}\psi_{(0)} - \chi_{(0)}\psi_{(1)})
  \nonumber \\ &&
  {}+ 2R_{(1)} (\chi_{(2)}\psi_{(0)} - \chi_{(0)}\psi_{(2)})
  \Bigr] x^5
  \nonumber \\ &&
  {}+ O(x^6), \\
{\frac{A}{4\pi q c_J}} &=&
  \frac{R_{(1)}}{24}
  (\chi_{(1)}\psi_{(0)} - \chi_{(0)}\psi_{(1)})x^2
  \nonumber \\ &&
  {}+ \frac{1}{180} \Bigl[
  7R_{(2)} (\chi_{(1)}\psi_{(0)} - \chi_{(0)}\psi_{(1)})
  \nonumber \\ &&
  {}+ 8R_{(1)} (\chi_{(2)}\psi_{(0)} - \chi_{(0)}\psi_{(2)})
  \Bigr]x^3
  \nonumber \\ &&
  {}+ O(x^4).
\end{eqnarray}

The expansion of $\Xi R$ is
\begin{eqnarray}
\label{XiRexpansion}
\Xi R &=& -\tfrac{1}{2}
  - \tfrac{\pi}{3} (\chi_{(1)}^2 + \psi_{(1)}^2) x^2
  \nonumber \\ &&
  {}+ \frac{2\pi}{45 R_{(1)}} \Bigl[
  R_{(2)} (\chi_{(1)}^2 + \psi_{(1)}^2)
  \nonumber \\ &&
  {}- 16 R_{(1)} (\chi_{(1)} \chi_{(2)}
  + \psi_{(1)} \psi_{(2)}) \Bigr] x^3
  \nonumber \\ &&
  {}+ O(x^4),
\end{eqnarray}
where we have already used the expansion of $R$ above. The expansion
of $R^2\Xi R$ takes the form
\begin{eqnarray}
\label{RXiRexpansion}
R^2\,\Xi R &=& -\frac{G_{(0)}^2}{2} x^2
  - \frac{G_{(0)} G_{(1)}}{2} x^3
  \nonumber \\ &&
  {}+ \frac{x^4}{72} \Bigl[ -9 G_{(1)}^2
  \nonumber \\ &&
  {}+ 8 G_{(0)} \bigl( -3 G_{(2)}
  + G_{(0)} \pi (\chi_{(1)}^2
  + \psi_{(1)}^2) \bigr) \Bigr]
  \nonumber \\ &&
  {}+ \mathcal{O}(x^5).
\end{eqnarray}
The two are of course equivalent for analytic functions, but we use
the latter in the code.

Next, we have the expansions for $R^{3/2}\hat{\Xi}\psi$ and
$R^{3/2}\hat{\Xi}\chi$, which appear in Eqs.~\eqref{Xipsieqn} and
\eqref{Xichieqn},
\begin{eqnarray}
R^{3/2}\hat{\Xi}\psi &=&
  \frac{\sqrt{R_{(1)}}}{2}\,\psi_{(1)}\,x^{3/2}
  \nonumber \\ &&
  {}+ \frac{3x^{5/2}}{20\sqrt{R_{(1)}}}
  \bigl(R_{(2)}\psi_{(1)} + 4R_{(1)}\psi_{(2)}\bigr)
  \nonumber \\ &&
  {}+ O(x^{7/2}), \\
R^{3/2}\hat{\Xi}\chi &=&
  \frac{\sqrt{R_{(1)}}}{2}\,\chi_{(1)}\,x^{3/2}
  \nonumber \\ &&
  {}+ \frac{3x^{5/2}}{20\sqrt{R_{(1)}}}
  \bigl(R_{(2)}\chi_{(1)} + 4R_{(1)}\chi_{(2)}\bigr)
  \nonumber \\ &&
  {}+ O(x^{7/2}).
\end{eqnarray}

Finally, we have
\begin{eqnarray}
\label{calHexpansion}
{\cal H} &=& \frac{2\pi}{R_{(1)}}
  (\chi_{(1)}^2 + \psi_{(1)}^2) x
  \nonumber \\ &&
  {}- \frac{2\pi}{5 R_{(1)}^2} \Bigl[
  3 R_{(2)} (\chi_{(1)}^2 + \psi_{(1)}^2)
  \nonumber \\ &&
  {}- 8 R_{(1)} (\chi_{(1)} \chi_{(2)}
  + \psi_{(1)} \psi_{(2)}) \Bigr] x^2
  \nonumber \\ &&
  {}+ O(x^3), \\
\label{Mexpansion}
{M} &=& \frac{c_M G_{(0)}^2 \pi}{3}
  \left( \chi_{(1)}^2 + \psi_{(1)}^2 \right) x^4
  \nonumber \\ &&
  {}+ \frac{2 c_M G_{(0)} \pi}{15} \Bigl[
  G_{(1)} (\chi_{(1)}^2 + \psi_{(1)}^2)
  \nonumber \\ &&
  {}+ 8 G_{(0)} (\chi_{(1)} \chi_{(2)}
  + \psi_{(1)} \psi_{(2)}) \Bigr] x^5
  \nonumber \\ &&
  {}+ O(x^6).
\end{eqnarray}


\section{The RN metric in 4+1 dimensions}
\label{RN}


The five-dimensional RN spacetime in static
coordinates is described by the line element
\begin{equation}
\label{RNmetric}
ds^2=-f(r)\,dt^2 + f(r)^{-1}\,dr^2 + r^2\,d\Omega_3^2,
\end{equation}
with the metric function
\begin{equation}
\label{fdef2}
f(r) := 1 - \frac{2\mathcal{M}_0}{c_M r^{2}}
+ \frac{c_Q^2 Q_0^{2}}{c_J r^{4}},
\end{equation}
where in this Appendix the real constants ${\cal M}_0>0$
and $Q_0$ are the mass and charge parameters of the RN solution. The
electromagnetic potential one-form, in our electromagnetic gauge, is
\begin{equation}
\label{RNpotential}
{A}_\mu=\left(-\frac{c_Q Q_0}{2r^2},\,0,\,0,\,0,\,0\right)
\end{equation}
The Hawking compactness of a sphere of areal radius $r$ satisfies
\begin{equation}
\label{fdef}
|\nabla r|^2=f(r)=1 - \frac{2\mathcal{M}_0}{c_M r^{2}}
+ \frac{c_Q^2 Q_0^{2}}{c_J r^{4}},
\end{equation}
The consistency condition between $c_M$, $c_J$ and $c_Q$ required for
$|Q_0|={\cal M}_0$ to represent an extremal black hole (where the
roots $r_\pm$ of $f(r)=0$ coincide) is
\begin{equation}
\label{extremalcond}
\frac{c_J}{c_M^2c_Q^2}=1.
\end{equation}
This seems to us an essential property of any choice of
convention. Assuming \eqref{extremalcond}, the event and Cauchy
horizons are located at
\begin{equation}
\label{appendix:r+-}
r_\pm =
  \sqrt{\frac{\mathcal{M}_0\pm\sqrt{\mathcal{M}_0^2-Q_0^2}}{c_M}}.
\end{equation}

From (\ref{Mdef}), $r_+=\sqrt{2M_0/c_M}$ on the event
  horizon. $M_0=r_+^2 c_M/2$ is also the irreducible mass of the
black hole. In the special case of an extremal black hole,
$r_+=r_-=\sqrt{2M_0/c_M}=\sqrt{{\mathcal M_0}/c_M}=\sqrt{Q_0/c_M}$ on
the event horizon.


\section{Derivation of the scaling of $\lambda_\text{trap}$}
\label{appendix:lambdatrapderivation}

The Raychaudhuri equation (\ref{Reqn}) along outgoing null cones can
be written in terms of the affine parameter $\lambda$ as
\begin{equation}
\label{Raychaudhuri}
R_{,\lambda\lambda}+{8\pi\over 3}|\phi_{,\lambda}|^2R=0,
\end{equation}
where we have used the definition (\ref{affine_parameter}). Note this
does not contain $Q$ or $q$, and that it holds in any spacetime
dimension, up to the constant factor in front of the second term. 

Assume that $\phi$ on the event horizon is mostly real (after a fixed
phase rotation), and that its dominant real part decays as a power,
that is
\begin{equation}
e^{-i\alpha}\phi\sim(\lambda-\lambda_0)^{-\sigma}
\end{equation}
for some universal constant $\sigma>0$ and some family-dependent
parameter $\alpha$. For exact fine-tuning to extremality
($p=p_1$), exactly on the event horizon ($u=u_\text{EH}$), and approximating the
undifferentiated $R$ in the second term in (\ref{Raychaudhuri}) as constant,
(\ref{Raychaudhuri}) then gives
\begin{equation}
\label{phifalloff}
R\simeq R_\text{EH}-c(\lambda-\lambda_0)^{-2\sigma},
\end{equation}
where $R_{\text{EH}}$ is the event horizon radius and $c>0$ is a
family-dependent constant, and hence
\begin{equation}
\label{Vfalloffextremal}
R_{,\lambda}\sim(\lambda-\lambda_0)^{-(2\sigma+1)}.
\end{equation}

More speculatively, assume further that the approximation
\begin{equation}
\label{Vfalloff}
R_{,\lambda}\simeq a(p-p_1)+b(\lambda-\lambda_0)^{-(2\sigma+1)}.
\end{equation}
holds {\em uniformly} over a 1-parameter family of solutions in the product
  of a fixed small interval of $p$ spanning $p=p_1$, a fixed small
interval of $u$ that includes the event horizon location, or
``would-be event horizon'' location \cite{GellesPretorius26}, and a
large interval of $\lambda$ (possibly including $\lambda\to\infty$),
with the family-dependent coefficients $a\ne0$ and $b>0$ depending
smoothly on $u$, $\lambda$ and $p$. We justify the first term in
$p-p_1$ in (\ref{Vfalloff}) by assuming that
$R_{,\lambda}(u,\lambda,p)$ depends smoothly and generically on $p$,
which means linearly to leading order in $p-p_1$, and that, by definition,
(\ref{Vfalloffextremal}) holds for $p=p_1$.

We note that $\lambda_\text{trap}(p)$ is defined by 
\begin{equation}
R_{,\lambda}[u_\text{trap}(p),\lambda_\text{trap}(p),p]=0,
\end{equation}
and substituting this into (\ref{Vfalloff}) and solving for $\lambda_\text{trap}(p)$ gives
\begin{equation}
\label{lambdatrapscaling}
\lambda_\text{trap}-\lambda_0\sim |p-p_1|^{-{1\over 2\sigma+1}}
\end{equation}
as $p\to p_{1e}$ (which is $p\to p_{1+}$ for $a<0$ and $p\to p_{1-}$
for $a>0$). 

We note that, up to a periodic wiggle, our numerical data in
Figs.~\ref{fig:psi_rot_chi_rot_power_scaling}-\ref{fig:lambda_trap_scaling}
perfectly fit (\ref{phifalloff}), (\ref{Vfalloffextremal}) and
(\ref{lambdatrapscaling}) for $\sigma=1/2$, and with the same
numerical choice of $\lambda_0>0$ in all three equations. 

$\sigma=1/2$ and the wiggle were also observed 
for (\ref{lambdatrapscaling}) in 3+1 dimensions by
\cite{GellesPretorius26}. Note again that our derivation holds for
arbitrary $D$, although $\sigma$ may depend on $D$.




\begin{figure*}[p]
\caption{Charge-to-mass ratio $Q/\mathcal{M}$ evaluated at the FMOTS
  as a function of $p$, for the 17 1-parameter families listed in
  Table~\ref{tab:perturbed_study}.  The multipliers $\hat p_1$ at
  extremality and the resulting value of
  $1-(Q/\mathcal{M})_{\text{FMOTS}}$ are also recorded there. The
  range of $p$ shown is chosen such that no trapped surfaces are
  present on the initial null slice $u=0$. Zoomed insets are provided
  where the threshold structure is not resolved in the full-scale
  plot. Where necessary, the sweep was repeated for the inset at a
  finer $\delta p$.}
\label{fig:allfamilies}
\begin{minipage}[b]{0.48\linewidth}
\includegraphics[width=\linewidth]{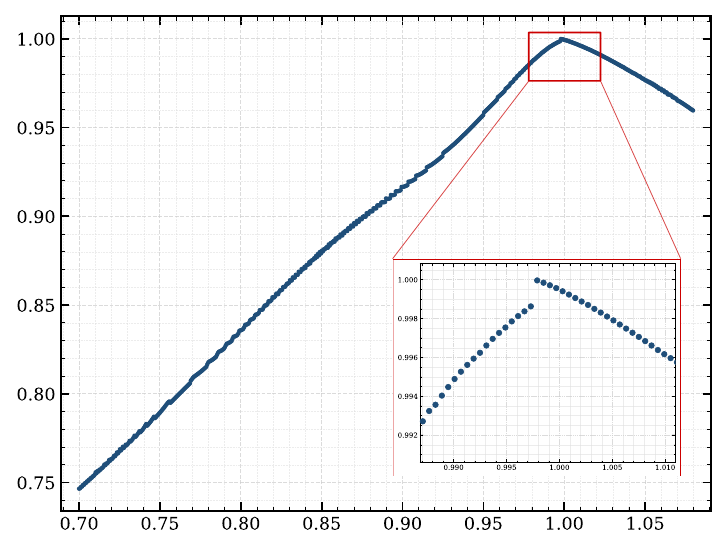}\\
        \vspace{0.1cm} $\mathcal{A}_2$ \label{fig:a2}
    \end{minipage}

    \vspace{0.5cm}
    
    \begin{minipage}[b]{0.48\linewidth}
        \centering
        \includegraphics[width=\linewidth]{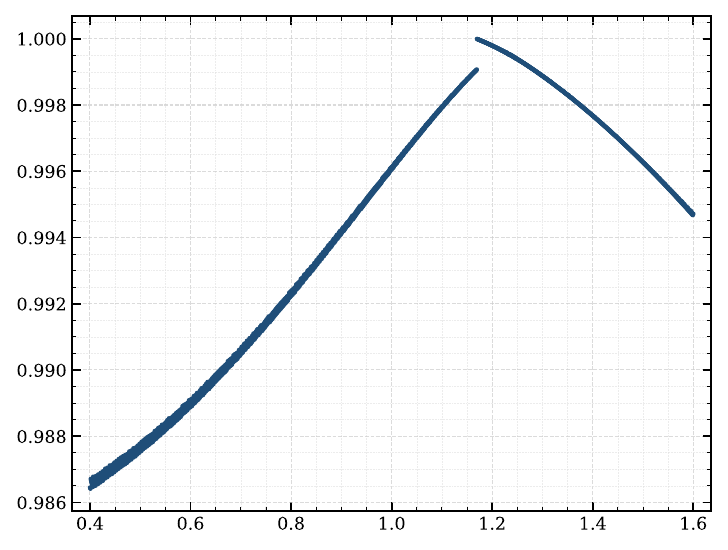}\\
        \vspace{0.1cm} $\mathcal{A}_1$ ($\mathcal{A}_2=0.0362$) \label{fig:a1_0.0362}
    \end{minipage}\hfill
    \begin{minipage}[b]{0.48\linewidth}
        \centering
        \includegraphics[width=\linewidth]{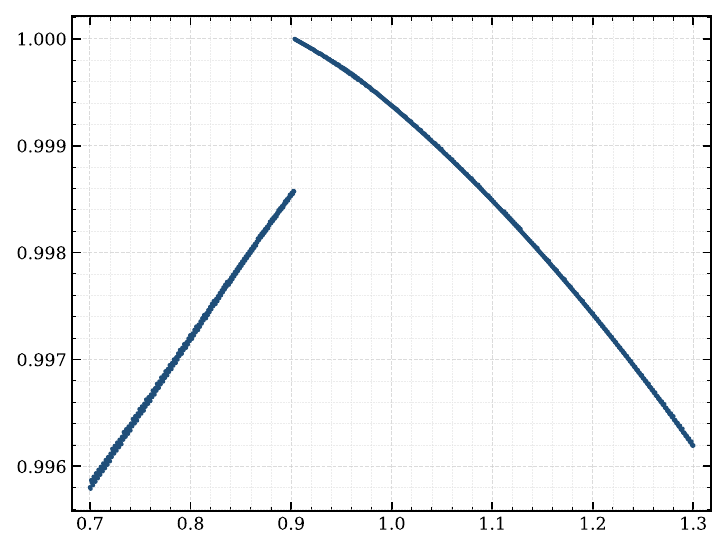}\\
        \vspace{0.1cm} $\mathcal{A}_1$ ($\mathcal{A}_2=0.0365$) \label{fig:a1_0.0365}
    \end{minipage}
    
    \vspace{0.5cm}
    
    \begin{minipage}[b]{0.48\linewidth}
        \centering
        \includegraphics[width=\linewidth]{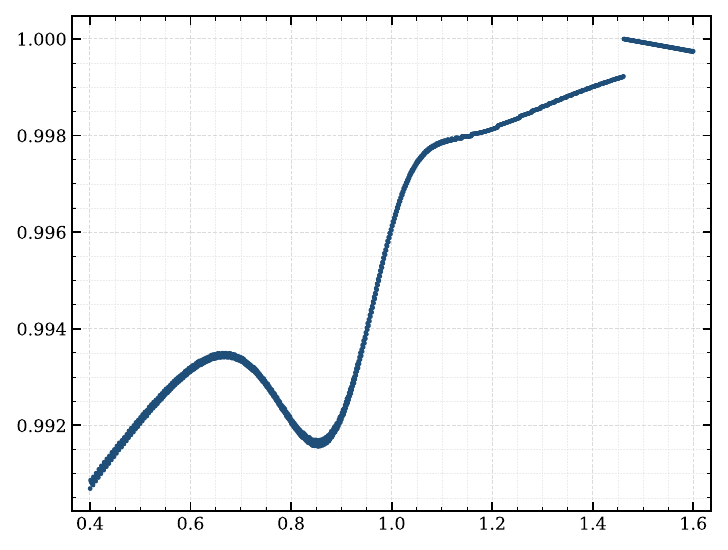}\\
        \vspace{0.1cm} $c_1$ ($\mathcal{A}_2=0.0362$) \label{fig:c1_0.0362}
    \end{minipage}\hfill
    \begin{minipage}[b]{0.48\linewidth}
        \centering
        \includegraphics[width=\linewidth]{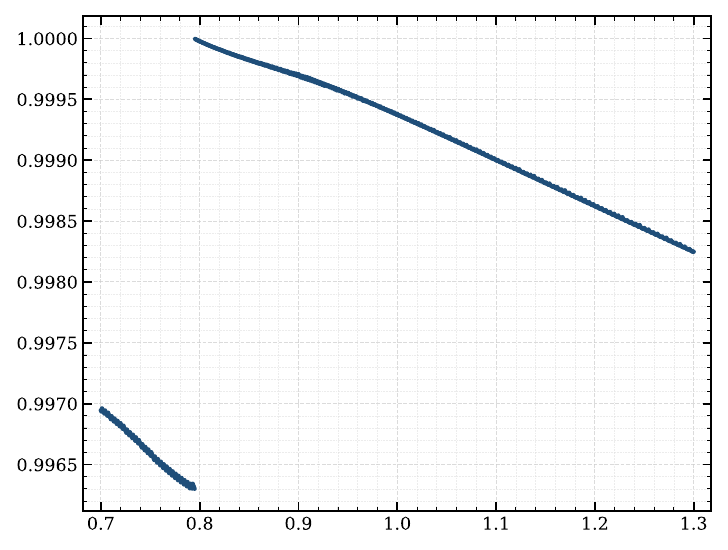}\\
        \vspace{0.1cm} $c_1$ ($\mathcal{A}_2=0.0365$) \label{fig:c1_0.0365}
    \end{minipage}
\end{figure*}

\addtocounter{figure}{-1} 
\begin{figure*}[p]
    \centering
    \vspace{0.3cm}
    
    \begin{minipage}[b]{0.48\linewidth}
        \centering
        \includegraphics[width=\linewidth]{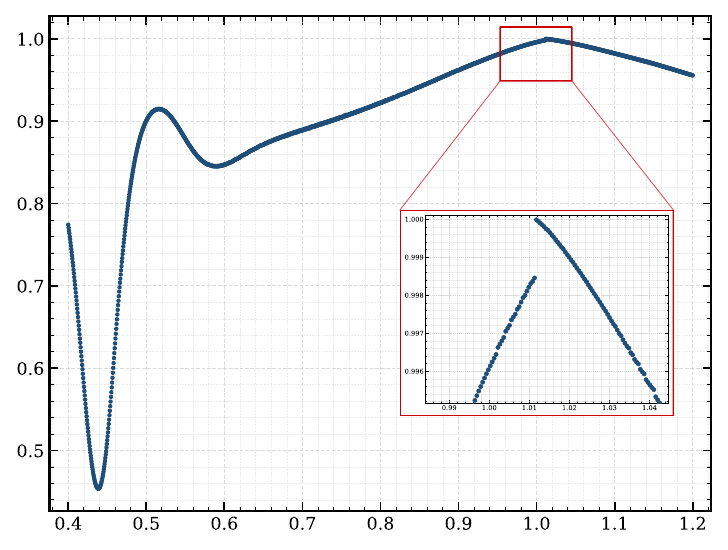}\\
        \vspace{0.1cm} $c_2$ ($\mathcal{A}_2=0.0362$) \label{fig:c2_0.0362}
    \end{minipage}\hfill
    \begin{minipage}[b]{0.48\linewidth}
        \centering
        \includegraphics[width=\linewidth]{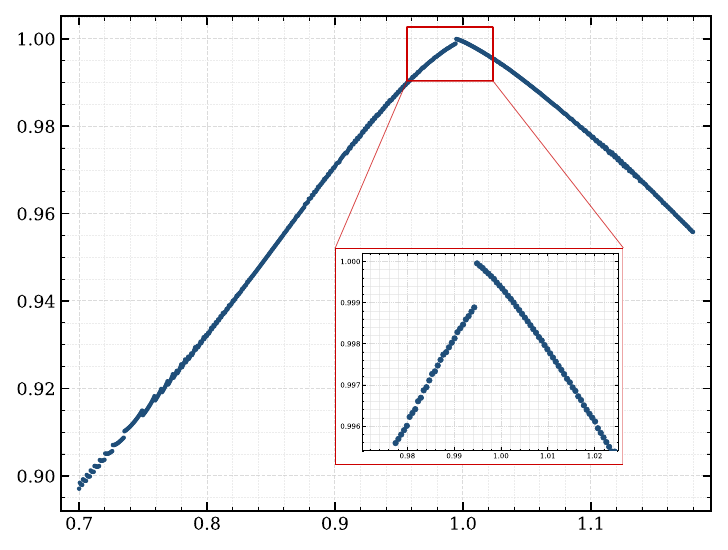}\\
        \vspace{0.1cm} $c_2$ ($\mathcal{A}_2=0.0365$) \label{fig:c2_0.0365}
    \end{minipage}
    
    \vspace{0.5cm}
    
    \begin{minipage}[b]{0.48\linewidth}
        \centering
        \includegraphics[width=\linewidth]{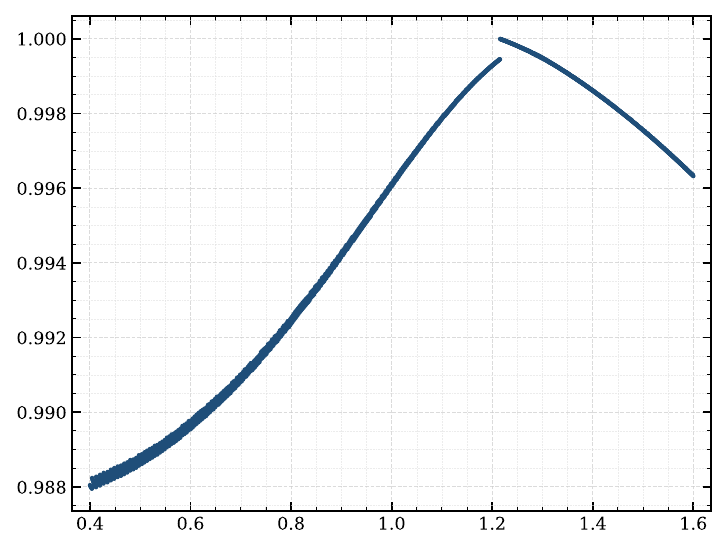}\\
        \vspace{0.1cm} $\omega_1$ ($\mathcal{A}_2=0.0362$) \label{fig:w1_0.0362}
    \end{minipage}\hfill
    \begin{minipage}[b]{0.48\linewidth}
        \centering
        \includegraphics[width=\linewidth]{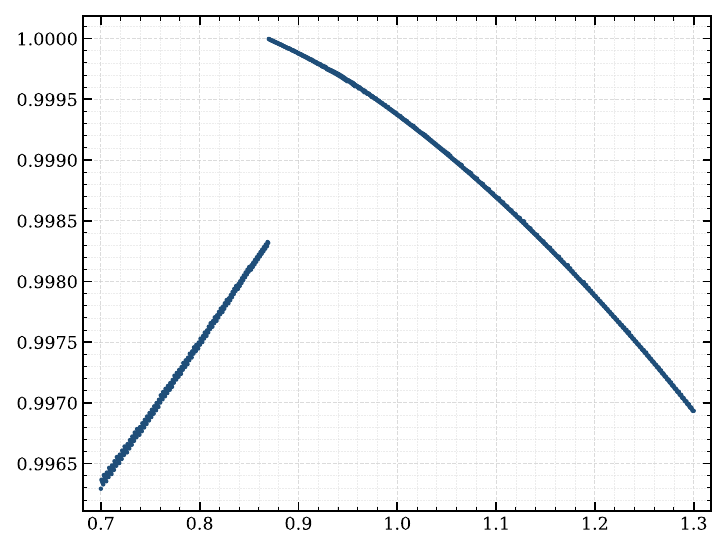}\\
        \vspace{0.1cm} $\omega_1$ ($\mathcal{A}_2=0.0365$) \label{fig:w1_0.0365}
    \end{minipage}

    \vspace{0.5cm}
    
    \begin{minipage}[b]{0.48\linewidth}
        \centering
        \includegraphics[width=\linewidth]{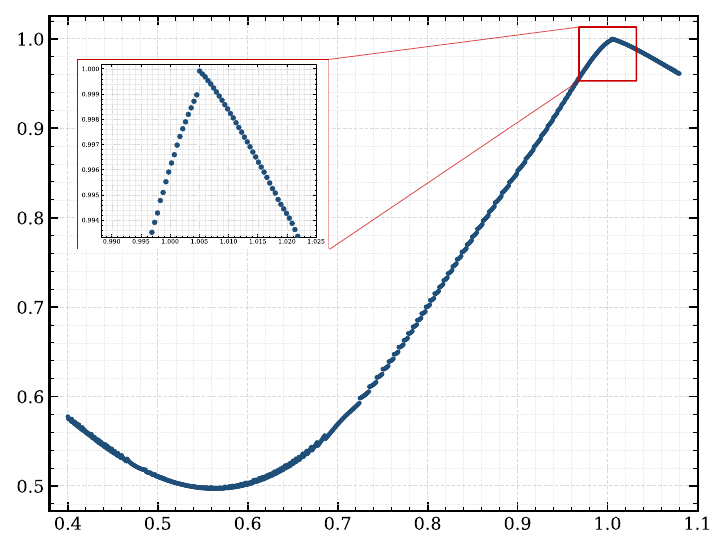}\\
        \vspace{0.1cm} $\omega_2$ ($\mathcal{A}_2=0.0362$) \label{fig:w2_0.0362}
    \end{minipage}\hfill
    \begin{minipage}[b]{0.48\linewidth}
        \centering
        \includegraphics[width=\linewidth]{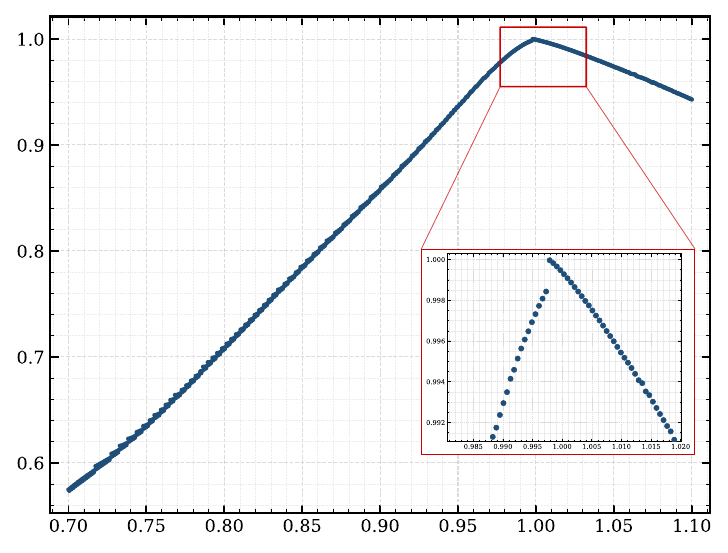}\\
        \vspace{0.1cm} $\omega_2$ ($\mathcal{A}_2=0.0365$) \label{fig:w2_0.0365}
    \end{minipage}
\end{figure*}

\addtocounter{figure}{-1} 
\begin{figure*}[p]
    \centering
    \vspace{0.3cm}

    \begin{minipage}[b]{0.48\linewidth}
        \centering
        \includegraphics[width=\linewidth]{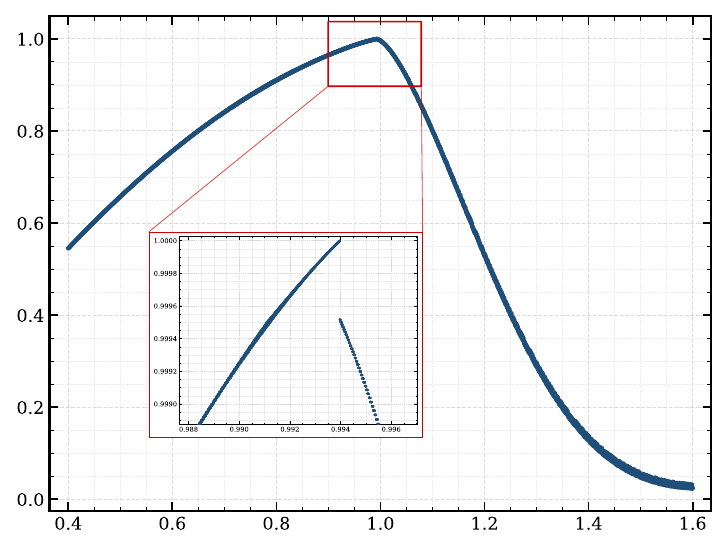}\\
        \vspace{0.1cm} $q$ ($\mathcal{A}_2=0.0362$) \label{fig:q_0.0362}
    \end{minipage}\hfill
    \begin{minipage}[b]{0.48\linewidth}
        \centering
        \includegraphics[width=\linewidth]{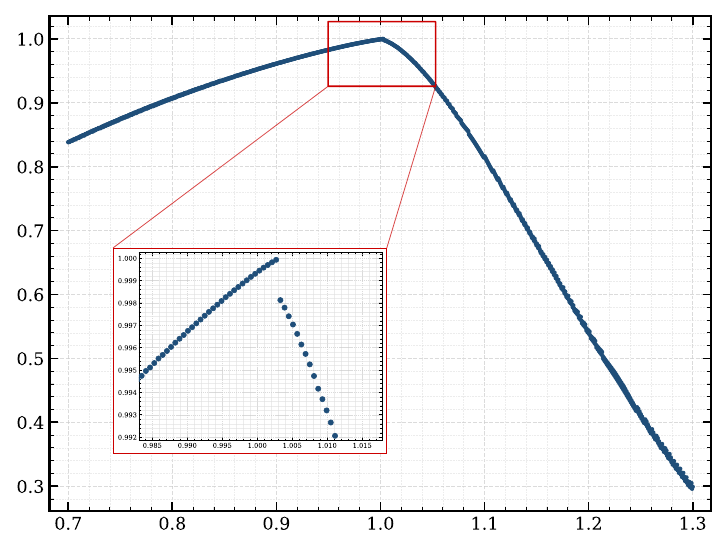}\\
        \vspace{0.1cm} $q$ ($\mathcal{A}_2=0.0365$) \label{fig:q_0.0365}
    \end{minipage}

    \vspace{0.5cm}

    \begin{minipage}[b]{0.48\linewidth}
        \centering
        \includegraphics[width=\linewidth]{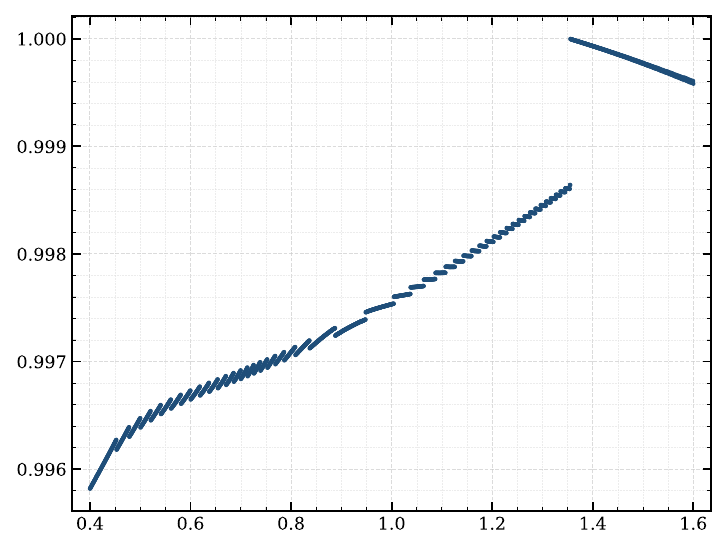}\\
        \vspace{0.1cm} $\sigma_1$ ($\mathcal{A}_2=0.0363$) \label{fig:sig1_0.0363}
    \end{minipage}\hfill
    \begin{minipage}[b]{0.48\linewidth}
        \centering
        \includegraphics[width=\linewidth]{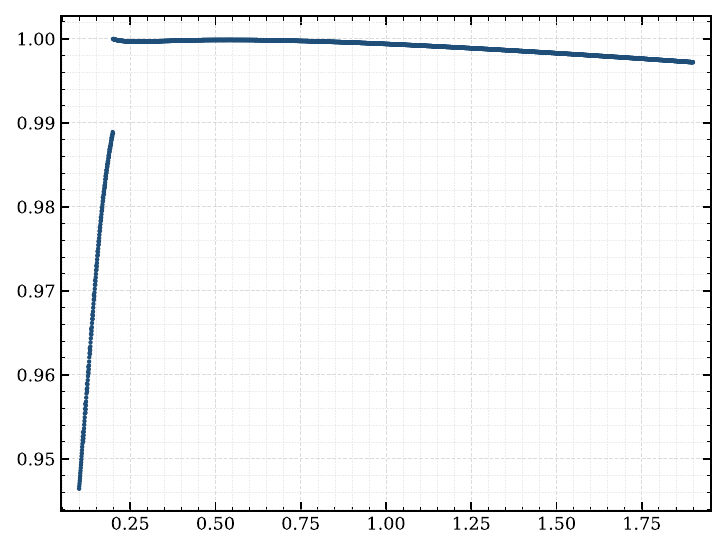}\\
        \vspace{0.1cm} $\sigma_1$ ($\mathcal{A}_2=0.0365$) \label{fig:sig1_0.0365}
    \end{minipage}

    \vspace{0.5cm}
    
    \begin{minipage}[b]{0.48\linewidth}
        \centering
        \includegraphics[width=\linewidth]{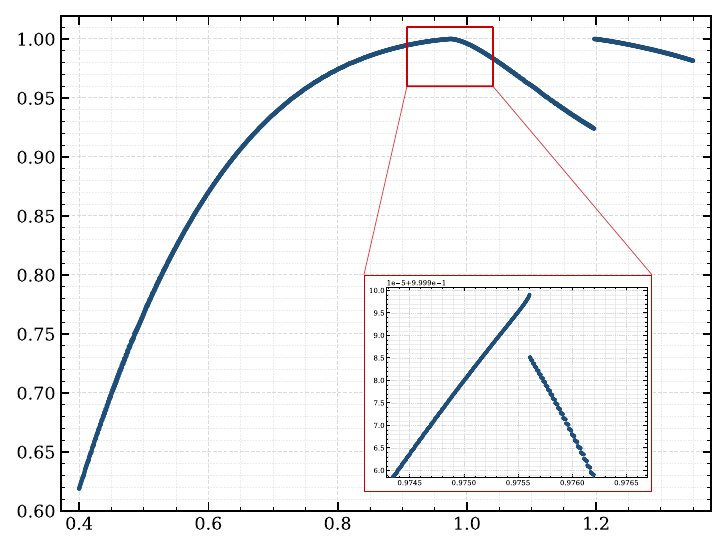}\\
        \vspace{0.1cm} $\sigma_2$ ($\mathcal{A}_2=0.0362$) \label{fig:sig2_0.0362}
    \end{minipage}\hfill
    \begin{minipage}[b]{0.48\linewidth}
        \centering
        \includegraphics[width=\linewidth]{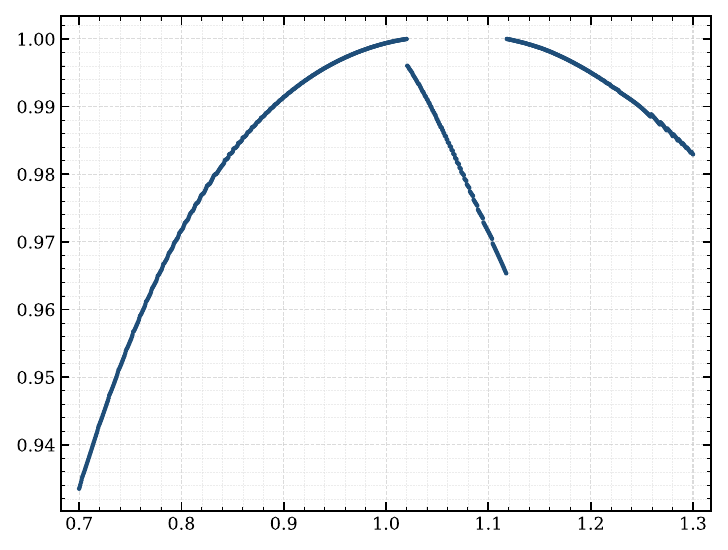}\\
        \vspace{0.1cm} $\sigma_2$ ($\mathcal{A}_2=0.0365$) \label{fig:sig2_0.0365}
    \end{minipage}
\end{figure*}


\end{document}